\documentclass[12pt]{article}
\usepackage[final]{ametsoc}
\usepackage{graphicx}
\usepackage{psfrag}
\usepackage{comment}
\usepackage{amstext}
\usepackage{amsmath}
\usepackage{amsthm}
\usepackage{amssymb}
\usepackage{color}
\renewcommand{\eqref}[1]{(\ref{eq:#1})}

\newcommand{\figref}[1]{Fig.~\ref{fig:#1}}

\newcommand{\secref}[1]{Sec.~\ref{sec:#1}}

\newcommand{\tabref}[1]{Tab.~\ref{tab:#1}}

\newcommand{\speedT}{{\Delta T_E}}
\newcommand{\be}[1]{{\widehat{#1}}}
\newcommand{\pbe}{{\be{p}}}
\newcommand{\qbe}{{\be{q}}}
\newcommand{\sibe}{{\be{\sigma}}}
\newcommand{\mube}{{\be{\mu}}}
\newcommand{\xibe}{{\be{\xi}}}
\newcommand{\lengthts}{{L}} \newcommand{\lengthbl}{{L_B}}

\title{Extreme Value Statistics of the Total Energy
  in an Intermediate Complexity Model of the Mid-latitude Atmospheric Jet.\\
  Part II: trend detection and assessment.}

\author{
  Mara Felici$^{1,2}$,
  Valerio Lucarini$^{1}$,
  Antonio Speranza$^{1}$,
  Renato Vitolo$^{1,}$
  \thanks{\textit{Corresponding author address:}
    Dr. Renato Vitolo, Department of Mathematics and
    Informatics, University of Camerino, via Madonna delle Carceri,
    62032 Camerino (MC), Italy. \newline
    E-mail: renato.vitolo@unicam.it}
}

\date{\today}

\begin{document}

\maketitle

\begin{center}
  \textit{
    $^1${
      PASEF -- Physics and Applied Statistics of Earth Fluids,\\
      Dipartimento di Matematica ed Informatica,
      Universit\`{a} di Camerino}\\[10pt]
    $^2${
      Dipartimento di Matematica U. Dini,
      Universit\`a di Firenze
    }
  }
\end{center}

\clearpage
\begin{abstract}
  A baroclinic model for the atmospheric jet at middle-latitudes is used as a
  stochastic generator of non-stationary time series of the total energy of
  the system.  A linear time trend is imposed on the parameter $T_E$,
  descriptive of the forced equator-to-pole temperature gradient and
  responsible for setting the average baroclinicity in the model.  The focus
  lies on establishing a theoretically sound framework for the detection and
  assessment of trend at extreme values of the generated time series. This
  problem is dealt with by fitting time-dependent Generalized Extreme Value
  (GEV) models to sequences of yearly maxima of the total energy.  A family of
  GEV models is used in which the location $\mu$ and scale parameters $\sigma$
  depend quadratically and linearly on time, respectively, while the shape
  parameter $\xi$ is kept constant. From this family, a model is selected by
  using diagnostic graphical tools, such as probability and quantile plots,
  and by means of the likelihood ratio test.  The inferred location and scale
  parameters are found to depend in a rather smooth way on time and,
  therefore, on $T_E$.  In particular, power-law dependences of $\mu$ and
  $\sigma$ on $T_E$ are obtained, in analogy with the results of a previous
  work where the same baroclinic model was run with fixed values of $T_E$
  spanning the same range as in this case. It is emphasized under which
  conditions the adopted approach is valid.
\end{abstract}

{PACS: 02.50.Tt, 02.70.-c, 47.11.-j, 92.60.Bh, 92.70.Gt}

\section{Introduction}
\label{sec:intro}
In the context of Climate Change, an intensely debated question is whether the
statistics of extreme meteo-climatic events is changing (and/or will change)
and, in case, how fast it is changing (and/or will change).  For example, the
role of time-dependence in the statistics of extreme weather events has been
at the heart of discussions about climate change since the work
by~\citet{KB92}.  In particular, the detection of trends in the frequency of
intense precipitation has been the object of much research, particularly at
regional level, see \emph{e.g.}~\citet{KKEQ96, KK98} for the USA
and~\cite{BMNN02,BBMMN04} for the Mediterranean area. The general relevance
of the problem has been highlighted in the 2002 release of a specific IPCC
report on \textit{Changes in extreme weather and climate events} (available at
http://www.ipcc.ch/pub/support.htm).  In fact, the emphasis laid on the
subject by the IPCC report reverberated in many countries the question ``is
the probability of major impact weather increasing?''.  This question reached
the big public almost everywhere and innumerable studies of trends in series
of ``extremes'' were undertaken.  These studies mainly deal with variables of
\emph{local character}, typically precipitation and temperature at specific
stations.  Moreover, most studies are regional: see \textit{e.g.}  the
proceedings of the Italia-USA meeting held in Bologna in 2004~\citep{DN06} for
the relevance of the extreme events in the Mediterranean Climates and the
INTERREG IIIB - CADSES project HYDROCARE,
\texttt{http://www.hydrocare-cadses.net}, for impacts of extreme events in the
hydrological cycle of the central-eastern Europe.

In a preceding, companion paper~\citep{FLSV06a} (which we refer to by Part I
in the sequel) we have addressed the problem of extreme value statistical
inference on statistically \emph{stationary} time series produced by a
dynamical system providing a minimal model for the dynamics of the
mid-latitudes baroclinic jet.  There reported is, from mathematical
literature, a suitable, rigorous, ``universal'' setting for the analysis of
the extreme events in stationary time series.  This is based on Gnedenko's
theorem~\citep{Gne43} according to which the distribution of the block-maxima
of a sample of independent identically distributed variables converges, under
fairly mild assumptions, to a member of a three-real parameter family of
distributions, the so-called Generalized Extreme Value (GEV)
distribution~\citep{Col01}.  The GEV approach to the analysis of extremse
requires that three basic conditions are met, namely the \emph{independence}
of the selected extreme values, the consideration of a \emph{sufficiently
  large} number of extremes, the selection of values that are \emph{genuinely}
extreme. This could be performed relatively easily for the case at hand.

Part I was originally motivated by the interest in weather having ``major
impact'' (on human life and property) in the Mediterranean area, in particular
intense precipitation and heat waves over Italy.  See, for
example,~\citet{BMNN02,BBMMN04,LNS04,LSN06,S06,ST06,TSDNBM06} and the MEDEX
Phase 1 report (available at \texttt{http://medex.inm.uib.es/}) for related
results and activities.  The study reported in Part I has revealed, among
other things, that diagnostics of extreme statistics can highlight interesting
dynamical properties of the analyzed system.  Properties which, thanks to the
``universality of the GEV'', can be investigated in a low dimensionality space
of parametric probability density functions, although at the expenses of the
total length of the observational record of the system in order to capture a
sufficient number of independent extremes.  A key role (that is presently
being explored elsewhere, in the context of general atmospheric circulation
theory) was played in Part I by the smoothness of variation of the extreme
statistics parameters (average, variance, shape factor) upon the external
(forcing) parameters of the system. In this paper, again, we devote attention
to exploring the statistics of extremes as a dynamical indicator, this time in
the framework of the (typically meteorological) statistical inference problem
of detecting trends in observations.

The definition of a rigorous approach to the study of extremes is much harder
when the property of stationarity does not hold.  One basic reason is that
there exists no universal theory of extreme values (such as \emph{e.g.} a
generalization of Gnedenko's theorem) for non-stationary stochastic processes.
Moreover, in the analysis of observed or synthetically generated sequences of
data of finite length, practical issues, such as the possibility of
unambiguously choosing the time scales which defines the statistical
properties and their changes, become of critical importance.  Nevertheless,
GEV-based statistical modeling offers a practical unified framework also for
the study of extremes in non-stationary time series.  In the applications, the
three parameters of the GEV distribution are taken as time-dependent and time
is introduced as a covariate in the statistical inference
procedure~\citep{Col01}.  The practical meaning of this assumption is that the
probability of occurrence (chance) of the considered extreme events evolves in
time pretty much as we are inclined to think in our everyday life.  However,
giving a scientific meaning to such an assumption is possible only in an
intuitive, heuristical fashion: in an ``adiabatic'' limit of infinitely slow
trends (but rigorously not even in such a limit). We adopt this point of view
not only because it is in line with the common practice and view of extremes,
but also because interesting dynamical properties can be inferred from
extremes, in analogy with the findings in Part I.

In the present paper we perform and assess time-dependent GEV inference on
non-stationary time series $E(t)$ of the total energy obtained by the same
simplified quasi-geostrophic model that was used in Part I.  The model
undergoes baroclinic forcing towards a given latitudinal temperature profile
controlled by the forced equator-to-pole temperature difference $T_E$;
see~\citet{LSV06,LSV06a} for a thorough description. We analyze how the
parameters of the GEV change with time when a linear trend is imposed on the
large scale macroscopic forcing $T_E$, that is, when $T_E$ is taken as a
(linear) function of time.  Since this functional relation is invertible, we
derive a parametrization relating the changes in the GEV to variations in
$T_E$ (instead of time).  One major goal here is to present a methodological
framework to be adopted with more complex models and with data coming from
observations, as well as an assessment of the performance of the GEV approach
for the analysis of trends in extremes in the somewhat \textit{grey area} of
non-stationary time series.  Methodologically, our set-up is somewhat similar
to that of~\citet{ZZL03} regarding the procedures of statistical inference.
However, in this case we face two additional problems:
\begin{enumerate}
\item
  as in Part I, the statistical properties of the time series $E(t)$
  cannot be selected \textit{a priori}: in the stationary case ($T_E$ constant
  in time) and much less in the non-stationary case there is no explicit
  formula for the probability distribution of the observable $E(t)$;
\item
  moreover, in the non-stationary case we even lack a \emph{definition}
  (and in fact a mere \emph{candidate}) of what might be the probability
  distribution of $E(t)$: certainly not a frequency limit for $t\to\infty$
  (and not by construction, as opposed to~\citet{ZZL03} who use genuinely
  stochastic generators).
\end{enumerate}
This also means that we have no hypothesis concerning the functional form of
the trend in the statistics of extremes of $E(t)$, resulting from the trend
imposed on the control parameter $T_E$. The lack of a general GEV theorem for
non-stationary sequences implies that the choice of the time-dependent GEV as
a statistical model is, in principle, arbitrary: other models might be equally
(or better) suitable.  Here comes into play the ``adiabaticity'' hypothesis
mentioned above, which leads to the central statement of this paper: if the
trend is sufficiently slow and if the statistical behaviour of the atmospheric
model has a sufficiently regular response with respect to variations in the
external parameters, the GEV remains a suitable model for inference of trend
in extremes.

The structure of the paper follows. In \secref{GEVproblems} we describe the
general problem of the characterization of statistical trends in deterministic
models, with both its conceptual and practical implications. Then in
\secref{GEVmethods} we describe how the GEV modeling can be applied to
non-stationary time-series and how the quality-check of the fits is performed.
In section \secref{data} we present the time series considered in this work
and the set-up of the numerical experiments performed with the atmospheric
model.  The inferences for various values of the trend in the forcing
parameter $T_E$ are presented in \secref{GEVinference} and a sensitivity
analysis is carried out in \secref{GEVassess}.  Comparison with the stationary
case analyzed in Part I is given in \secref{smooth}.  In \secref{conclusions},
finally, we summarize the main findings of this work, highlighting
the future research developments.

\section{Statistical trends: the theoretical problem}
\label{sec:GEVproblems}
The stochastic generator used in this paper to produce the time series is a
deterministic model (an ordinary differential equation), whose dynamics, for
the considered range of values of $T_E$, is chaotic in the sense that it takes
place on a strange attractor $\Lambda$ in phase space~\citep{ER}.
See~\cite{LSV06,LSV06a} for a study of the properties of this attractor,
including sensitivity with respect to initial conditions.  The statistical
behaviour of this type of time series is determined by the Sinai-Ruelle-Bowen
(SRB) probability measure $\mu$~\citep{ER}: this is a Borel probability
measure in phase space which is invariant under the flow $f^t$ of the
differential equation, is ergodic, is singular with respect to the Lebesgue
measure in phase space and its conditional measures along unstable manifolds
are absolutely continuous, see~\cite{LSY} and references therein.  Moreover,
the SRB measure is also a \emph{physical measure}: there is a set $V$ having
full Lebesgue measure in a neighbourhood $U$ of $\Lambda$ such that for every
continuous observable $\phi: U \to \mathbb{R}$, we have, for every $x\in V$,
the frequency-limit characterization
\begin{equation}
  \label{eq:physical}
  \lim_{t\to\infty}\frac{1}{t}\int_{0}^{t}\phi(f^t(x))dt=\int\phi d\mu.
\end{equation}
Existence and uniqueness of an SRB measure $\mu$ have been proved only for
very special classes of flows $f^t$ (in particular, for flows that possess an
Axiom-A attractor, see~\cite{LSY}). However, existence and uniqueness of $\mu$
are \emph{necessary} to define a stationary stochastic process associated to
an observable $\phi$. In turn, this allows to consider a given time series of
the form ${\phi(f^t(x)): t>0}$ as a realization of the stationary process,
justifying statistical inference on a solid theoretical basis. In part I, we
conjectured existence an uniqueness of an SRB measure for the atmospheric
model, providing the theoretical foundation to the application of GEV models
in the inference of extreme values.

In certain cases of non-autonomous ordinary differential equations (for
example, if the dependence on time is periodic), it still is possible to
define, at least conceptually, what an SRB measure is.  However, in the
present case, due to the form of time-dependence adopted for the parameter
$T_E$, the atmospheric model admits \emph{no invariant measure}.  This means
that there is no (known) way to associate a stochastic process to the time
series of the total energy. In other words, it is even in doubt what we mean
by ``statistical properties'' of the time series, since it is
\emph{impossible} to define a probability distribution.  This conceptual
problem has a very serious practical consequence: the ``operational''
definition of probability as a frequency limit (as in~\eqref{physical})
\emph{is not valid in the non-stationary case}, since the time evolution is
not a sampling of a unique probability distribution.  Even if one assumes the
existence of a sequence of distinct probability distributions, one for each
observation, one realization (the time series) does not contain sufficient
statistical information, since each distribution is very undersampled (with
only one observation).

Despite all these problems, the results in Part I suggest a framework which
is, for the moment, formulated in a heuristic way. Suppose you evolve an
initial condition $x$ in phase space by the flow $f^t$ of the
\emph{autonomous} atmospheric model, that is the system in which $T_E$ is kept
fixed to some value $T_E^0$.  After an initial time span (transient), say for
$t$ larger than some $t_0>0$, the evolution $f^t(x)$ may be thougth to take
place on the attractor $\Lambda$ and time-averages of the form
\begin{equation}
  \label{eq:time-average}
  \frac{1}{t-t_0}\int_{t_0}^{t}\phi(f^t(x))dt
\end{equation}
may be considered as approximations of
\begin{equation}
  \label{eq:SRB-average}
  \int\phi d\mu_0,
\end{equation}
which is the average of $\phi$ by the SRB measure $\mu_0$
existing at the value $T_E=T_E^0$ (the ``attractor average'' at $T_E^0$).
Now suppose that at some $t_1>t_0$ the parameter $T_E$ is abruptly changed
to some value  $T_E^1>T_E^0$: there will be some transient interval,
call it $[t_1,t_2]$, during which the evolution $f^t(x)$ approaches
the \emph{new} attractor of the atmospheric model, that is
the attractor existing for $T_E$ fixed at $T_E^1$.
After that time span, the evolution may be considered to take place
on the new attractor.

In our case, though $T_E$ varies continuously (linearly) with time, if the
trend magnitude is low, then $T_E$ may be considered constant (with good
approximation) during time spans that are sufficiently long in order to have
both convergence to the ``new'' attractor and good sampling of the ``new'' SRB
measure, in the sense sketched above.  Though admittedly heuristic, this
scenario allows to clarify under which condition it is still reasonable to
speak of ``statistical properties'' of a time series generated by a
non-autonomous system: namely, the closeness to a stationary situation for
time spans that are sufficiently long. This is the ``adiabatic'' hypothesis
which we mentioned in the introduction. An essential ingredient for this to
hold is that the statistical properties of the autonomous model \emph{do not
  sensibly depend on the external parameter $T_E$}, in the sense that no
abrupt transitions (bifurcations) should take place as $T_E$ is varied.  This
was indeed checked for the system at hand in Part I.  Notice that the validity
of the ``adiabatic'' hypothesis also has a useful practical consequence: one
can use the statistics of the stationary system as a reference against which
the results for the non-stationary case can be assessed.  Having this scenario
in mind, we proceed to the description of the time-dependent GEV approach in
the next section.

\section{GEV modelling for non-stationary time series}
\label{sec:GEVmethods}

\subsection{Stationary case}
\label{sec:GEVmethodsstationary}
The GEV approach for sequences of independent, identically distributed
(i.i.d.) random variables is by now rather
standard~\citep{Cas88,Col01,EKM97,FHR94,Gal78,Gum58,Jen55,LLR83,RT01,Tia84}.
The foundation is Gnedenko's theorem~\citep{Gne43,FT28}.  Consider a time
series, assumed to be a realization of a sequence of i.i.d.  random variables.
The time series is divided into $m$ consecutive time-frames (data blocks),
each containing $n$ subsequent observations, equally spaced in time.  Denote
by $z_{1},\ldots,z_{m}$ the sequence of the block maxima taken over each
time-frame.  Under fairly mild assumptions, the distribution of the
block-maxima converges, in a suitable limit involving a rescaling, to the GEV
distribution, defined as
\begin{equation}
  \label{eq:GEV}
  G(x;\mu,\sigma,\xi)= \exp \left\{ - \left[
      1+ \xi\left( \frac{x-\mu}{\sigma} \right) \right]^{-1/ \xi}\right\}
\end{equation}
for all $x$ in the set $\{ x:\, 1+\xi(x-\mu)/\sigma \,>\,0\}$
and $G(x)=0$ otherwise, where $\sigma >0$ and $\xi\neq0$.
If $\xi=0$ the limit distribution is the Gumbel distribution
\begin{equation}
  \label{eq:Gumbel}
  G(x;\mu,\sigma,0)= \exp \left( - \exp \left( -
  \frac{x-\mu}{\sigma} \right) \right), \quad x\, \in \mathbb{R}.
\end{equation}
GEV inference consists in estimating the distributional parameters
$(\mu,\sigma,\xi)$ (called  location, scale and shape parameter,
respectively) from the available data. A widely used technique
consists in numerically maximizing the \emph{log-likelihood} function
$l(\mu,\sigma,\xi)$. For $\xi\neq0$, this is defined as
\begin{multline}
  \label{eq:logllGEV}
  l(\mu,\sigma,\xi)=\\
  -m\,\log \, \sigma -\left(1+ \frac{1}{\xi}\right)
  \sum_{t=1}^m \left \{
    \log \, \left[ 1+ \xi \left (\frac{z_{t}-\mu}{\sigma} \right) \right]
    - \left[  1+ \xi \left (\frac{z_{t}-\mu}{\sigma} \right)
    \right]^{-\frac{1}{\xi}}
  \right \},
\end{multline}
while an analogous formula holds  for $\xi=0$~\citep{Col01}.

A generalization of the GEV theorem holds for time series
that are realizations of stationary stochastic processes
such that the long-range dependence is weak
at extreme levels~\citep{Lea74,Lea83}.
In the applications, this property is assumed to hold
whenever the block maxima are uncorrelated for sufficiently
large block sizes.
In this case, GEV inference and assessment is carried out
by the same tools (maximum likelihood, diagnostic plots, etc.)
used in the i.i.d. context, see~\citet{Col01}.

\subsection{Time-dependent case}
\label{sec:GEVmethodstime}

If stationarity of the time series does not hold, then the limiting
distribution function is no longer bound to be the GEV or any other
family~\eqref{GEV}: no theories of extreme values exist in this context.  Some
exact results are known only in certain very specialized types of
non-stationarity~\citep{Hus86,LLR83}, but it is very unlikely that a general
theory can be established.  However, GEV-based statistical modeling of extreme
values can be performed also in the case of time-dependent phenomena by
adopting a pragmatic approach, where the GEV distribution~\eqref{GEV} is used
as a template: time-dependent parameters $\mu(t)$ and $\sigma(t)$ are
considered, yielding a GEV model of the form
\begin{equation}
  \label{eq:GEVt}
  G(x;\mu(t),\sigma(t),\xi).
\end{equation}
Usually $\xi$ is kept time-independent in order to avoid numerical problems,
since it is usually the most delicate parameter to estimate~\citep{Col01}.
Different kinds of time-dependence can be imposed for $(\mu(t),\sigma(t))$.
In this paper, we adopt a simple polynomial family of models:
\begin{equation}
  \label{eq:musigmat}
  \mu(t)=\mu_0+\mu_1 t+\mu_2 t^2,\qquad
  \sigma(t)=\sigma_0+\sigma_1 t,
\end{equation}
with $\mu_{0,1,2}$ and $\sigma_{0,1} \in \mathbb{R}$.
GEV models in family~\eqref{musigmat} are denoted by $G_{p,q}$,
with $0\le p\le 2$ and $0\le q\le 1$, where $p$ and $q$ denote the
maximum degree of $t$ in $\mu(t)$ and in $\sigma(t)$, respectively.
The time-dependent GEV model~\eqref{GEVt} constructed in this way
is a generalization of~\eqref{GEV} (the latter is obtained
by setting $\mu_1=\mu_2=\sigma_1=0$ in~\eqref{musigmat}).
For model~\eqref{GEVt} with parameters~\eqref{musigmat},
GEV inference amounts to estimating the parameter vector
\begin{equation}
  \label{eq:beta}
  \boldsymbol{\beta}=[\mu_0,\mu_1,\mu_2,\sigma_0,\sigma_1,\xi]
\end{equation}
by including time $t$ as a covariate.

Suppose we have a non-stationary dataset, from which a sequence
of block maxima $z_{t},$ with $t=1,\ldots,m$, is constructed.
A log-likelihood function for the case $\xi\neq0$ is defined as
\begin{multline}
  \label{eq:logllGEVt}
  l(\boldsymbol{\beta})= - \displaystyle
  \sum_{t=1}^m \left\{ \log \, \sigma(t)
    +(1+ 1/ \xi) \log \, \left[ 1+ \xi \left (\frac{z_t
          -\mu(t)}{\sigma(t)} \right) \right]+ \right.\\
  \left. +\left[ 1+ \xi \left (\frac{z_t -\mu(t)}{\sigma(t)} \right)
    \right]^{-1/\xi}\right\}
\end{multline}
(compare with~\eqref{logllGEV}), provided that
\begin{equation}\label{eq:domainlogllGEVt}
  1+ \xi \left(\frac{z_t - \mu(t)}{\sigma(t)} \right)\; > 0, \quad
  i=1,\ldots,m.
\end{equation}
If $\xi=0$, an alternative log-likelihood function,
derived from the Gumbel distribution, must be used~\citep{Col01}.
Numerical procedures are used to maximize the selected
log-likelihood function, yielding the maximum likelihood estimate
of the parameter vector $\boldsymbol{\beta}$.
Confidence intervals for $\boldsymbol{\beta}$ may be computed
by the expected or observed information matrix~\citep{Col01}.

Of course, all of the above procedure is performed in the spirit of ``pure''
inference, that is determining the likelihood of the adopted parametric
hypothesis and \emph{not} its truth which, in the absence of a supporting
theorem, remains unknown. Moreover, it should be kept in mind that several
different models might fit the observations with similar reliability
(likelihood). In this case, as no \emph{universal} model is suggested or
enforced (as opposed to the stationary case), there is no reason to
prefer the one above the other.

\subsection{Assessment of statistical models}
\label{sec:GEVmethodstimeassess}
In the non-stationary context the analysis starts from a list of models
($G_{p,q}$ in our case, see~\eqref{musigmat}) which we fit to the data
searching for the most adequate one.
Assessment and comparisons of the inferences are based on
standard graphical tools such as \emph{probability plot, quantile plot},
and the \emph{likelihood ratio test}.
However, for the graphical model-checking the non-stationarity must
be taken into account. Reduction to \emph{Gumbel scale} is a practical
way to treat this problem~\citep{Col01}.

Let $z_t$, $t=1,\ldots,m$ be a sequence of block maxima
extracted from a non-stationary time series,
from which the time-dependent GEV model $G(\mube(t),\sibe(t),\xibe\,)$
has been fitted as described in the previous section.
The sequence of maxima is transformed according to
\begin{equation}
  \label{eq:Gumbelscale}
  \widetilde{z}_t=\log\left[\left(1+\xibe\left(
        \frac{z_t-\mube(t)}{\sibe(t)}\right)\right)^
    {-\frac{1}{\xibe}}\right], \qquad t=1,\ldots,m.
\end{equation}
If $Z_t$ are random variables with distribution $G(\mube(t),\sibe(t),\xibe)$,
then the transformation~\eqref{Gumbelscale} produces variables
$\widetilde{Z}_t$ that have the standard Gumbel distribution:
\begin{equation}
  \label{eq:GumbelDF}
  P(\widetilde{Z}_t<x)=\exp\,(-e^{-x}),\quad x\in \mathbb{R},
\end{equation}
which is the GEV with parameters $(\mu,\sigma,\xi)=(0,1,0)$. Therefore,
transformation~\eqref{Gumbelscale} attempts to remove the time-dependence from
the sequence of maxima, bringing it as close as possible to the common scale
given by the standard Gumbel distribution~\eqref{GumbelDF}. This way, the
distribution function and the quantiles of the transformed sequence of maxima
$\widetilde{z}_t$ can be compared with the empirical ones derived
from~\eqref{GumbelDF}. The probability plot is a graph of the pairs
\begin{equation}
  \label{eq:PPplot}
  \left(\frac{j}{m+1};\exp\,(-e^{-\widetilde{z}_{(j)}} ) \right ),\quad
  j=1,\ldots, m,
\end{equation}
where
$\widetilde{z}_{(1)}\le\widetilde{z}_{(2)}\le\dots\le\widetilde{z}_{(m)}$ is
the order statistics of the transformed sequence of maxima $\widetilde{z}_t$.
The quantile plot is given by the pairs
\begin{equation}
  \label{eq:QQplot}
  \left(-\log\left(-\log\left(\frac{j}{m+1}\right)\right);\;\widetilde{z}_{(j)}
  \right ),\quad j=1,\ldots, m.
\end{equation}
For both plots, displacement of points from the diagonal indicates low quality
of the inference.

The \textbf{likelihood ratio test} is used to compare the goodness-of-fit of
two \emph{nested models}, that is, two models such that one of them is a
sub-model (a particular case) of the other one. Our family $G_{p,q}$ of models
is nested: for example $G_{1,0}$ is a sub-model of $G_{2,1}$, obtained by
setting $\mu_2$ and $\sigma_1$ to zero in the parameter vector
$\boldsymbol{\beta}$ defined in~\eqref{beta}. Given integers $0\le p_1\le
p_2\le 2$ and $0\le q_1\le q_2\le 1$, let $l_1$ and $l_2$ be the maximized
values of the log-likelihood~\eqref{logllGEVt} for the nested models
$G_{p_1,q_1}$ and $G_{p_2,q_2}$, respectively, and define the \textbf{deviance
  statistic} as
\begin{equation}
  \label{eq:devstat}
  \mathcal{D}=2\{l_{2}-l_{1}\}.
\end{equation}
The likelihood ratio test states that the simpler model
$G_{p_1,q_1}$ is to be rejected at the $\alpha$-level of confidence
in favor of $G_{p_2,q_2}$ if $\mathcal{D}>c_{\alpha}$, where $c_{\alpha}$
is the $(1-\alpha)$-quantile of the $\chi_k^2$ distribution and
$k$ is the number of parameters that belong to $G_{p_2,q_2}$
and that are null in $G_{p_1,q_1}$ ($k=2$ in our example above).

Although the number of time-dependent models one may choose from is
virtually infinite, parsimony in the construction is
reccommended~\citep{Col01}: too many coefficients in the functions
$(\mu(t),\sigma(t))$ would result in unacceptably large uncertainties,
especially if few data are available.
The search of the best model is carried out by trial-and-error:
the choice of a more complex model should be strongly justified
on theoretical grounds or by a significantly higher accuracy
(that is, $\mathcal{D}$ exceedingly larger than $c_{\alpha}$ for nested
models).
However, the convergence of the above described procedure is
\emph{by no means} a guarantee of good estimate of the ``true''
probability density function: the latter is conceptually \emph{undefined}.
See our remarks at the end of \secref{GEVmethodstime}.

\section{The time series: Total Energy of the Atmospheric Jet Model
  with a trend in average baroclinicity}
\label{sec:data}

We consider here the same baroclinic jet model used in Part I, where
the spectral order $JT$ is set to $32$.  The model temperature is relaxed
towards a given equator-to-pole profile which acts as baroclinic forcing.  The
statistical properties of the model radically change when the parameter $T_E$,
determining the forced equator-to-pole temperature gradient, is varied.  A
physical and dynamical description of the model is given
in~\citet{SM88,MTS90,LSV06,LSV06a}.

In Part I we performed an extreme value analysis of the system's response with
respect to variations in $T_E$.  Several stationary time series of the total
energy $E(t)$ were used as a basis for GEV inference.  Each time series was
generated with $T_E$ fixed at one value within a uniform grid on the interval
$[10,50]$, with spacing of 2 units.  We recall that, given the
non-dimensionalization of the system, $T_E=1$ corresponds to
$3.5\,\mathrm{K}$, 1 unit of total energy corresponds to roughly
$5\times10^{17} \mathrm{J}$, and $t=0.864$ is one day,
see~\citet{LSV06,LSV06a}.  In that case, all parameters of the system being
kept fixed, after discarding an initial transient each time series of the
total energy could be considered as a realization of a stationary stochastic
process having weak long-range dependence.  Therefore, the classical framework
for GEV modeling was applied (see \secref{GEVmethodsstationary}).

In the present setting, a specific linear trend is imposed on $T_E$: starting
at time $t=0$, the model is run with a the time-dependent forcing parameter
\begin{equation}
  \label{eq:TEt}
  T_E(t)=(T_E^0-1)+t\,\speedT, \qquad t\in[0,t_0],
\end{equation}
with $T_E^0=10$.
Three values are chosen for the trend intensity $\speedT$:
$2$ units every $\lengthbl=$ $1000$, $300$, and $100$ years,
yielding three time series for the total energy $E(t)$.
The range swept by $T_E(t)$ during integration is kept fixed
in all three cases to the interval $[9,51]$, so that the total
length of the time series depends on $\speedT$.
Each time series is split into 21 data blocks $B^i$, $i=1,\dots,21$.
The length $\lengthbl$ of each block corresponds to a time interval $I^i$
such that, as $t$ varies within $I^i$, the baroclinicity parameter $T_E(t)$
by~\eqref{TEt} spans the interval
\begin{equation}
  \label{eq:TEintervals}
  [T_E^i-1,T_E^i+1],
\end{equation}
which is 2 units wide and centered around one of the
values $T_E^i$ considered in Part I:
\begin{equation}
  \label{eq:TEi}
  (T_E^0,T_E^1,\dots,T_E^{21})=
  (10,12,14,\dots,50).
\end{equation}
Therefore, the total length $\lengthts$ of the time series
depends on the trend intensity, so that we have
$\lengthts$ $=$ $21\times2/\speedT$ $=$ $21\lengthbl$.
Moreover, since the time-span over which the maxima are computed
is kept fixed to one year, the number of maxima in
each data block $B^i$ also depends on $\speedT$:
in fact, it is equal to $\lengthbl$, see \tabref{data}.

Such a selection of the intervals as in~\eqref{TEintervals}
allows for a direct comparison of the present results
with those obtained for stationary time series
in Part I.
Moreover, our choices regarding block length and other factors
are based on the indications provided in Part I,
where the goodness-of-fit assessments
performed by a variety of means showed that:
\begin{itemize}
\item
  the adopted block length of one year makes sure that
  the extremes are uncorrelated and \emph{genuinely extreme};
\item
  the minumum length (100 data) used for the sequences of maxima
  yields robust inferences.
\end{itemize}

\section{Time-dependent GEV Analysis of the Total Energy}
\label{sec:GEVinference}

\begin{figure}[hpt]
  \centering
  \begin{tabular}{cccc}
    $\qquad(p,q)=(0,0)$&$\qquad(p,q)=(1,0)$&$\qquad(p,q)=(1,1)$&$\qquad(p,q)=(2,1)$\\[3pt]
    $\qquad l=-8017.32$&$\qquad l=-7384.38$&$\qquad l=-7382.56$&$\qquad l=-7655.70$\\[3pt]
    \includegraphics[scale=0.18]{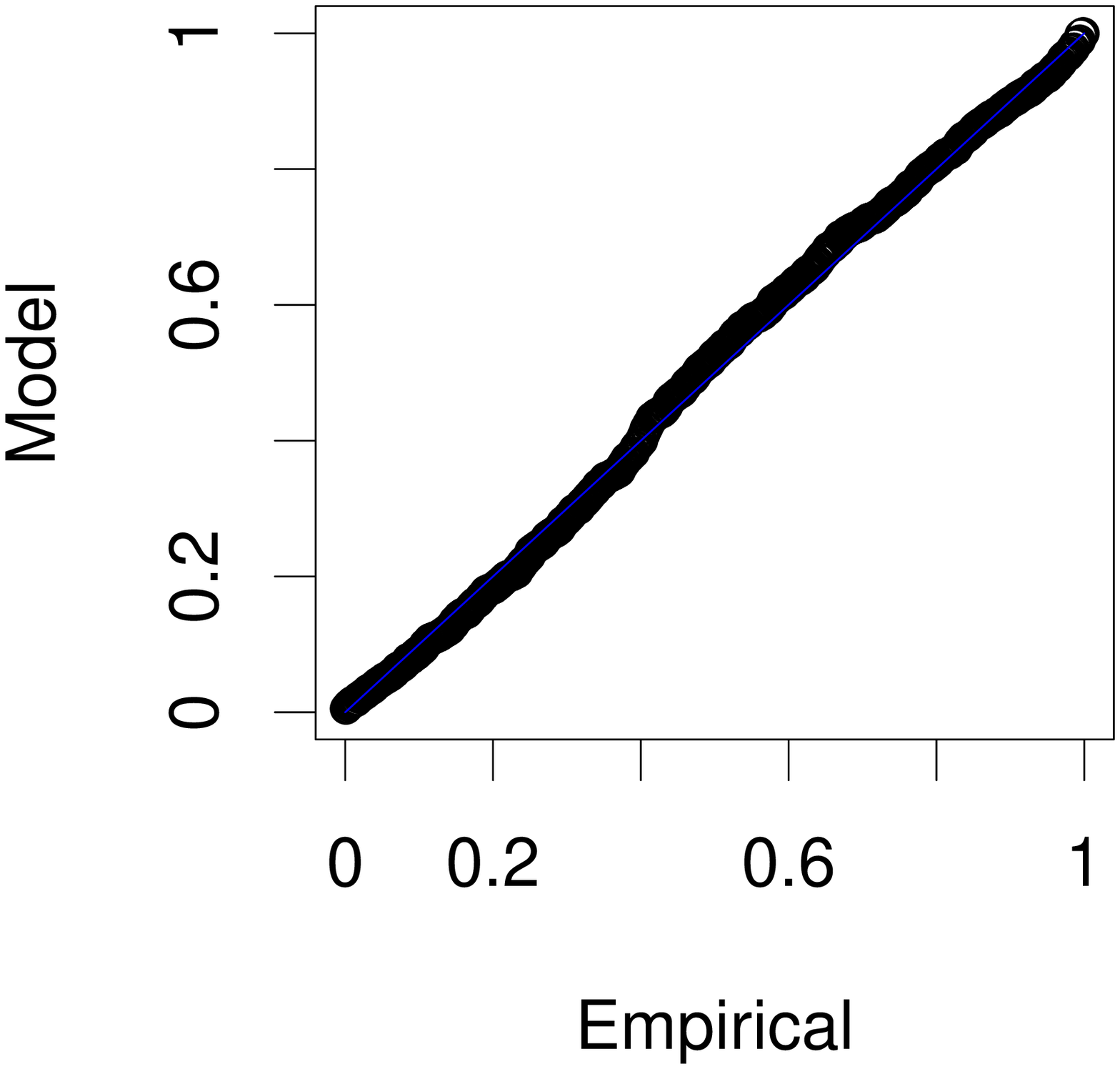}&
    \includegraphics[scale=0.18]{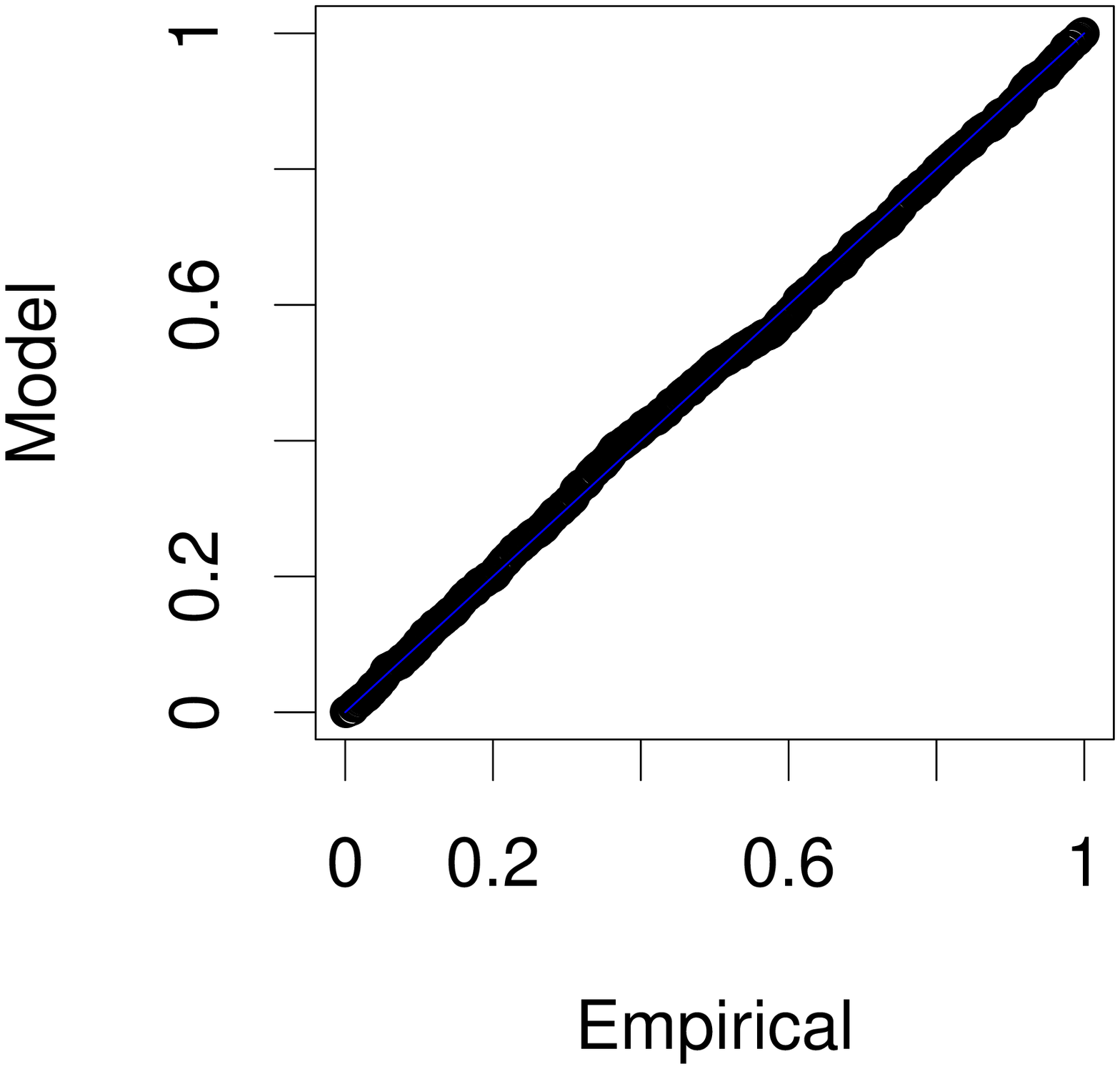}&
    \includegraphics[scale=0.18]{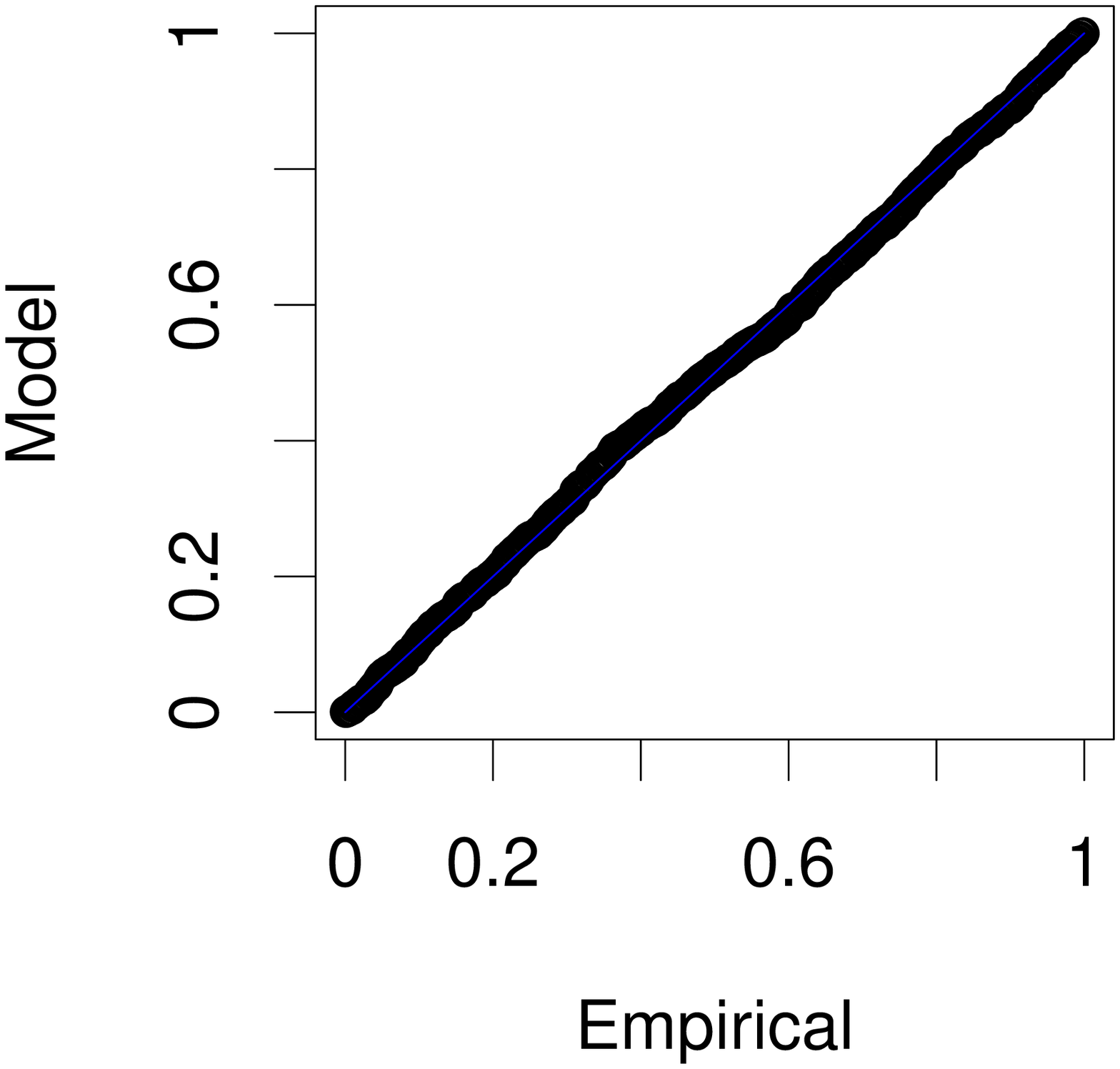}&
    \includegraphics[scale=0.18]{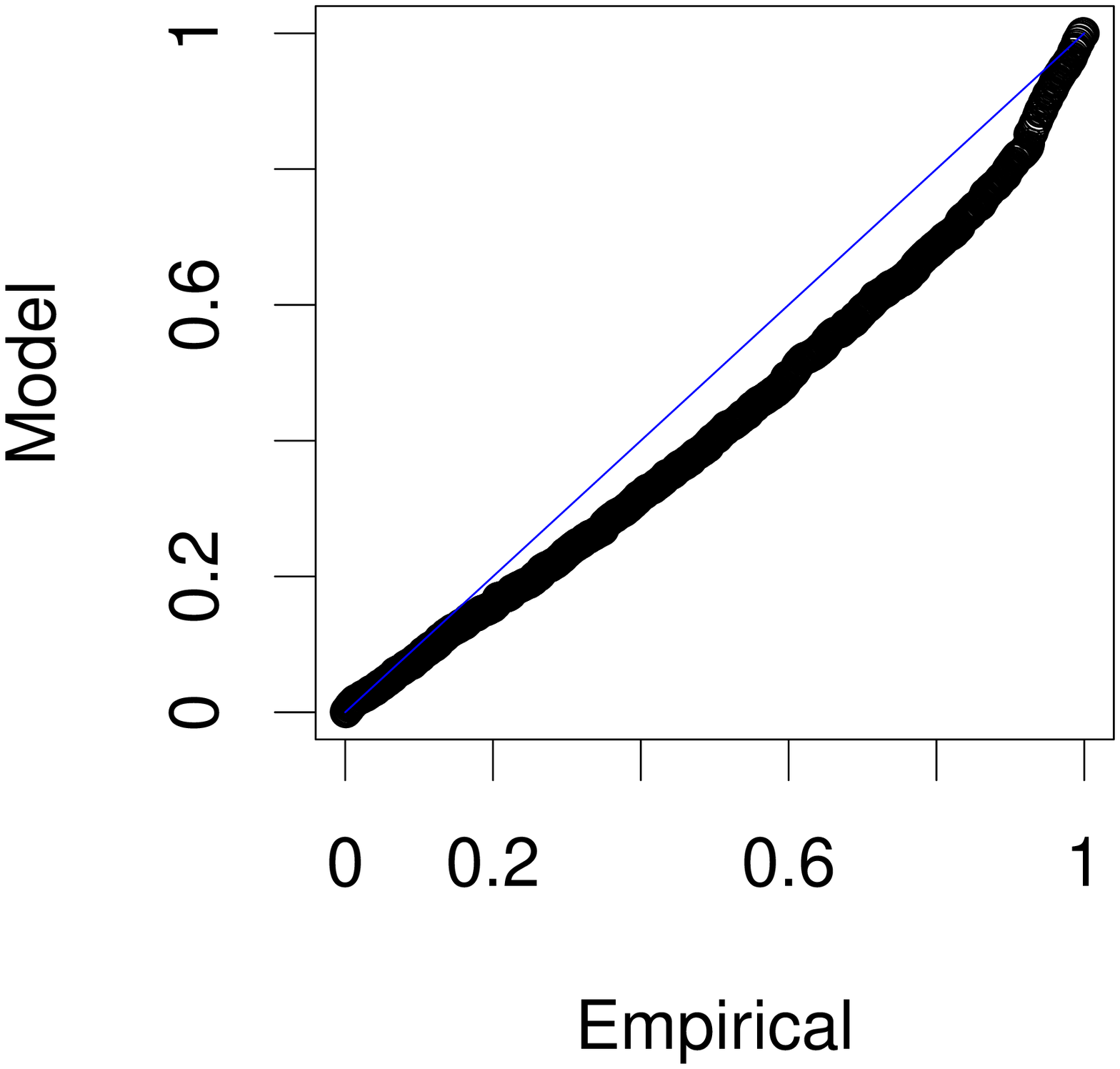}\\[5pt]
    \includegraphics[scale=0.18]{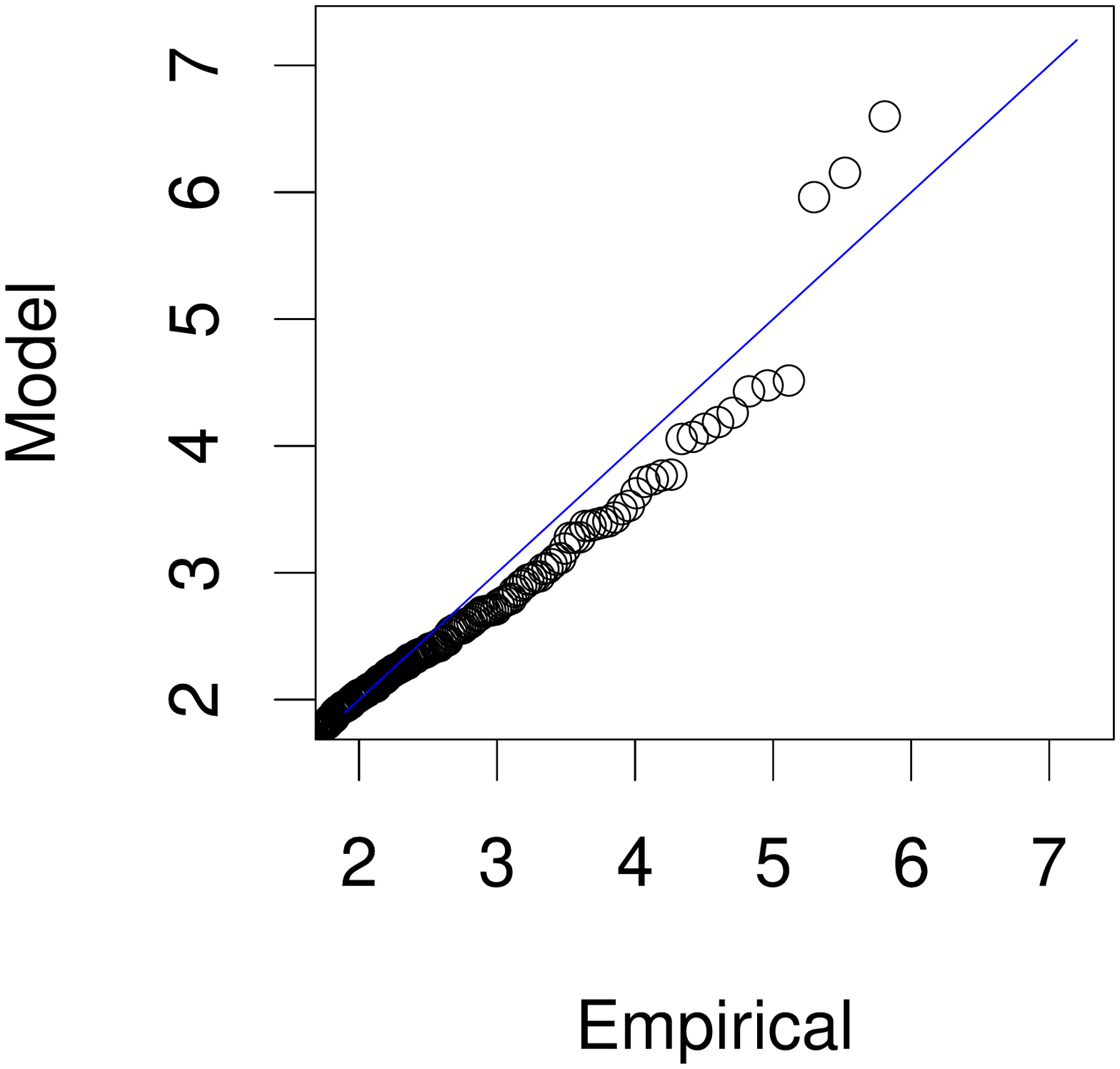}&
    \includegraphics[scale=0.18]{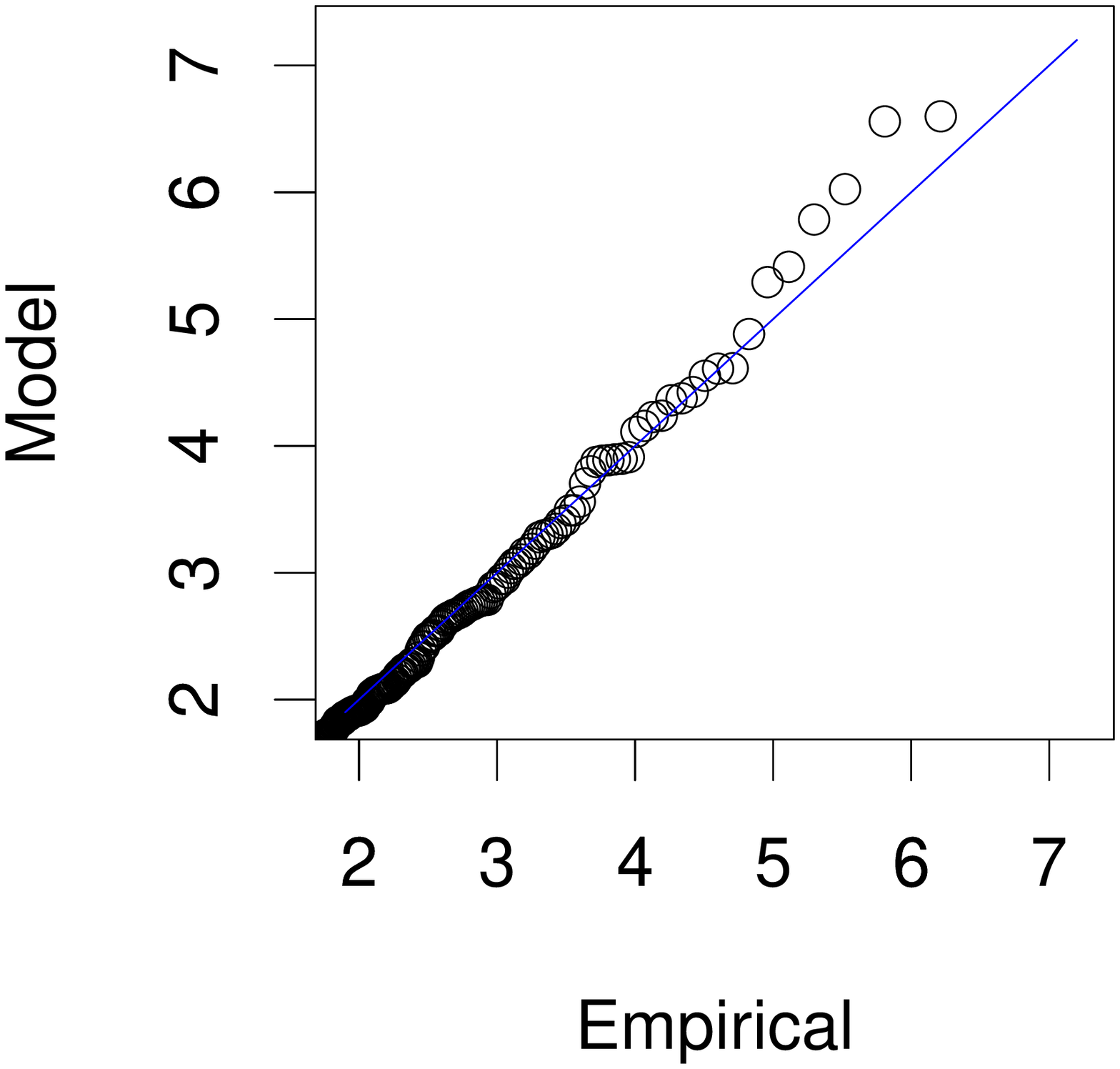}&
    \includegraphics[scale=0.18]{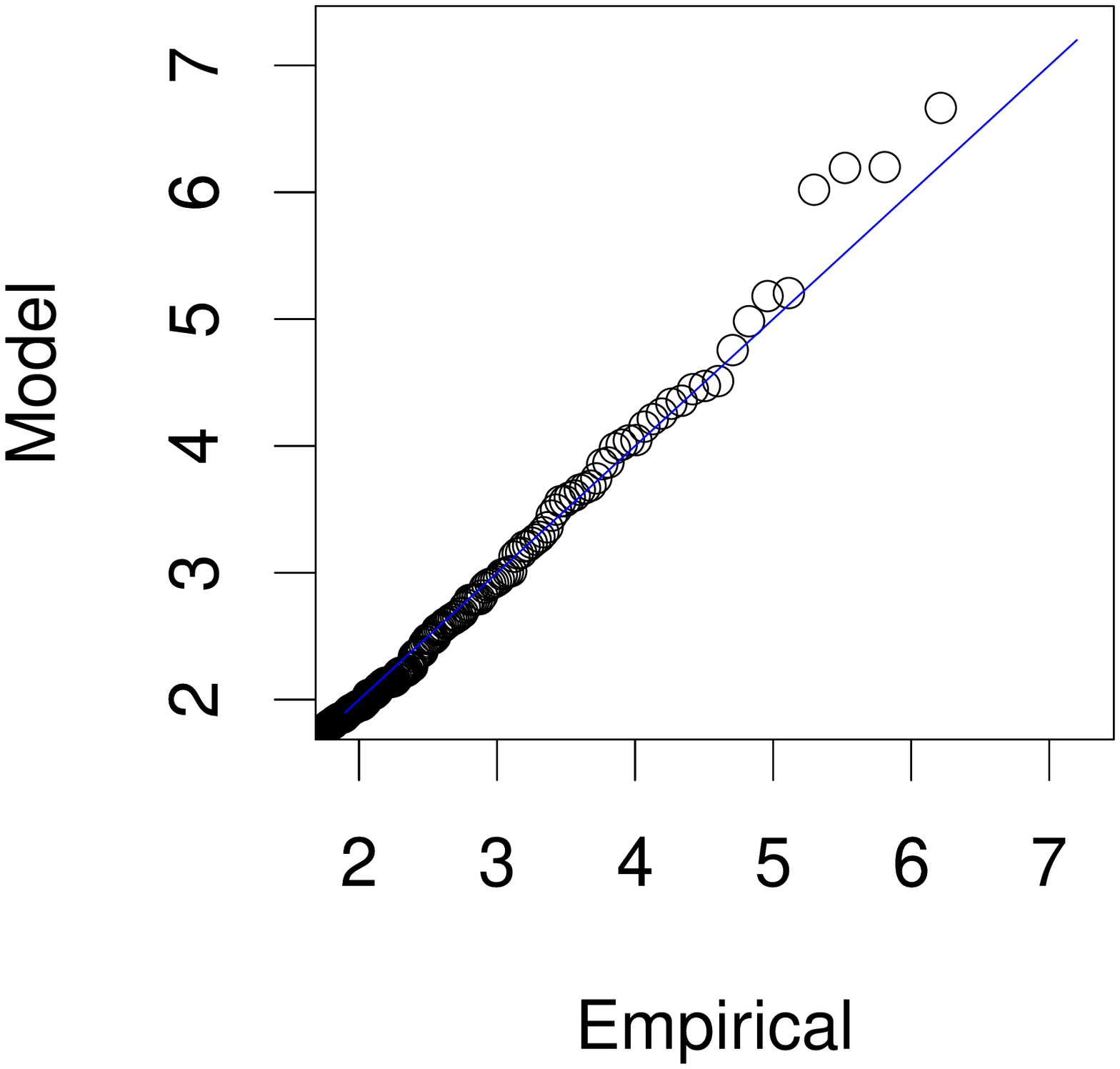}&
    \includegraphics[scale=0.18]{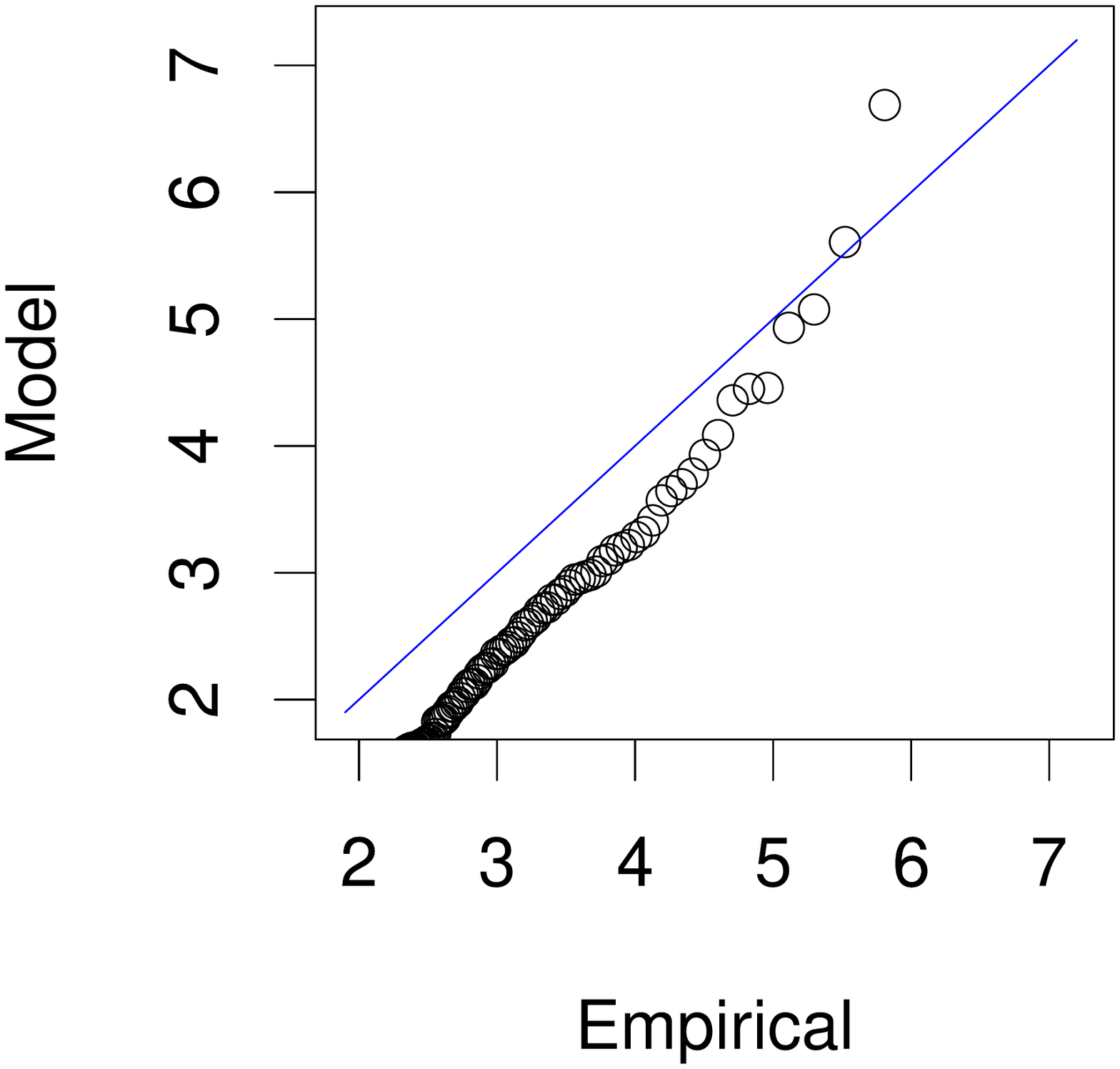}
  \end{tabular}
  \caption{\baselineskip=24pt
    Diagnostic plots of GEV inferences with model~\eqref{GEVt}
    and parameters as in~\eqref{musigmat}, for block $B_i$
    with $i=8$ (corresponding to $T_E^i=24$)
    and $\speedT=2/(1000\, \textrm{years})$.
    Top row: probability plots. Bottom row: quantile plots.
    From left to right column: plots for models $G_{p,q}$
    (see~\eqref{musigmat}), with $(p,q)$ and the corresponding
    log-likelihood $l$ (see~\eqref{logllGEV}) indicated on top.
    }
  \label{fig:diagnotrend}
\end{figure}

For each data block $B^i$, $i=1,\ldots,21$, we first extract
a sequence of yearly maxima $z_t^i$, with $t=1,\dots,\lengthbl$.
For compactness, each sequence is denoted in vector form as
$\boldsymbol{z}^i=(z_1^i,z_2^i,\dots,z_\lengthbl^i)$.
One GEV model of the form $G_{p,q}$ (see~\eqref{musigmat})
is fitted from each of the sequences $\boldsymbol{z}^i$.
For each $i=1,\dots,21$, the analysis follows three main steps:
\begin{enumerate}
\item
  nested models $G_{p,q}$, for $0\le p\le 2$ and $0\le q\le 1$,
  are fitted on the $i$-th sequence of maxima $\boldsymbol{z}^i$;
\item
  models with too low maximum likelihood are discarded and
  the deviance test is applied to the others to select the
  best estimate model;
\item
  the best estimate model is graphically checked by examining
  the probability and quantile plots,
  and it is possibly rejected in favor of a model having less
  nonzero parameters in the vector $\boldsymbol{\beta}$ as in~\eqref{beta}.
\end{enumerate}
Following the above procedure, for each time interval $I^i$, $i=1,\ldots,21$
time-dependent GEV models $G_{\pbe^i,\qbe^i}(\boldsymbol{z}^i)$ with
parameters $(\mube^i(t),\sibe^i(t),\xibe^i)$ are inferred from the data block
$\boldsymbol{z}^i$.  Model $G_{\pbe^i,\qbe^i}(\boldsymbol{z}^i)$ (denoted for
shortness $G_{\pbe^i,\qbe^i}$ in the rest of this section) is the best
estimate for the $i$-th data block, relative to the family of models
$G_{p,q}$.  Choosing a model with different orders $(p,q)$ would either give
poor results in the graphical checks, or fail to pass the likelihood ratio
test.  An example is given in \figref{diagnotrend}, for the data block $i=8$
in the time series with $\speedT=2/(1000\, \textrm{years})$.  The best
estimate model has orders $(\pbe^i,\qbe^i)=(1,0)$.  Models $G_{0,0}$ and
$G_{2,1}$ are rejected, since they have too small likelihood and since the fit
quality is very low, as it is illustrated by the probability and quantile
plots.  On the basis of the diagnostic plots, models $G_{1,0}$ and $G_{1,1}$
are both acceptable.  However, the deviance statistic satisfies
$\mathcal{D}=2\{l_{1,1}-l_{1,0}\}=3.64<c_{0.5}=3.84$ which is the
0.95-quantile of the $\chi^2_1$-distribution.  Therefore, as there is no
strong support for selecting model $G_{1,1}$, according to the likelihood
ratio test the more parsimonious model $G_{1,0}$ is preferred.

\begin{figure}[hb]
  \centering
  \psfrag{TE}{$T_E$}
  \psfrag{mu}{$\mu$}
  \psfrag{sg}{$\sigma$}
  \psfrag{xi}{$\xi$}
  \includegraphics[scale=0.27]{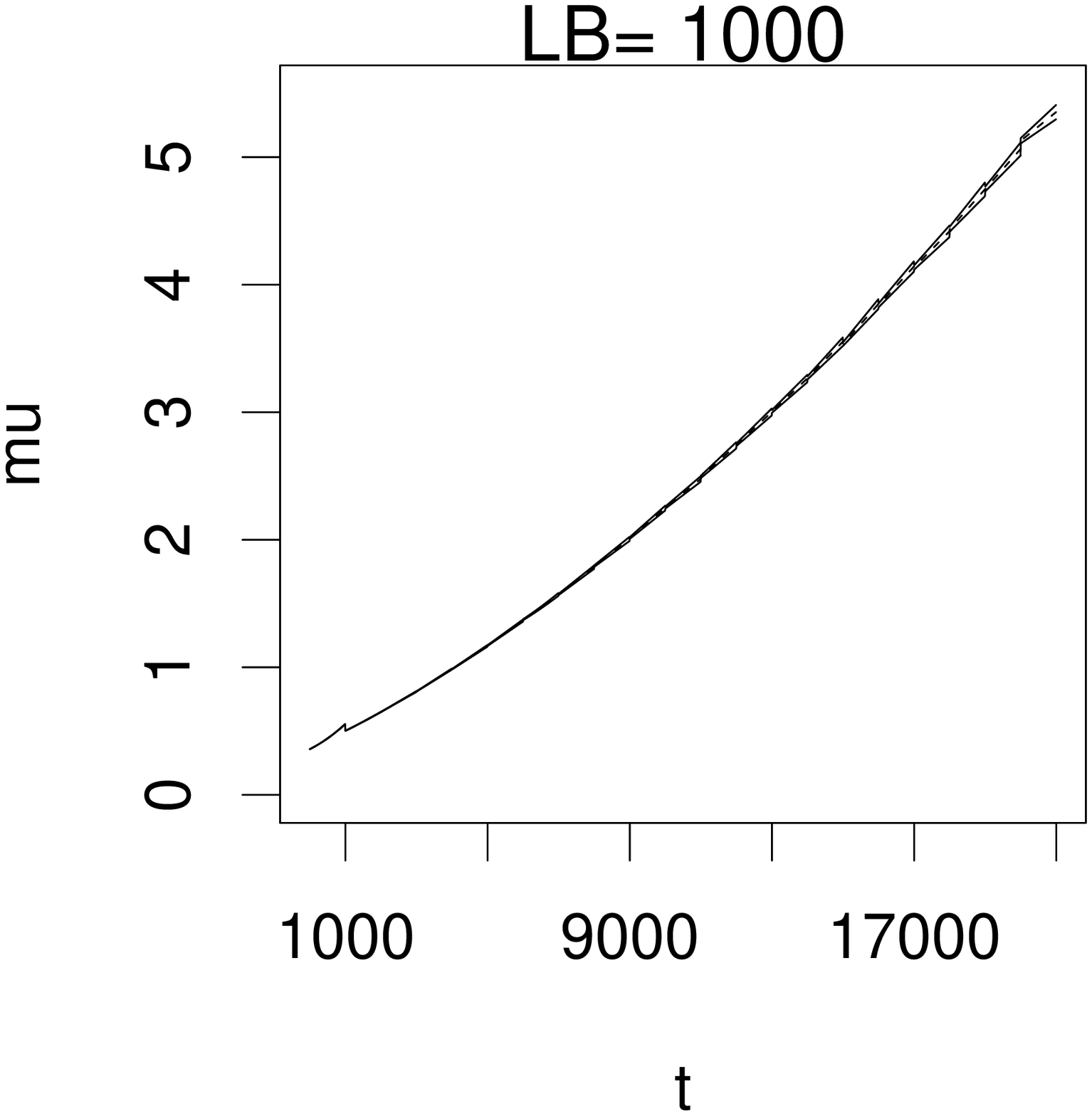}
  \includegraphics[scale=0.27]{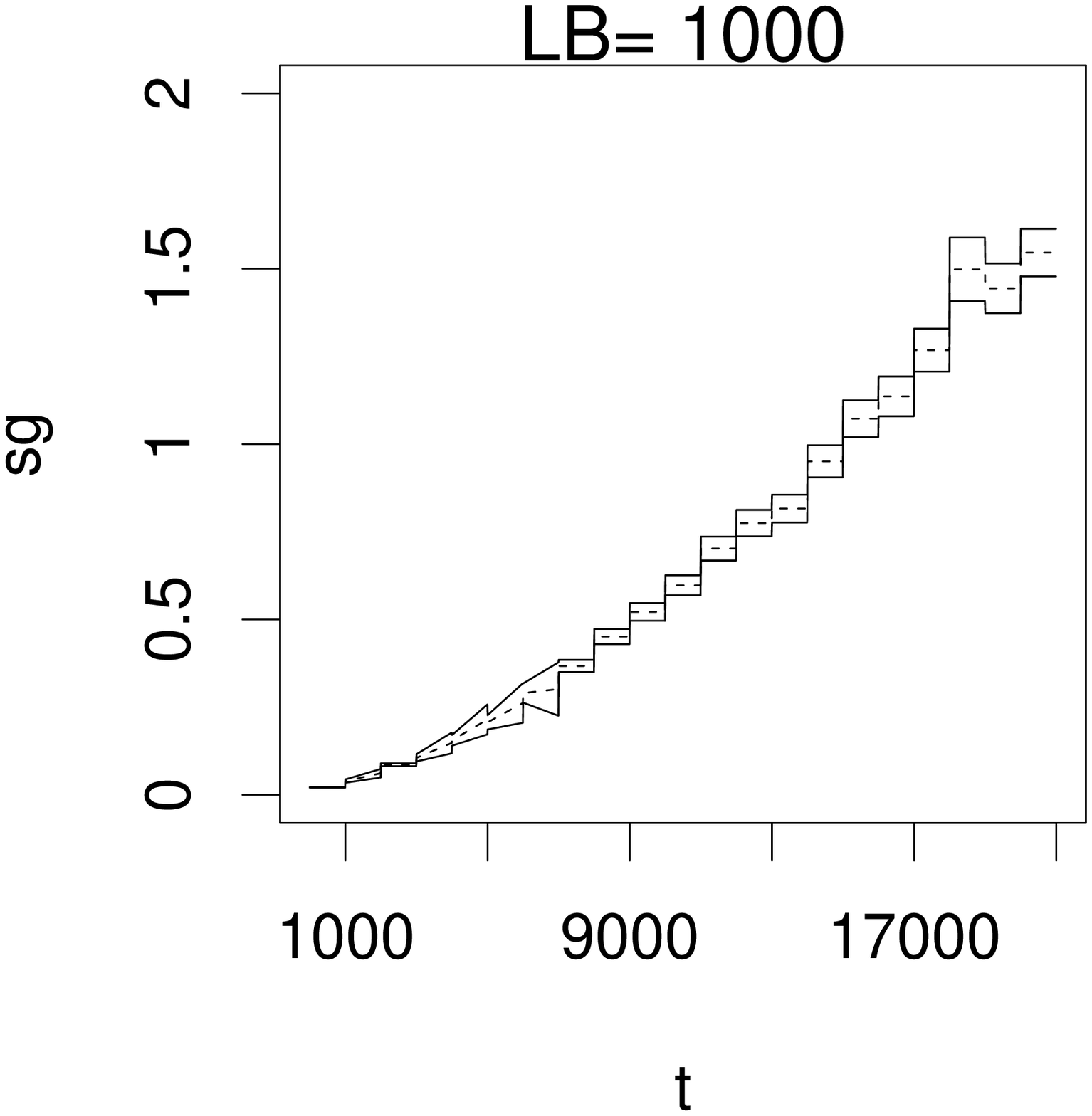}
  \includegraphics[scale=0.27]{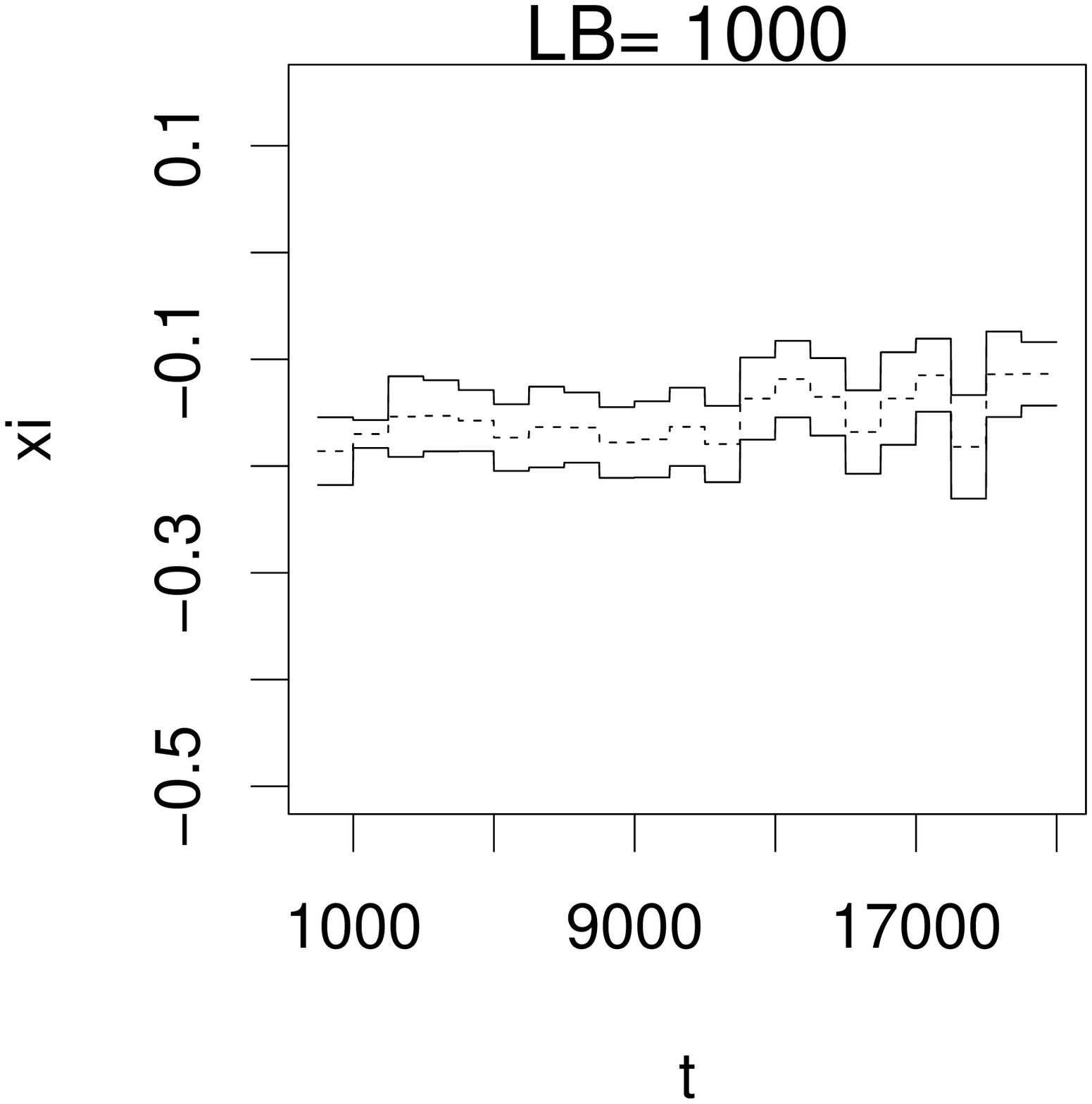}\\[5pt]
  \includegraphics[scale=0.27]{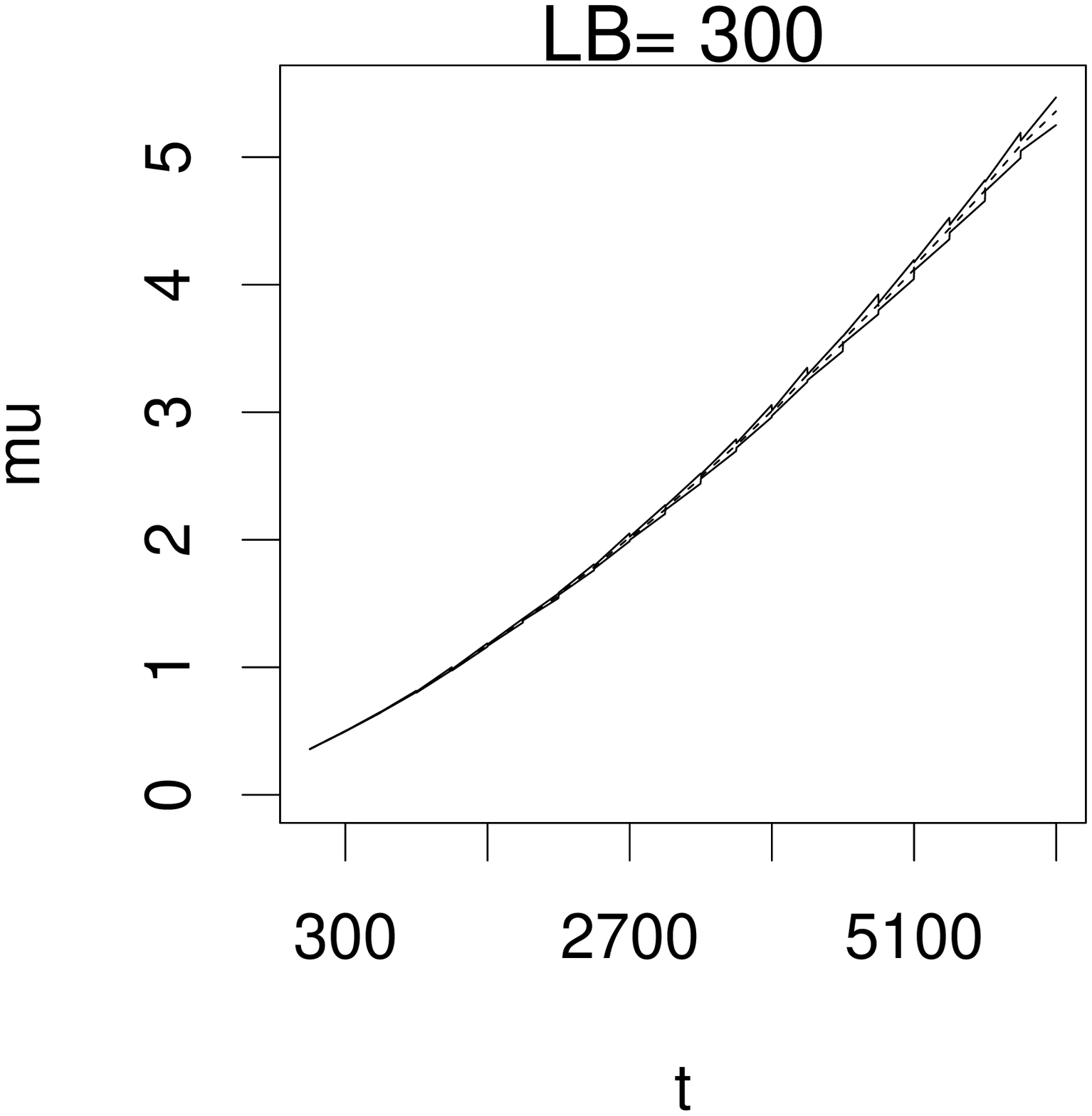}
  \includegraphics[scale=0.27]{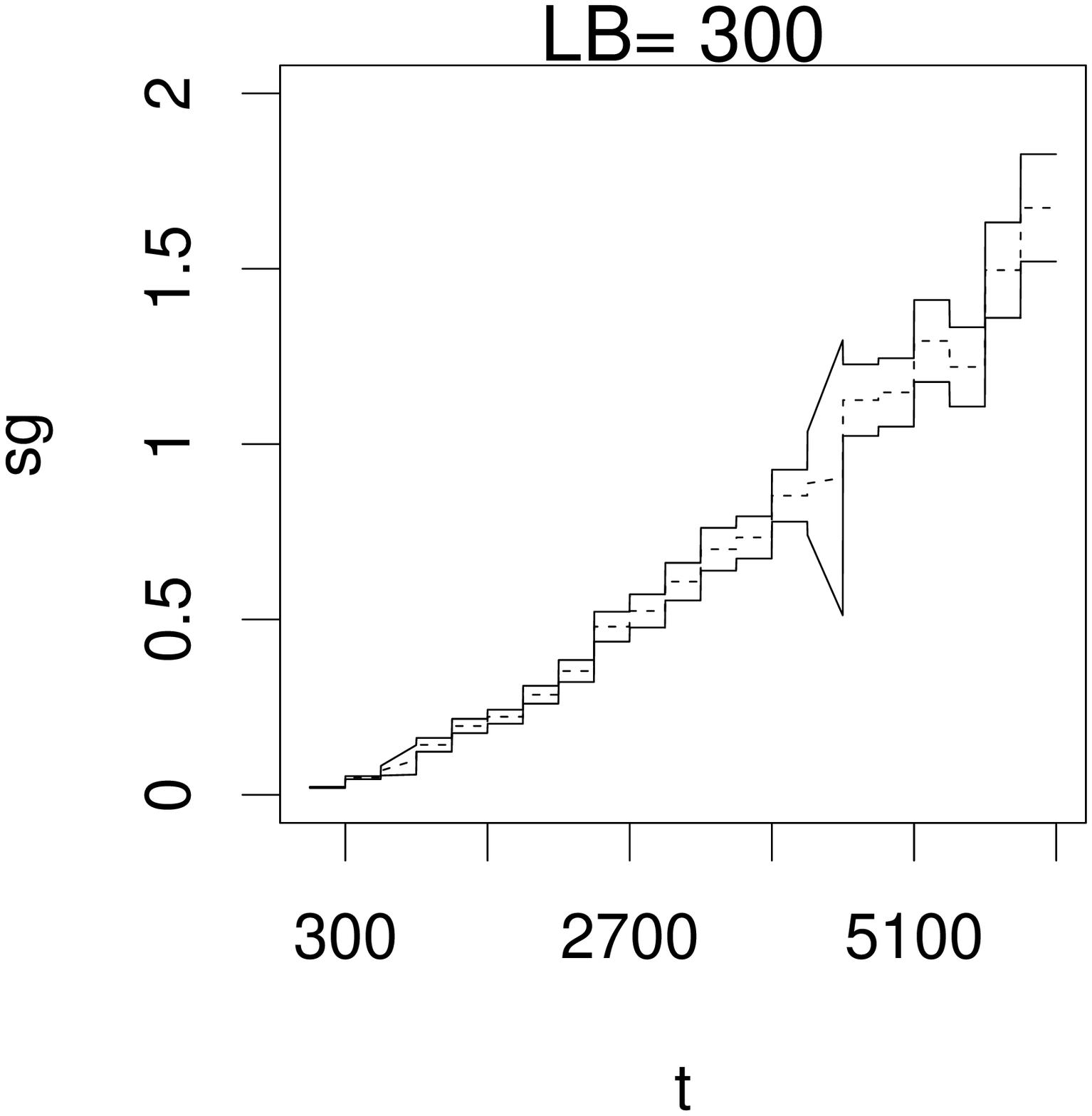}
  \includegraphics[scale=0.27]{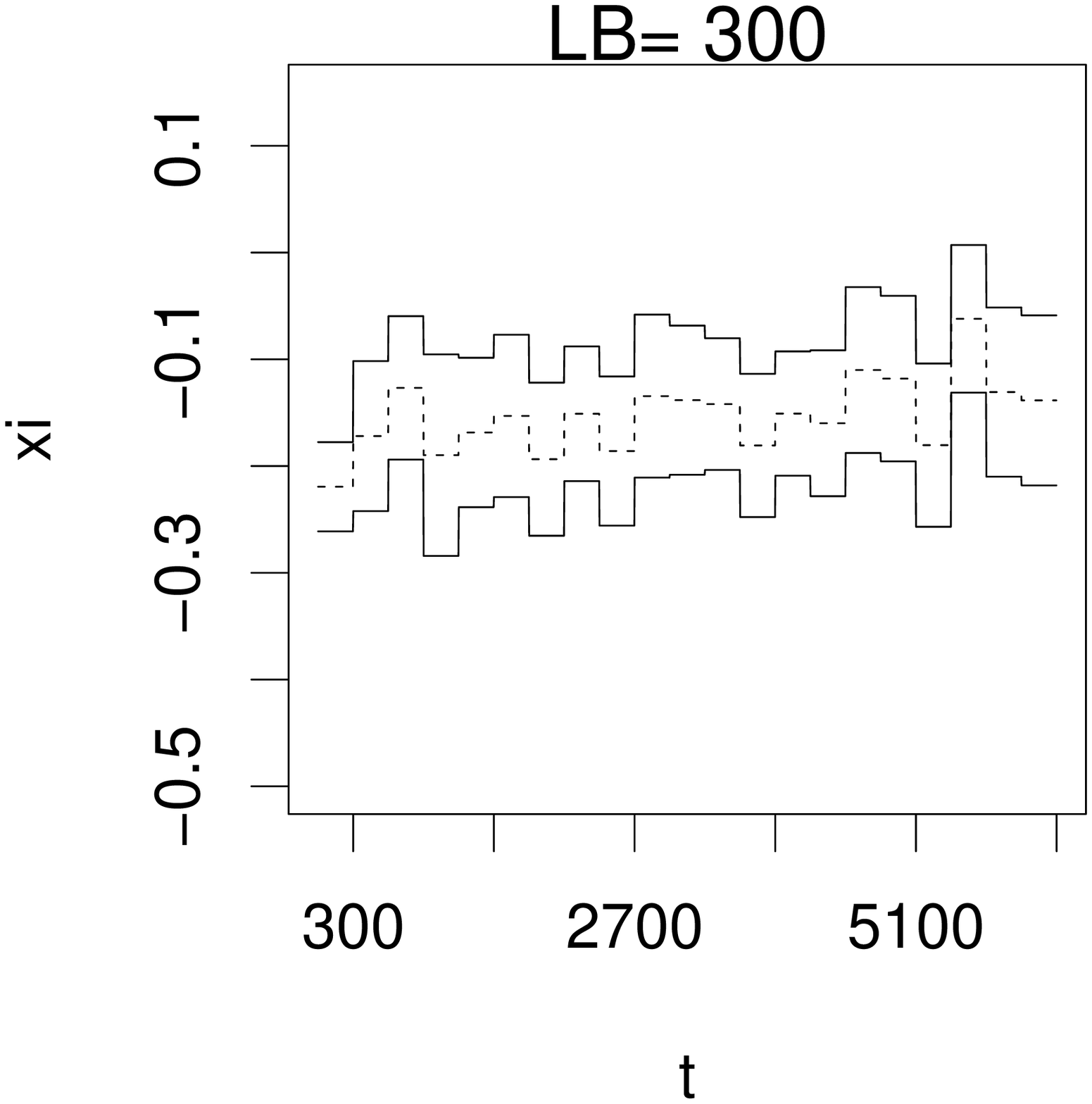}\\[5pt]
  \includegraphics[scale=0.27]{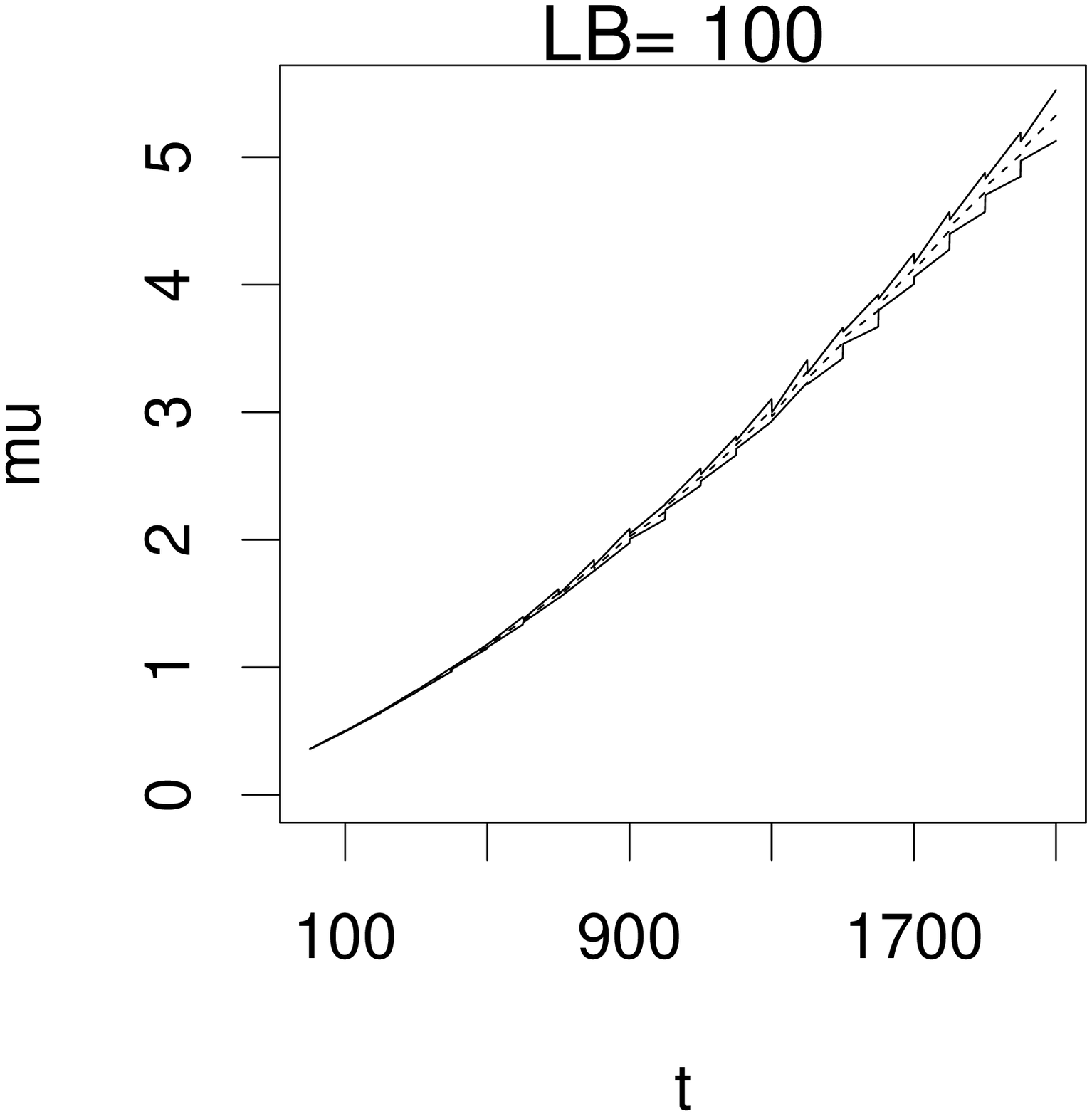}
  \includegraphics[scale=0.27]{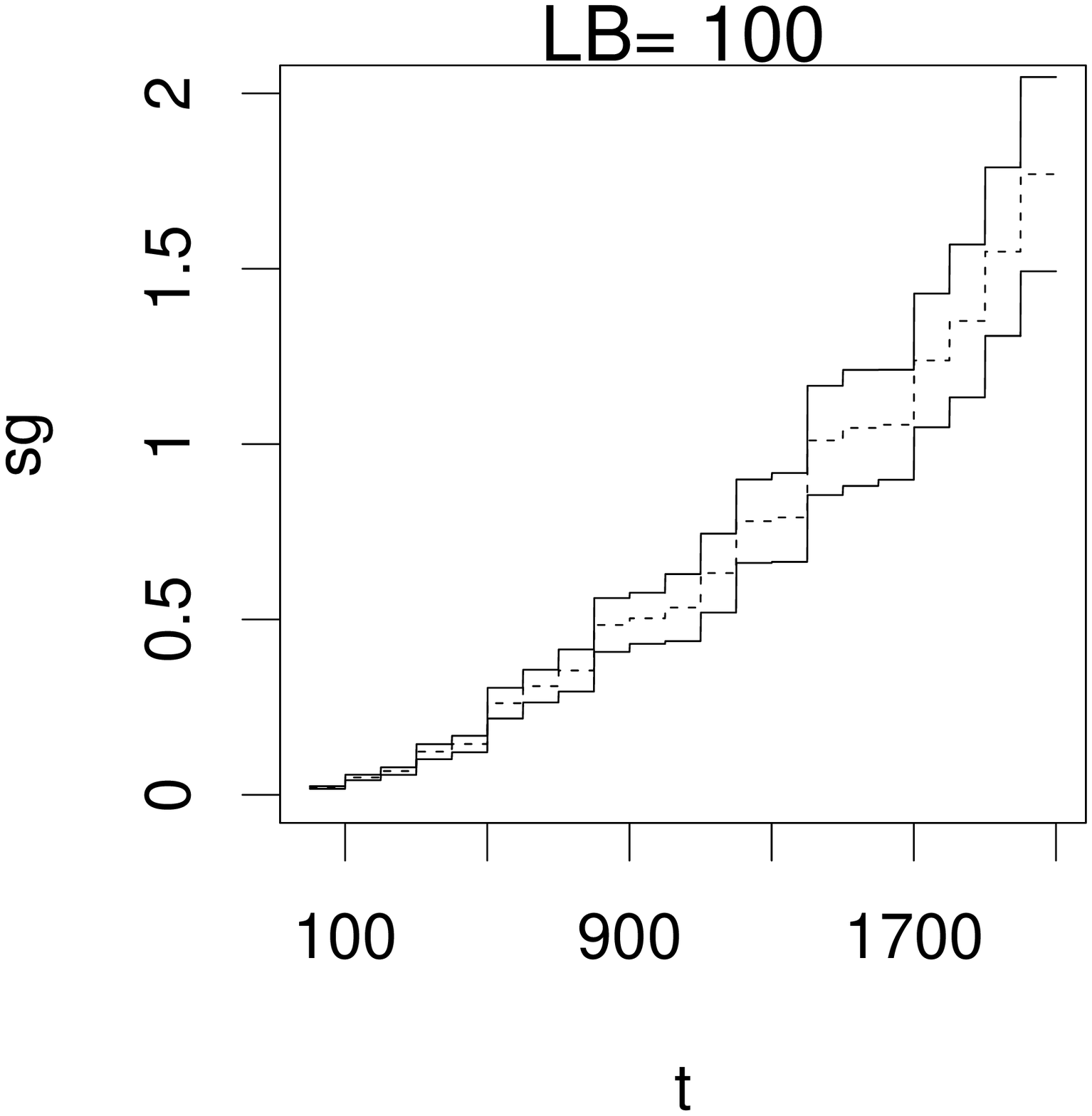}
  \includegraphics[scale=0.27]{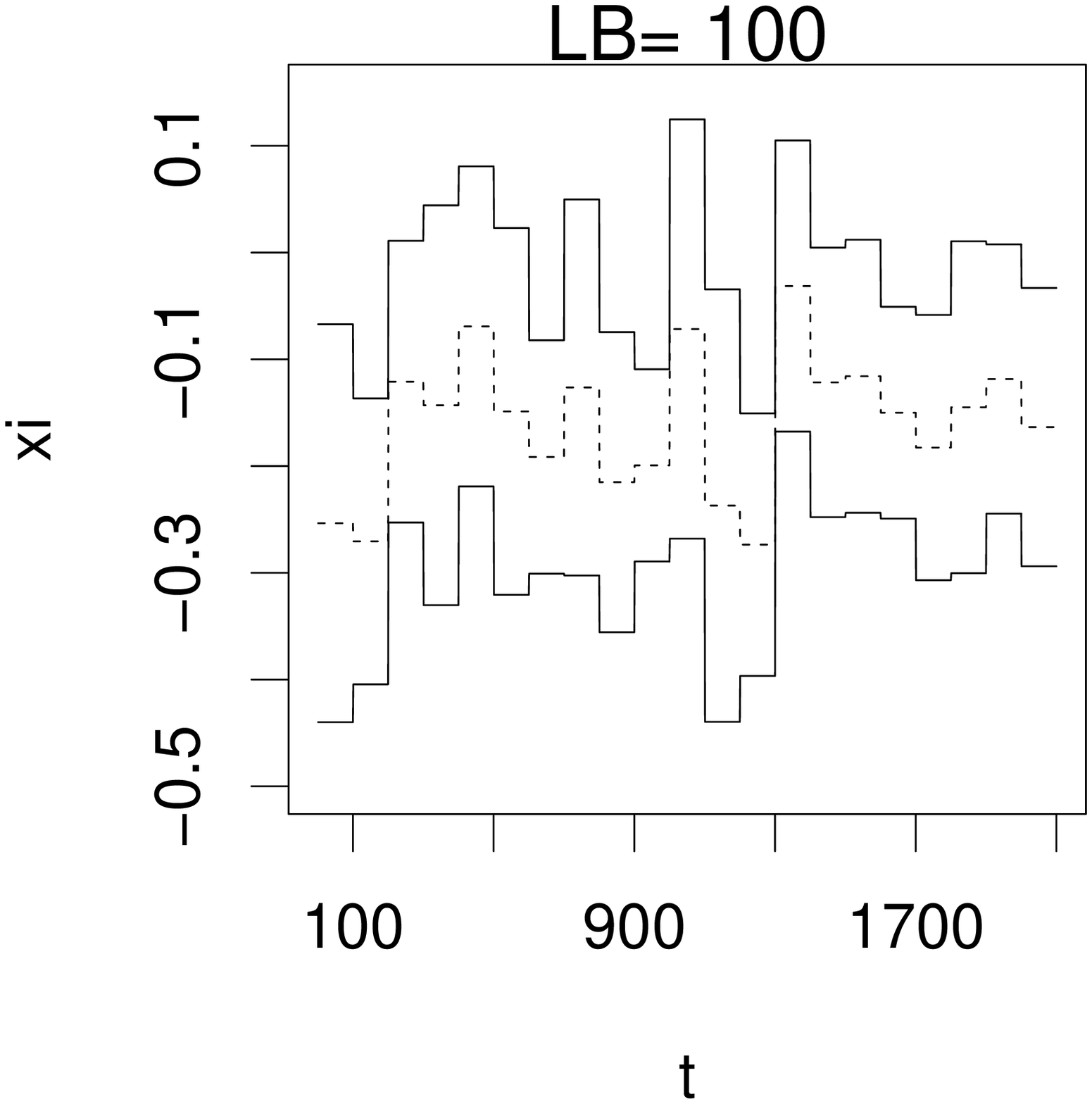}
  \caption{\baselineskip=24pt
    GEV parameters as functions of time, for the three considered values
    of trend intensity $\speedT$: from top to bottom,
    $\speedT=2/(1000\, \textrm{years})$,
    $2/(300\, \textrm{years})$, and $2/(100\, \textrm{years})$, respectively.
    For each trend intensity the inferred time-dependent parameters
    $(\mu(t),\sigma(t),\xi)$ (left, center, right column, respectively)
    of the best estimate model $G_{\pbe^i,\qbe^i}(\boldsymbol{z}^i)$ are  plotted.
  }
  \label{fig:GEVtime}
\end{figure}

Plots of the best estimate parameters $(\mube^i(t),\sibe^i(t),\xibe^i)$
as functions of time are proposed in \figref{GEVtime}.
Confidence intervals are computed as the \emph{worst case}
estimates: suppose that a model is chosen with $p=1$, that is,
for the best estimate $\mu(t)$ is linear $\mube^i(t)=\mu^i_0+\mu^i_1t$.
Let $\sigma_{\mu^i_0}$ and $\sigma_{\mu^i_1}$ be the uncertainties
in $\mu^i_0$ and $\mu^i_1$, respectively, provided by the
observed information matrix (see \secref{GEVmethods}).
Then the confidence interval at time $t$ is computed as
\begin{equation}
  \label{eq:worstcase}
  [\mube^i(t)-2(\sigma_{\mu^i_0}+\sigma_{\mu^i_1}t),
  \mube^i(t)+2(\sigma_{\mu^i_0}+\sigma_{\mu^i_1}t)].
\end{equation}
For most of the time intervals $I^i$, the best estimate model is such that
$\mube^i(t)$ and $\sibe^i(t)$ are respectively linear and constant in time,
that is, $(\pbe^i,\qbe^i)=(1,0)$.  This is so for all models inferred with the
fastest trend intensity $\speedT=2/(100\, \textrm{years})$ (see
\tabref{best100}), whereas for $\speedT=2/(300\, \textrm{years})$ there are
two exceptions: intervals $i=3$ and $i=8$, for which also $\sibe^i(t)$ grows
linearly in time ($\qbe^i=1$, see \tabref{best300}).  For the slowest trend
intensity we obtain $\qbe^i=1$ for $i=2, 4, 5, 6, 7$ and zero otherwise,
whereas $\pbe^i=2$ for $i=1,2,7$ and $\pbe^i=1$ otherwise, see
\tabref{best100}.  Summarizing, the best fits are mostly achieved by lowest
order non-stationary models of the form $(\pbe^i,\qbe^i)=(1,0)$. For slower
trends, however, in some cases the best fit is of the form
$(\pbe^i,\qbe^i)=(1,1)$ or even $(\pbe^i,\qbe^i)=(2,1)$.  These cases
typically occur for low $i$, that is, in the first portion of the time series.
This is due to the fact that, for small $T_E$ (corresponding to small $t$
through equation~\eqref{TEt}), although the hypothesis of smoothness,
described in \secref{GEVproblems}, may still be considered valid, the rate of
variation of the SRB measure with respect to variations in $T_E$ is
comparatively larger.  To put it in simple words, the statistical behaviour
(the attractor) of the baroclinic model is rather sensitive with respect to
changes in $T_E$.  Therefore, the variation of the statistical properties in
time is not quite ``adiabatic'', in the sense specified in
\secref{GEVproblems}.  Correspondingly, a statistical model of enhanced
complexity (more parameters) is needed to achieve goodness-of-fit.

In concluding this section, we emphasize that the convergence of the numerical
procedure used in the maximization of the likelihood function is here
considerably more problematic than in the stationary case studied in Part I.
Indeed, in the present case it is often necessary to choose a good starting
point for the maximization procedure in order to achieve convergence. For
example, in several cases, after achieving convergence for a GEV model with
order, say $(p,q)=(2,1)$, by using the inferred values of the parameter vector
$\boldsymbol{\beta}$ in~\eqref{beta} as starting values for the maximization,
a better fit (larger likelihood) of \emph{lower order} is obtained.  In fact,
this has allowed us to reduce the total number of inferences with $p=2$ and/or
$q=1$.

\section{Trend assessment }
\label{sec:GEVassess}
\begin{figure}[hb]
  \centering
  \psfrag{mu}{$\mu(t)$}
  \includegraphics[scale=0.37]{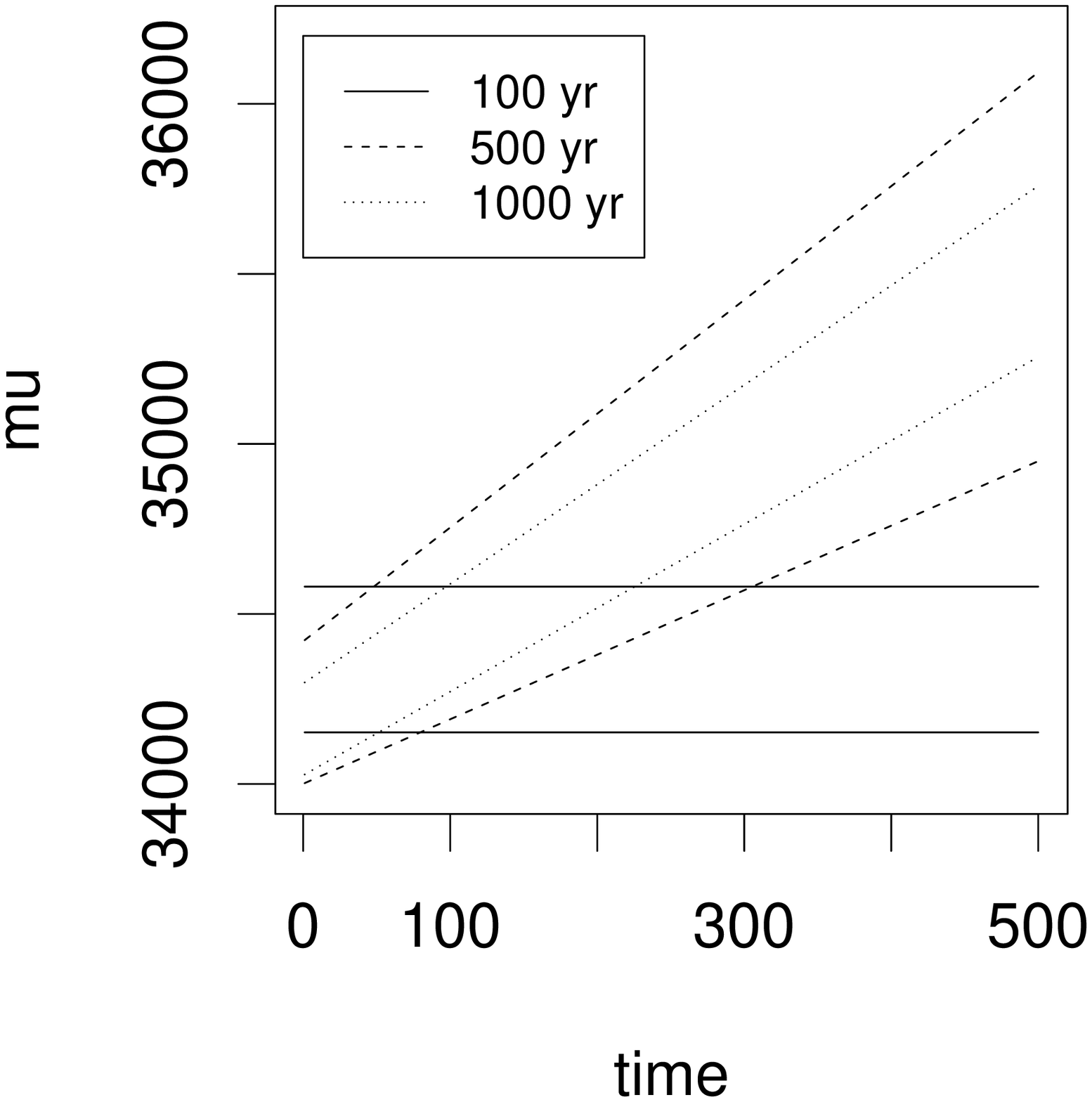}
  \includegraphics[scale=0.37]{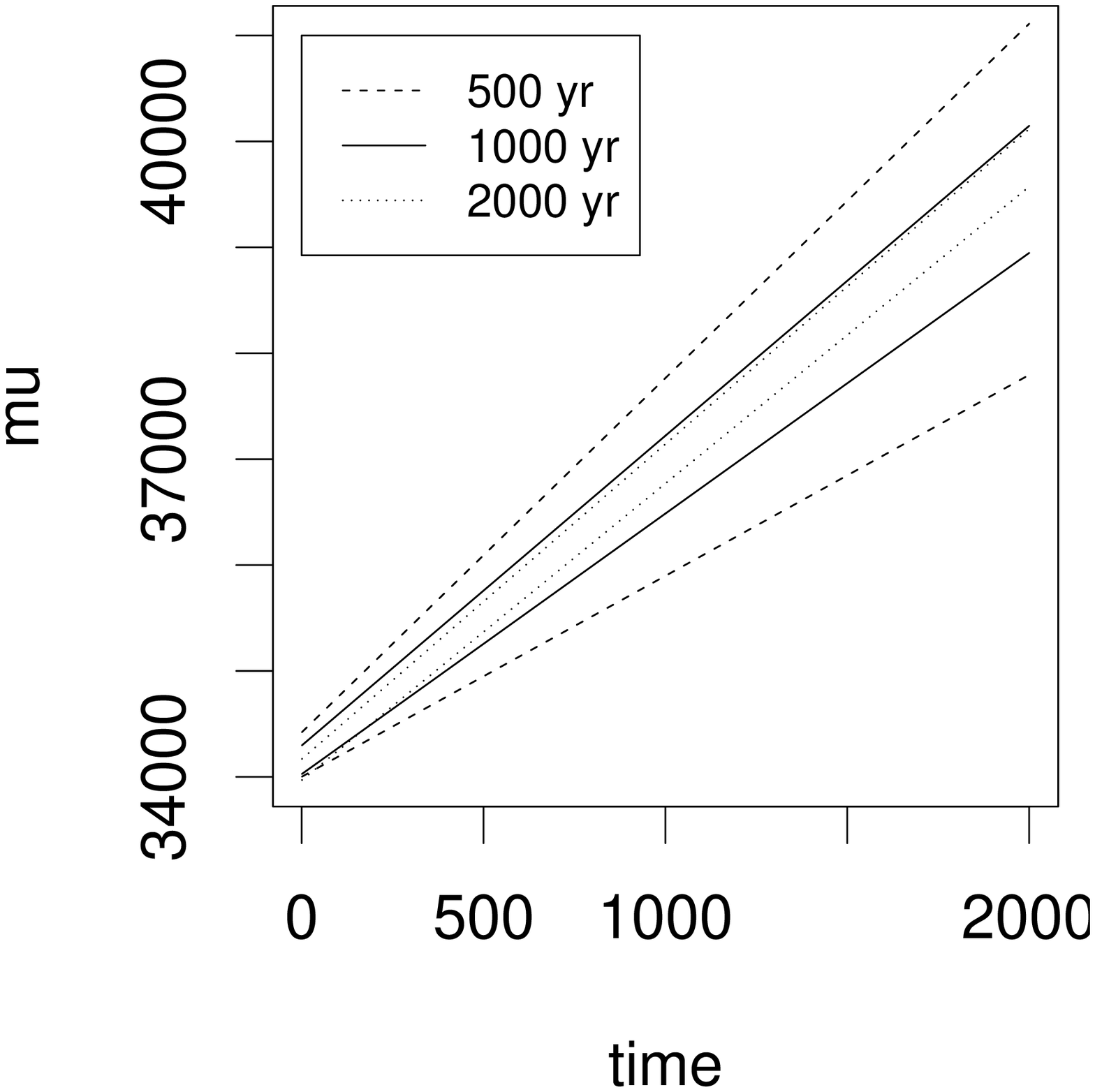}
  \caption{\baselineskip=24pt
    Parameter $\mu(t)$ of the best estimate GEV inferences
    as a function of time. The time series with slowest trend intensity
    $\speedT=2/(1000\, \textrm{years})$ has been used, taking
    yearly maxima over a data block starting at year $14001$.
    For legibility, only the confidence intervals have been plotted.
    Left: inferences obtained with 100, 500, and 1000 yearly maxima.
    The best estimate fit based on 100 data is stationary
    ($\mu_1=0$) and it has been extrapolated to 500 years.
    Right: inferences obtained with 500, 1000, and 2000 yearly maxima.
    In all cases, the best estimate GEV model has $p=1$,
    that is, $\mu(t)=\mu_0+\mu_1t$ is a linear function of time.
  }
  \label{fig:GEVassess}
\end{figure}

When dealing with non-stationary data, the problem of assessing the
sensitivity of trend inferences is particularly delicate.  Beyond the serious
conceptual problems explained in \secref{GEVproblems}, one is confronted with
several practical issues. Most of the sensitivity tests in Part I
were based on examining a shorter portion of same time series or on
calculating the maxima on data blocks of different lengths.  In the present
non-stationary context, both operations would result in an alteration of the
statistical properties of the sample (exactly because of the non-stationarity)
and this makes comparisons somewhat ambiguous.  An example is provided in
\figref{GEVassess}, where we compute the best estimate GEV fits using
sequences of yearly maxima having different lengths -- but starting at the
same instant (year 14500) -- extracted from the time series with the slowest
trend intensity $\speedT=2/(1000\, \textrm{years})$.  Notice that the best fit
obtained by taking 100 yearly maxima is stationary.  The corresponding
extrapolations in time are, of course, completely wrong.  By using 500 and
1000 maxima, the best estimates obtained (not shown) fall inside the
confidence band of the 2000 years-based estimate for most values of time.

The above example illustrates the \emph{trend dilemma}: on the one hand, in
order to be detected, a statistical trend has to be sufficiently fast with
respect to the length of the record of observations; on the other hand, if the
trend is too fast then the ``adiabatic hypothesis'' discussed in
\secref{GEVproblems} is no longer valid: one is left with no ``reference
statistics'' against which the inferred models can be compared.

Moreover, when considering large time spans a further practical complication
arises: due to the nonlinear dependence of the statistical properties with
respect to the external parameter $T_E$, a functional relation between the GEV
parameters and time might require many parameters to achieve goodness-of-fit.
Therefore, one faces the problem of large uncertainties in the parameter
estimates or even lack of convergence.  This has indeed been observed for the
present time series: if we consider a long record, such that the change in
$T_E$ is large, the model family $G_{p,q}$ with parameters as
in~\eqref{musigmat} becomes inadequate to catch the time dependence of the
statistics of extremes. A first indication of this was reported at the end of
\secref{GEVinference}.  As a further example, we have examined a data block of
length 5000 starting at year 14500 in the time series with $\speedT=2/(1000\,
\textrm{years})$.  Inspection of graphical diagnostics (probability and
quantile plots) reveals that no model in the family $G_{p,q}$ produces an
acceptable inference. It should be emphasized that goodness-of-fit is achieved
for the \emph{same} time series using blocks of length 1000, that is,
performing inferences that are more localized in time.  So in this case the
problem is not the failure of the ``adiabatic'' or smoothness hypothesis, but
the nonlinear dependence of the attractor on the parameter $T_E$, which
manifests itself on sufficiently large time intervals.

\section{Smooth dependence on the forcing}
\label{sec:smooth}
\begin{figure}[hb]
  \centering
  \psfrag{TE}{$T_E$}
  \psfrag{mu}{$\mu$}
  \psfrag{sg}{$\sigma$}
  \psfrag{xi}{$\xi$}
  \includegraphics[scale=0.27]{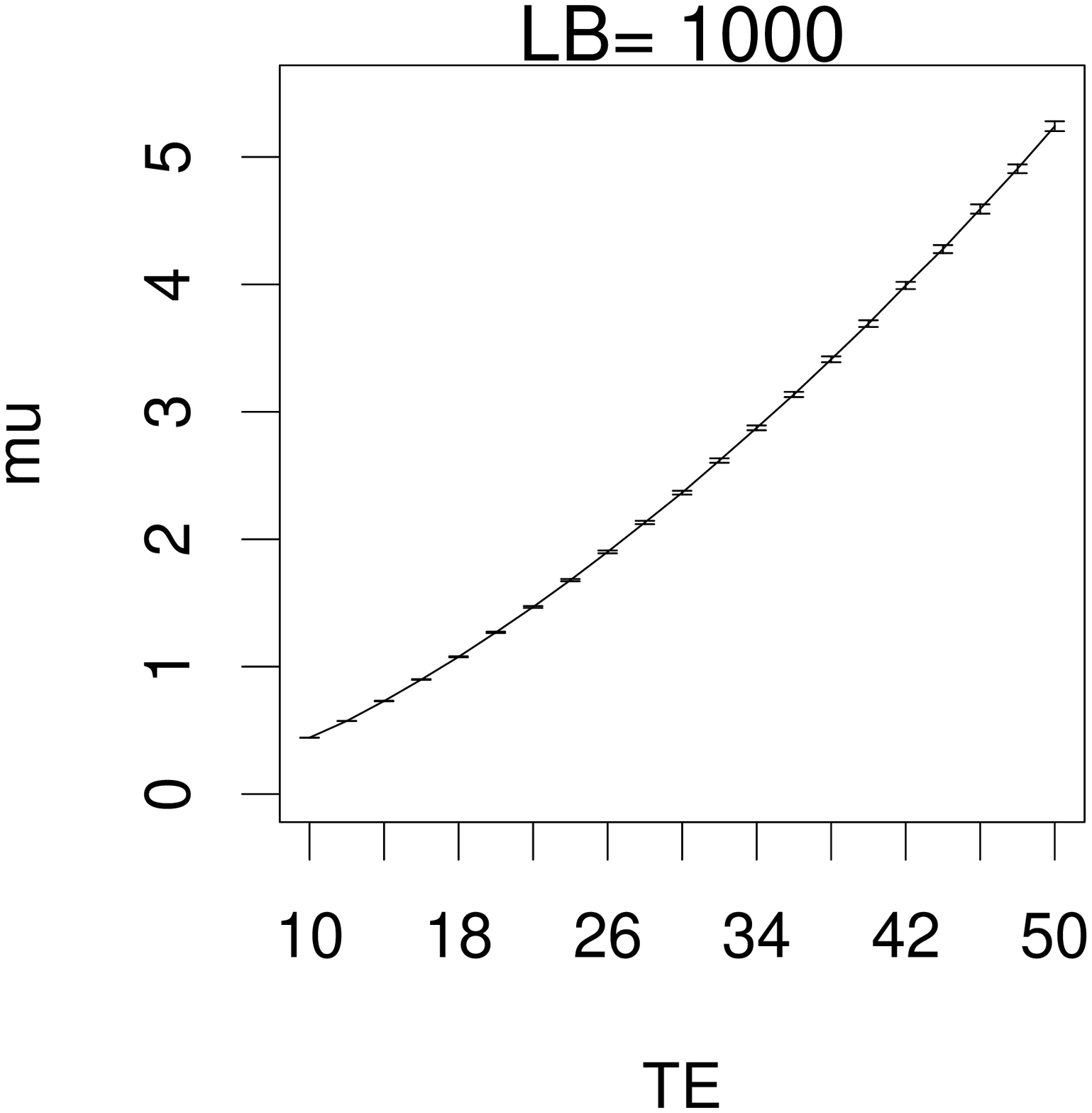}
  \includegraphics[scale=0.27]{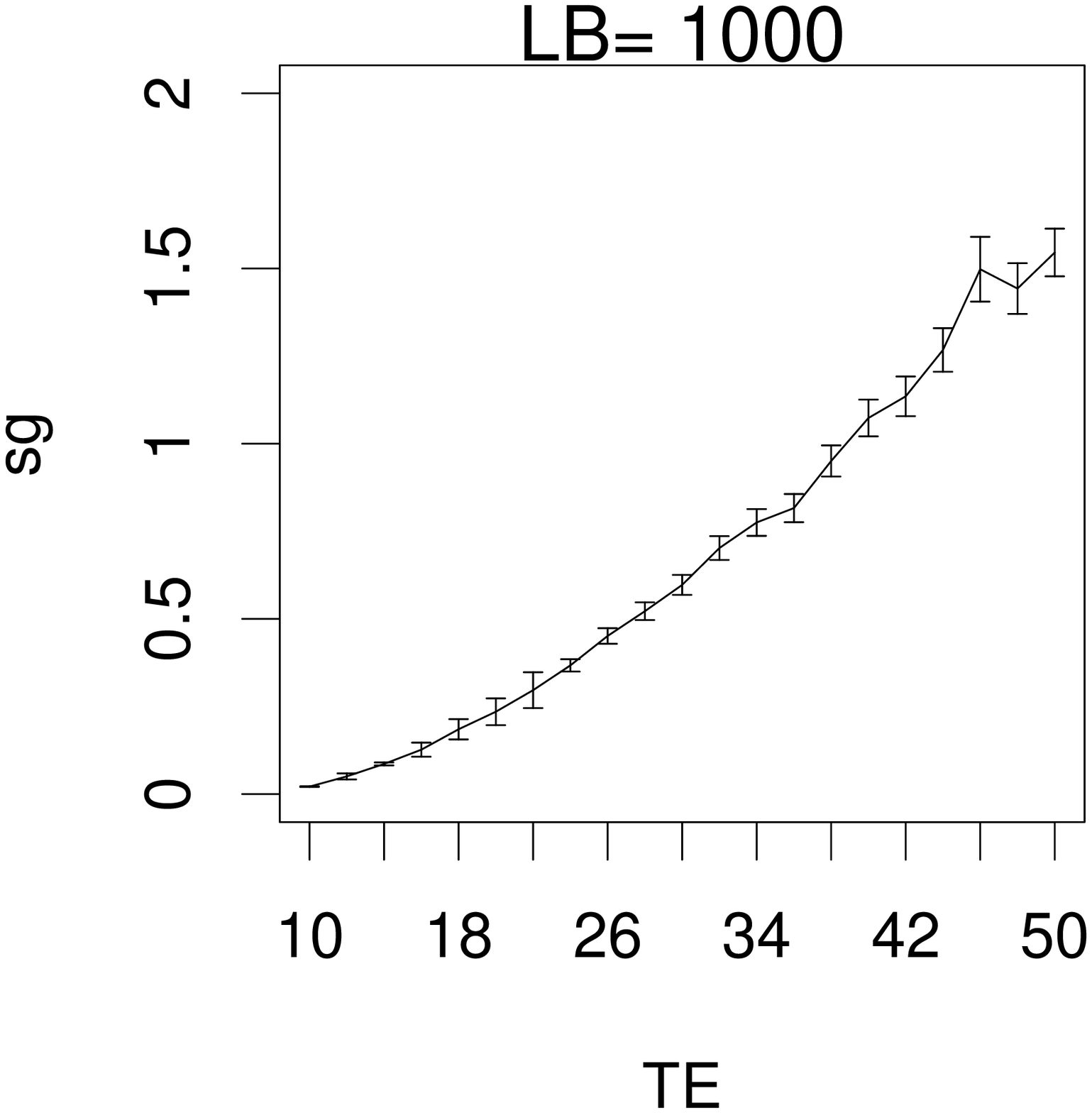}
  \includegraphics[scale=0.27]{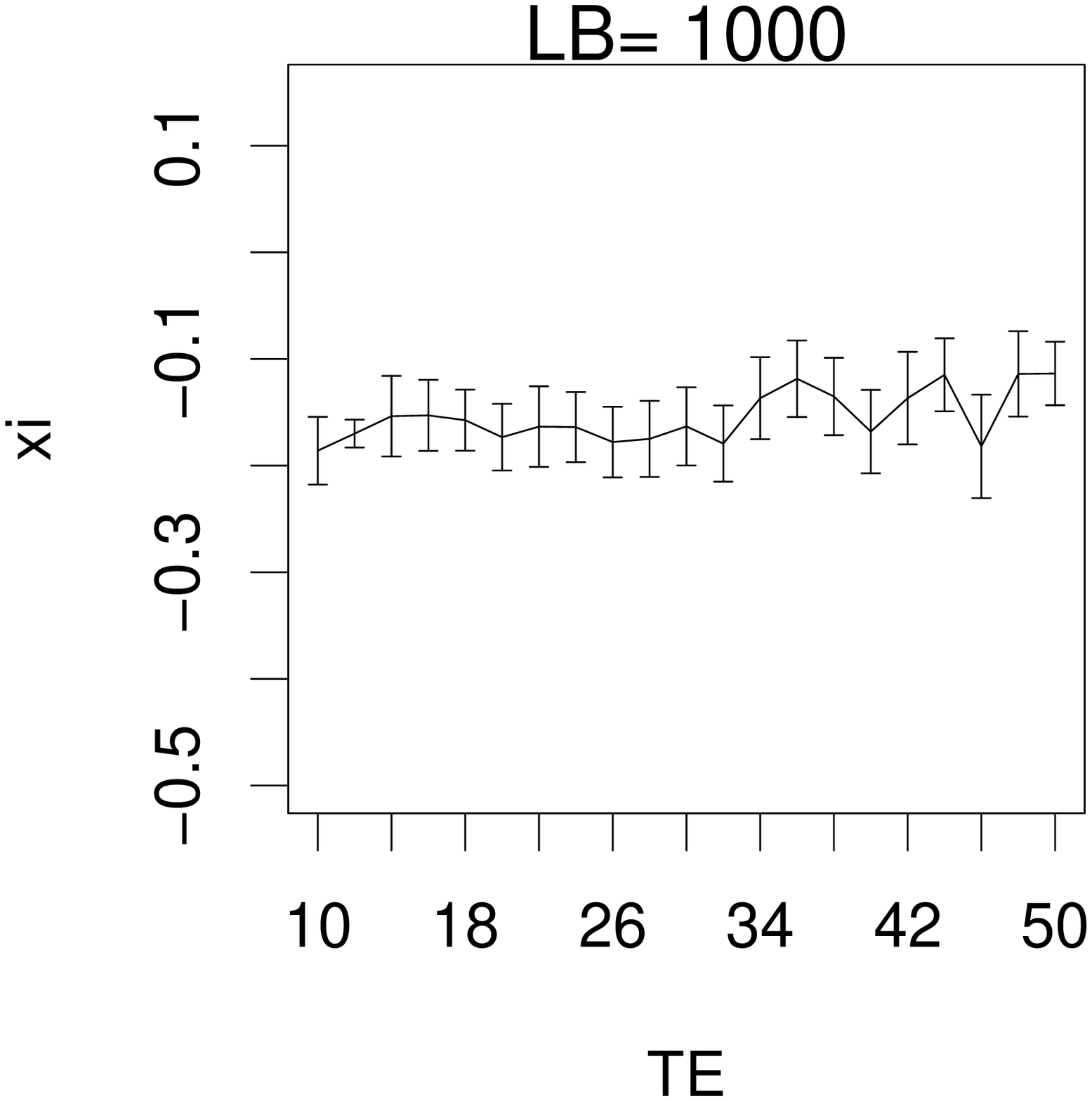}\\[5pt]
  \includegraphics[scale=0.27]{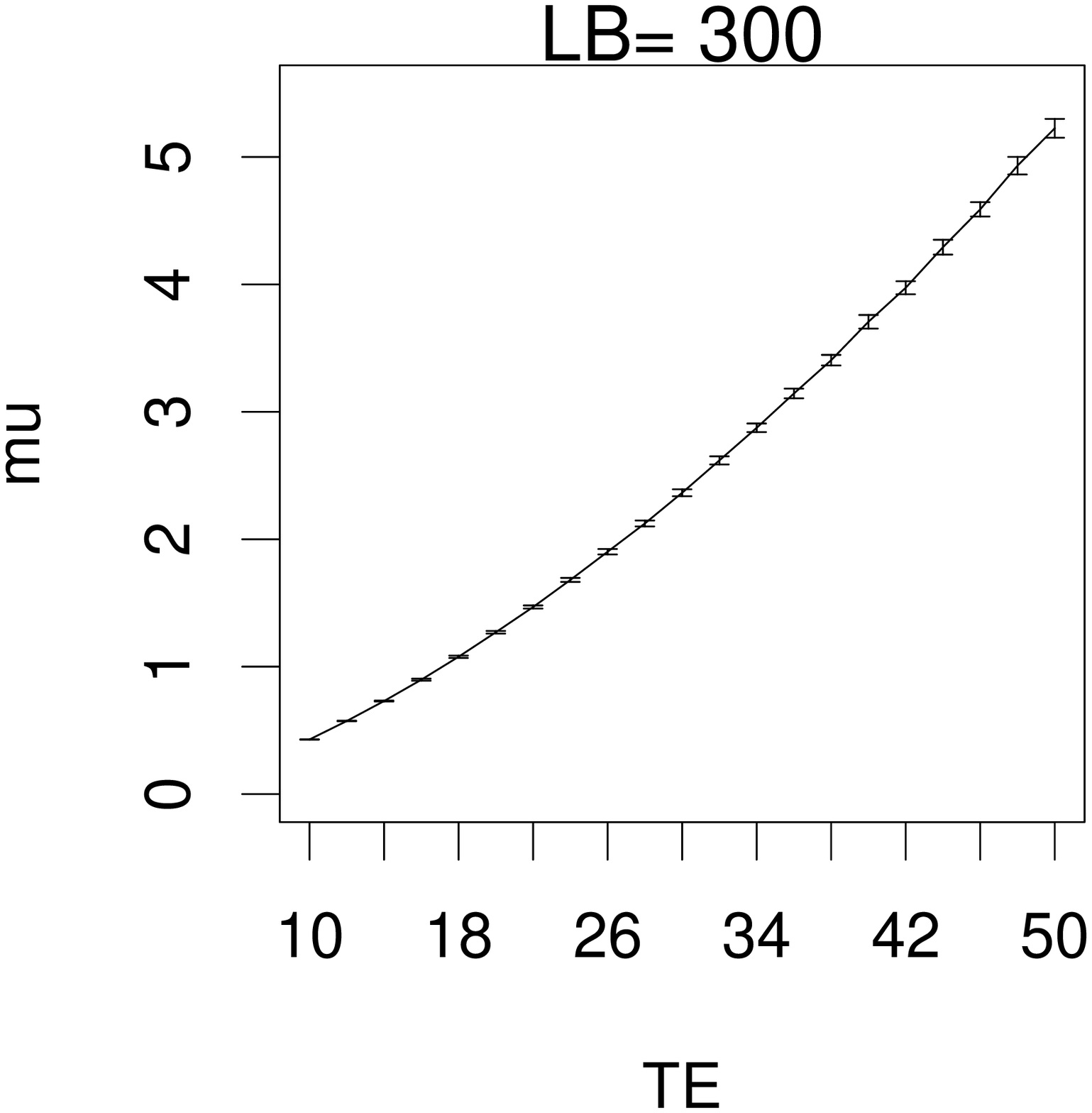}
  \includegraphics[scale=0.27]{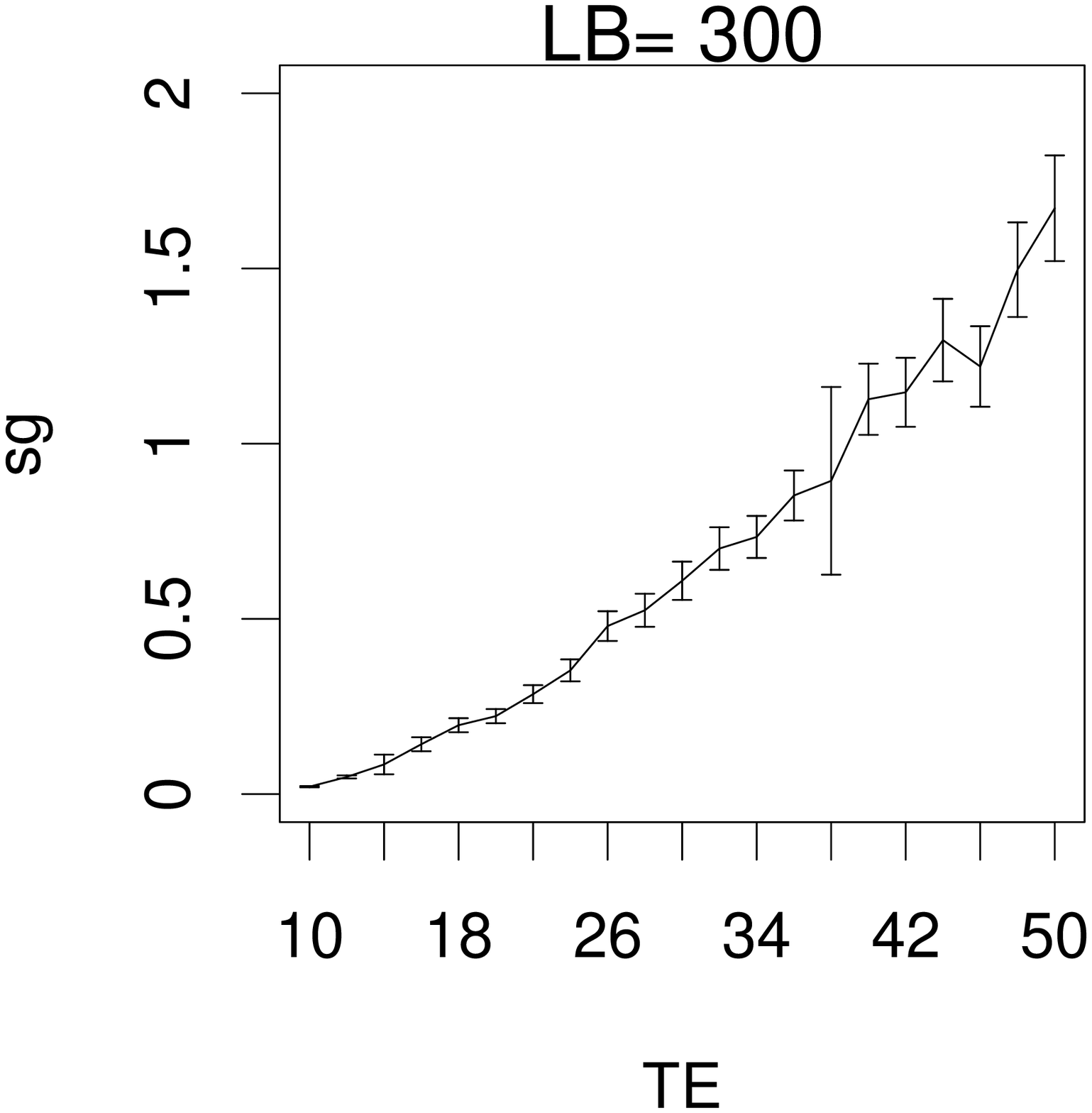}
  \includegraphics[scale=0.27]{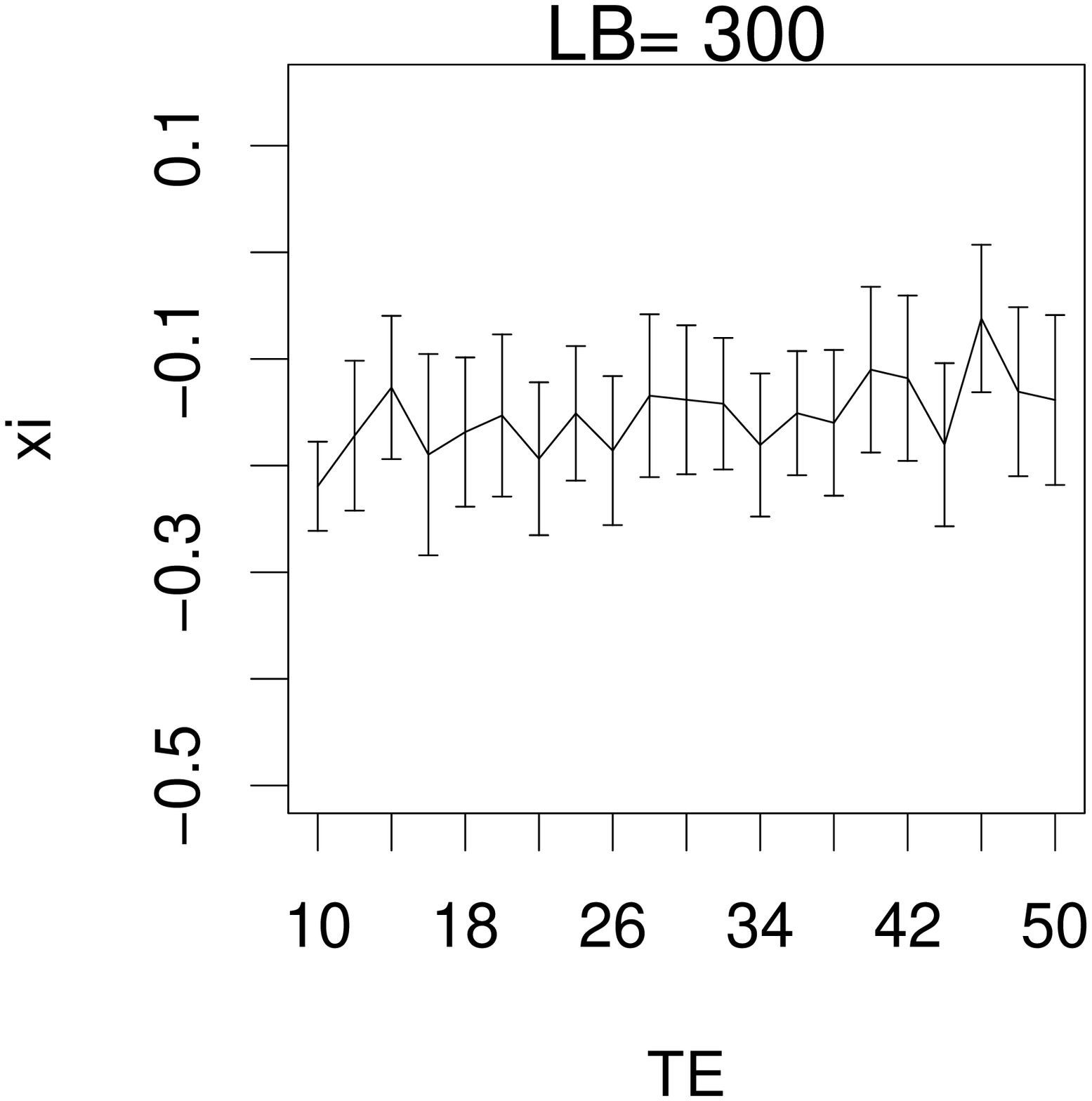}\\[5pt]
  \includegraphics[scale=0.27]{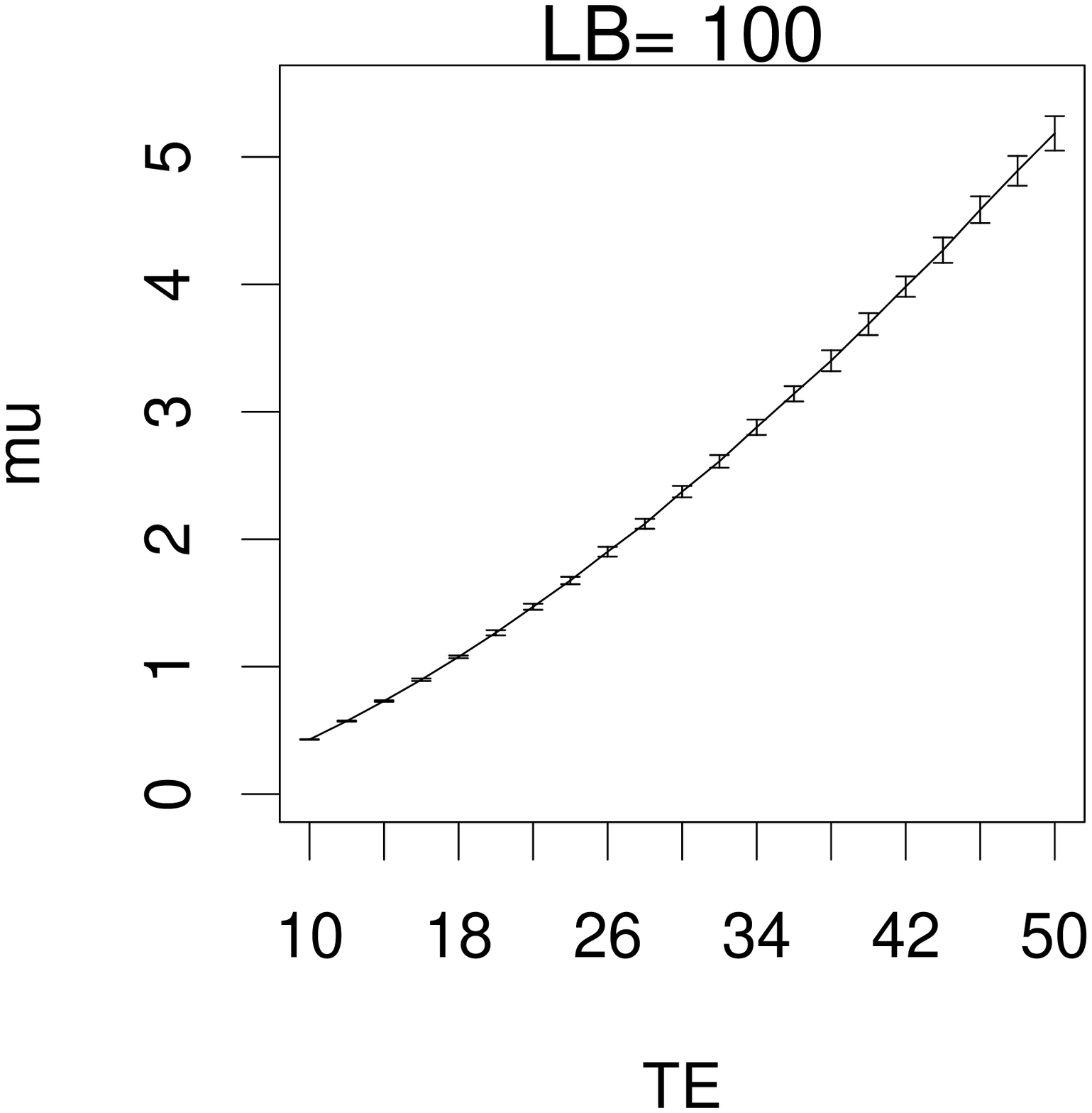}
  \includegraphics[scale=0.27]{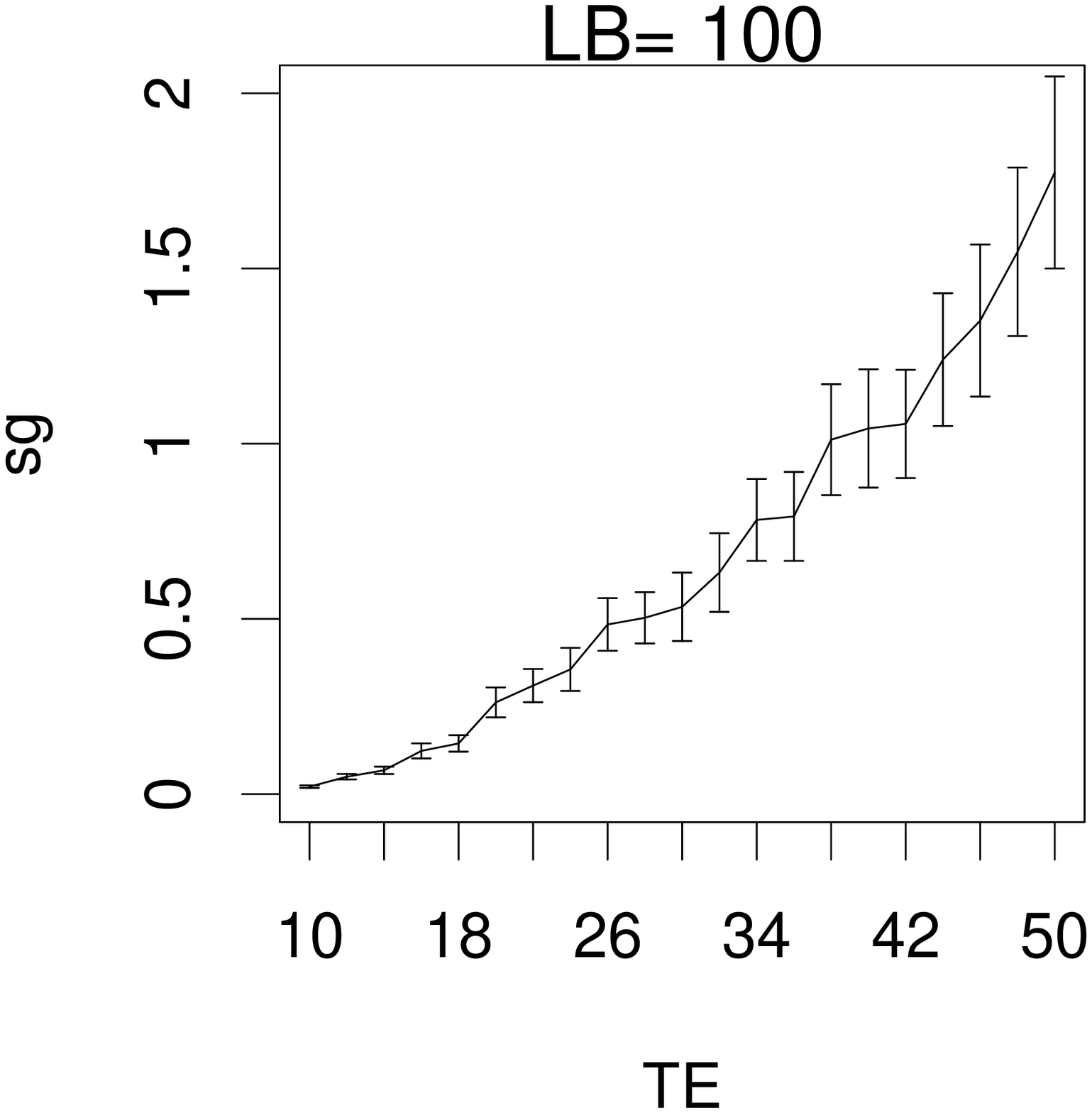}
  \includegraphics[scale=0.27]{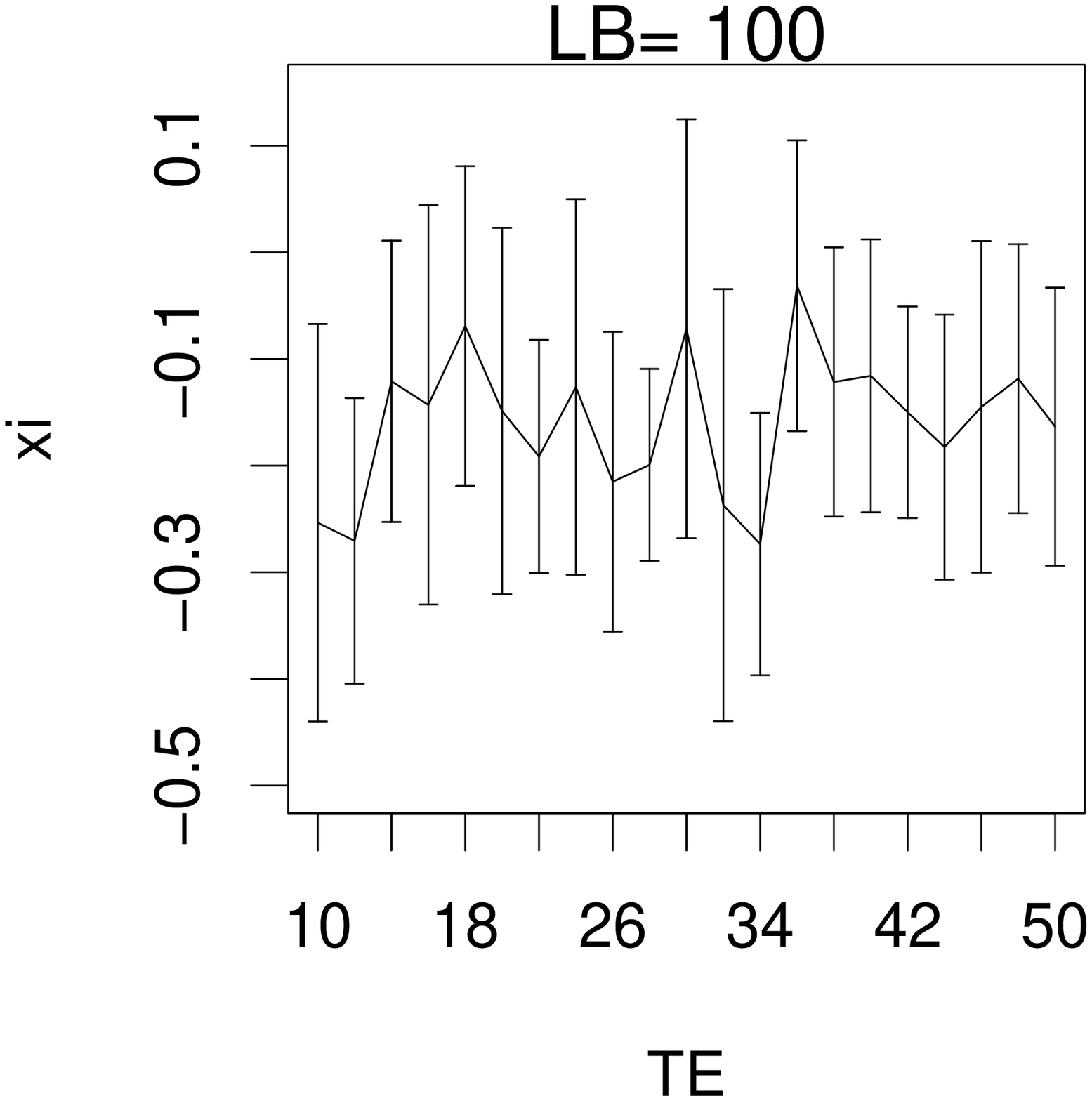}
  \caption{\baselineskip=24pt
    Parameters $(\mube^i(T_E^i),\sibe^i(T_E^i),\xibe^i)$
    (from left to right, respectively)
    of the best estimate GEV model $G_{\pbe^i,\qbe^i}(\boldsymbol{z}^i)$
    evaluated at the central point $T_E^i$ of each of
    the 21 intervals~\eqref{TEintervals}.
    From top to bottom the trend intensity $\speedT$ is equal to
    $2/(1000\, \textrm{years})$,
    $2/(300\, \textrm{years})$,
    and $2/(100\, \textrm{years})$, respectively.
  }
  \label{fig:GEV1000}
\end{figure}

The set-up of the present analysis (see \secref{data}) has been chosen
to allow comparison of the non-stationary GEV inferences with the results
of Part I, obtained from statistically stationary time series.
To perform the comparison, for each $i=1,\ldots,21$
the best estimate parameters $(\mube^i(t),\sibe^i(t),\xibe^i)$
inferred from data block $B^i$ are first expressed as functions of $T_E$
inside the interval~\eqref{TEintervals}.
This is achieved by inverting the trend formula~\eqref{TEt},
(writing time as a function of $T_E$):
\begin{equation}
  \label{eq:tTE}
  t(T_E)=\frac{T_E-T_E^0+1}{\speedT}, \qquad T_E\in[9,51].
\end{equation}
and inserting this into the expression of $(\mube^i(t),\sibe^i(t),\xibe^i)$.
This yields functions that are denoted as
$(\mube^i(T_E),\sibe^i(T_E),\xibe^i)$.  These are evaluated at the central
point $T_E^i$ of the interval of definition and plotted in \figref{GEV1000}.
Confidence intervals are given by the same worst case
estimate~\eqref{worstcase} used for \figref{GEVtime}.  A rather smooth
dependence on $T_E$ is observed, especially for the GEV parameters $\mu$ and
$\sigma$.  The location parameter $\mu$ turns out to be not very sensitive to
changes in the trend intensity, being much more sensibly dependent on
variations in $T_E$.  Moreover its confidence intervals are always very small
(relatively to the size of $\mu$).

\begin{figure}[hb]
  \centering
  \psfrag{TE}{$T_E$}
  \psfrag{mu}{$\mu$}
  \psfrag{sg}{$\sigma$}
  \psfrag{xi}{$\xi$}
  \includegraphics[scale=0.27]{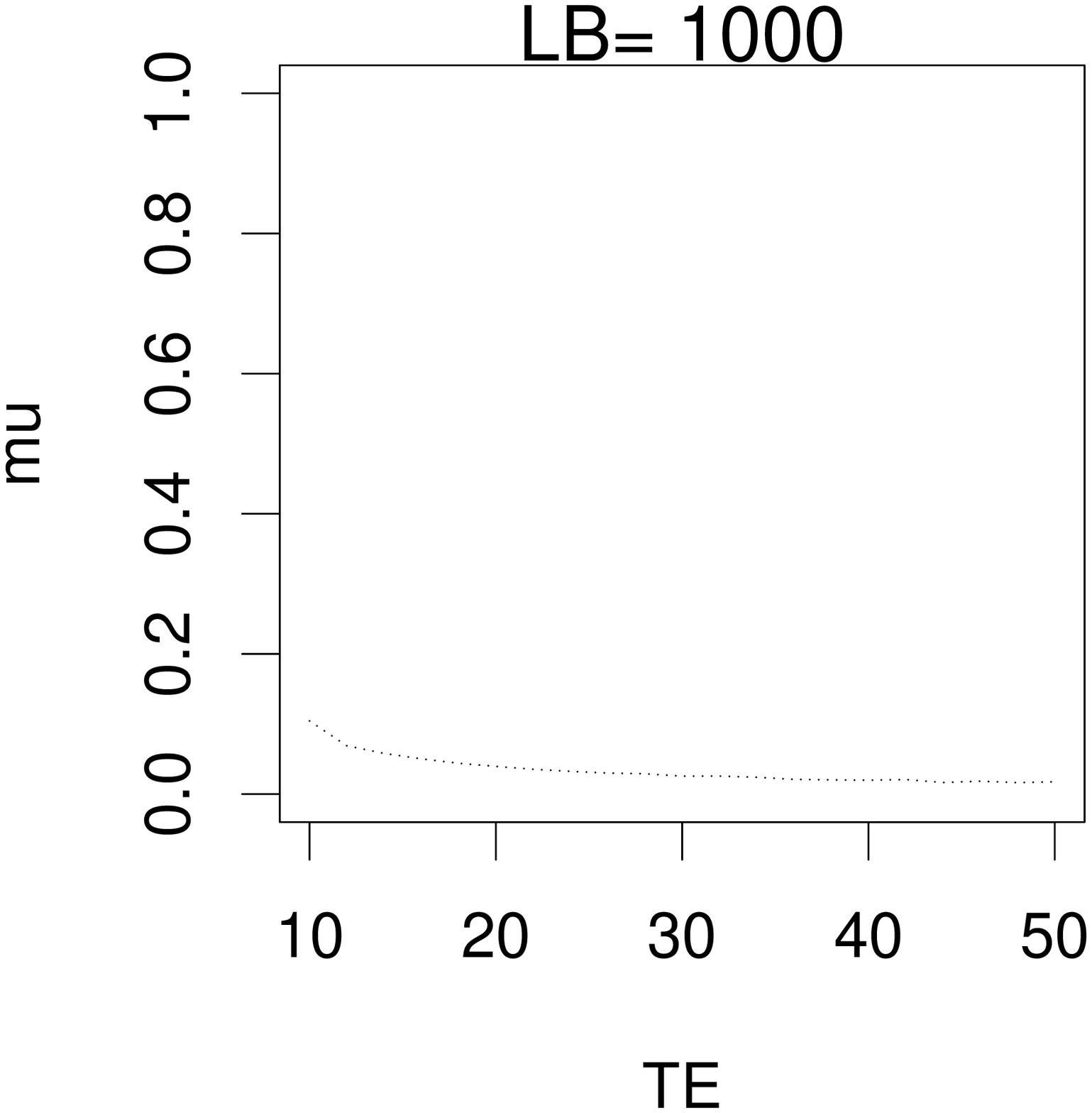}
  \includegraphics[scale=0.27]{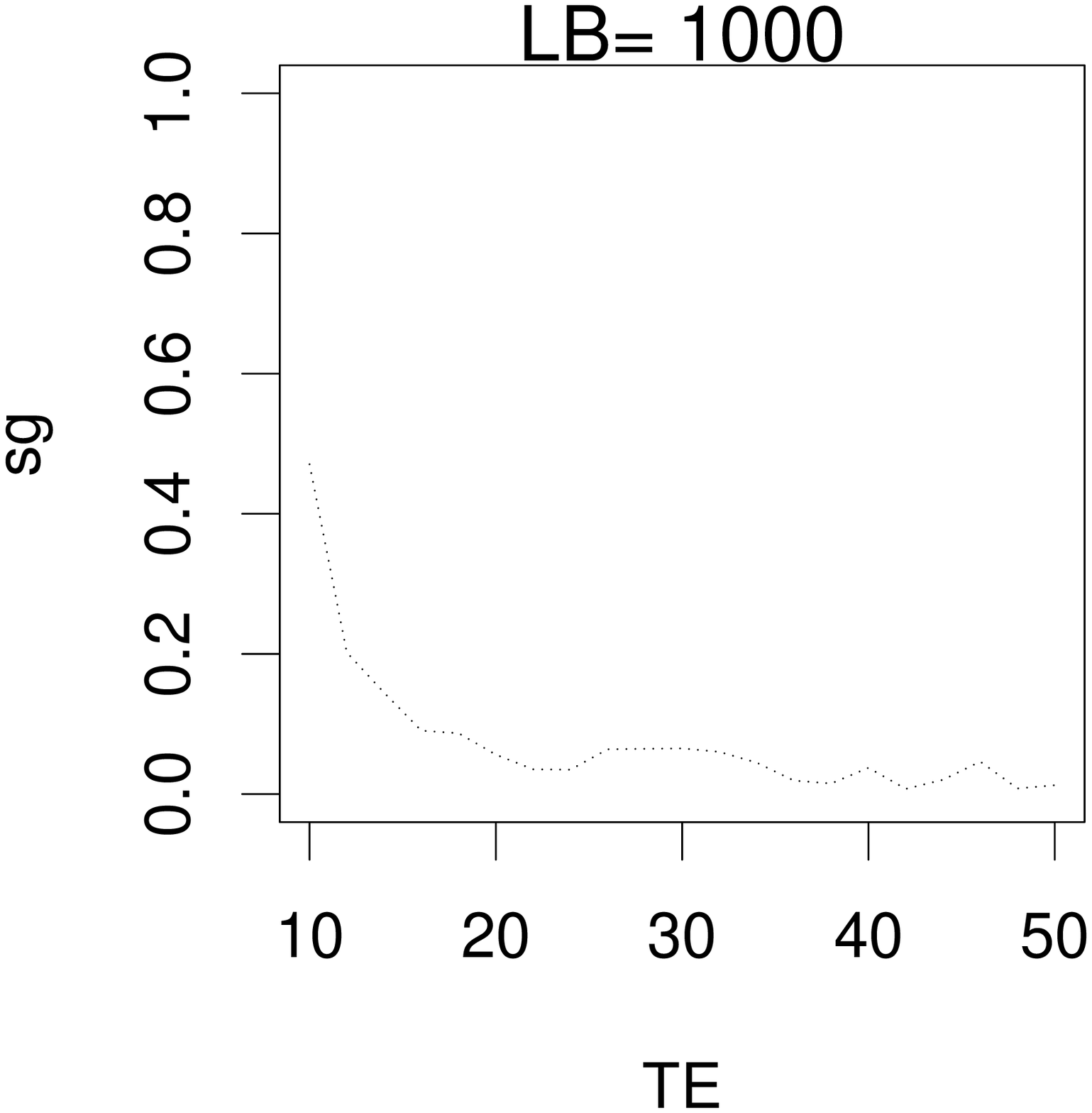}
  \includegraphics[scale=0.27]{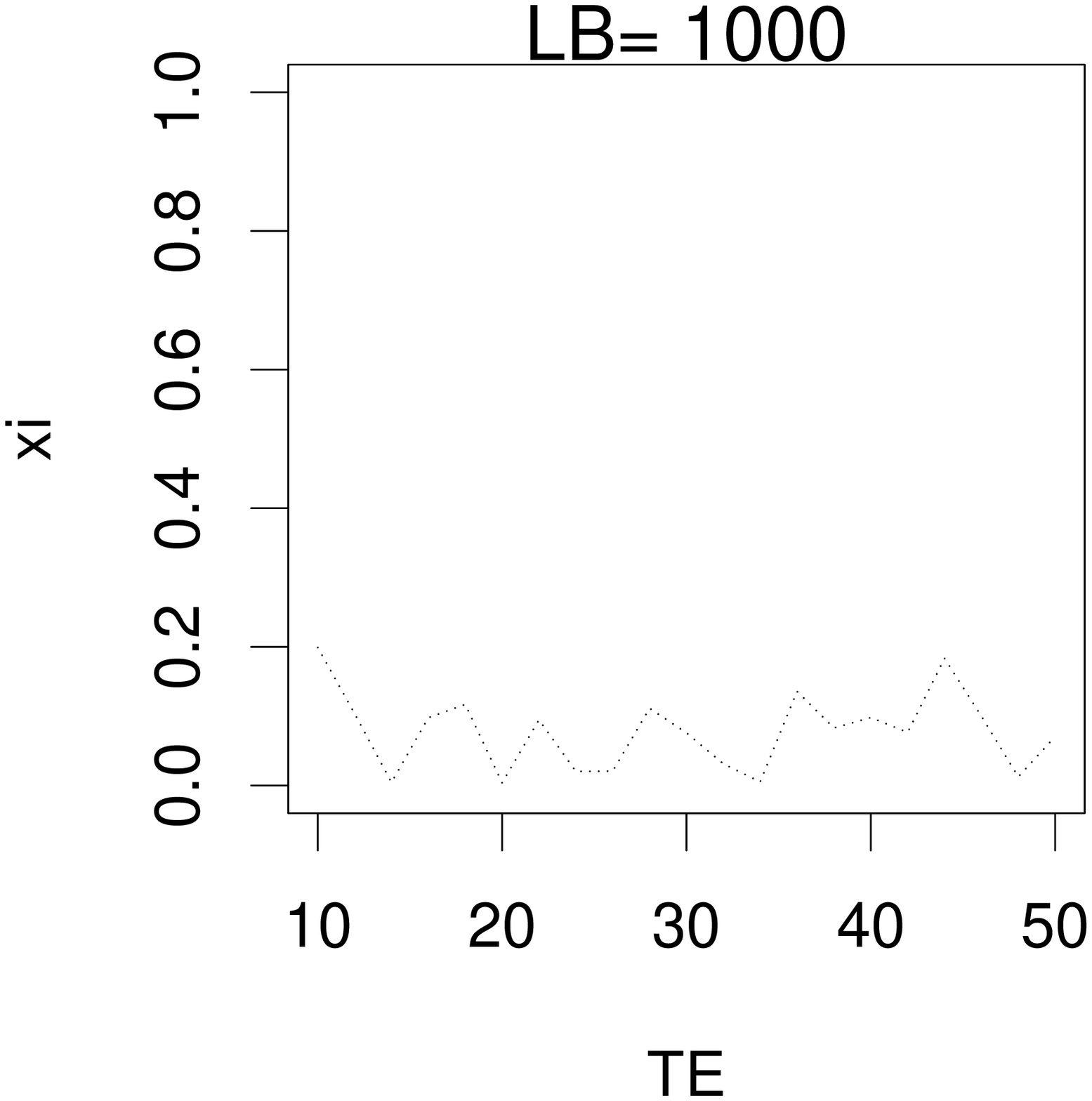}\\[5pt]
  \includegraphics[scale=0.27]{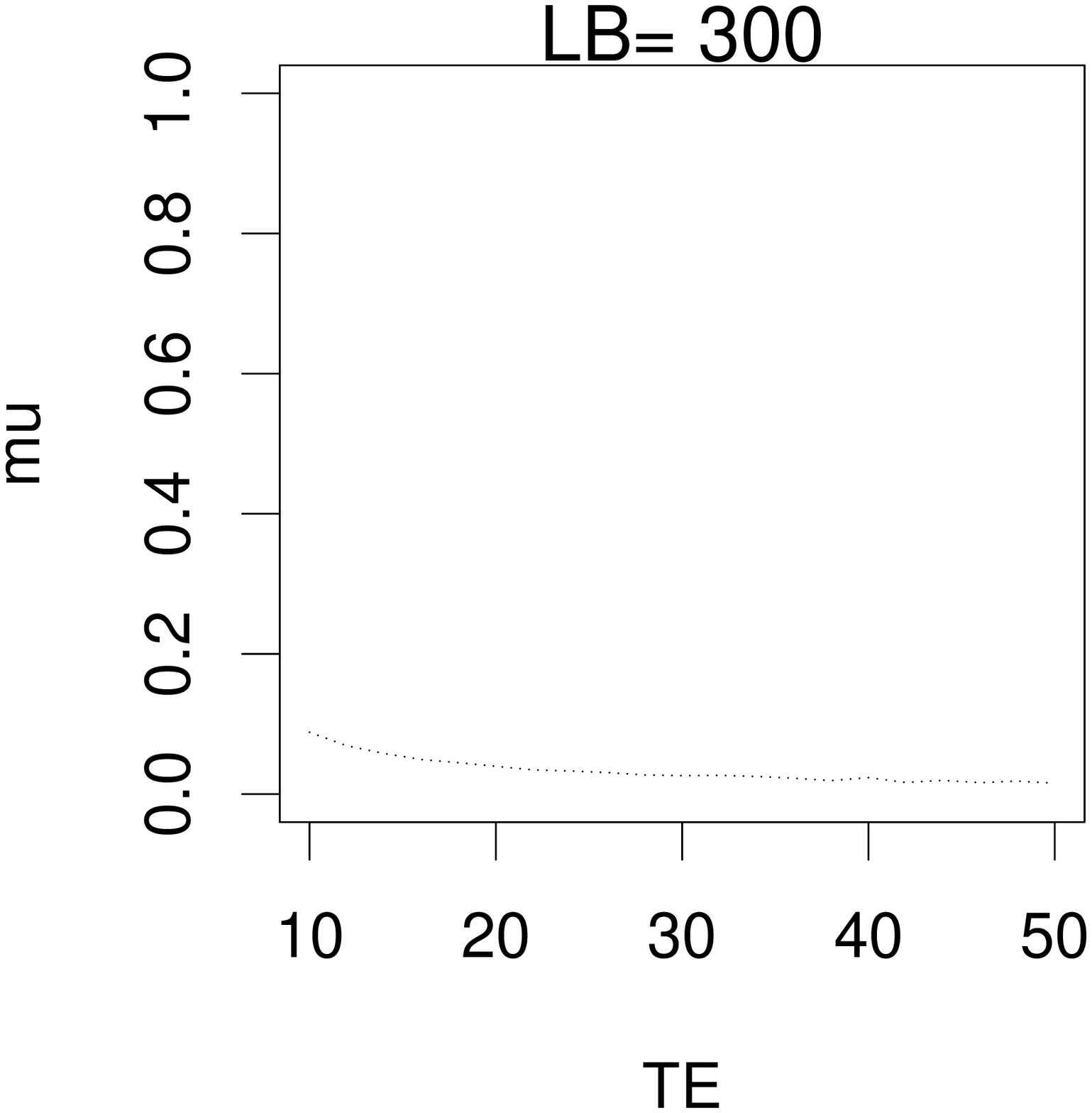}
  \includegraphics[scale=0.27]{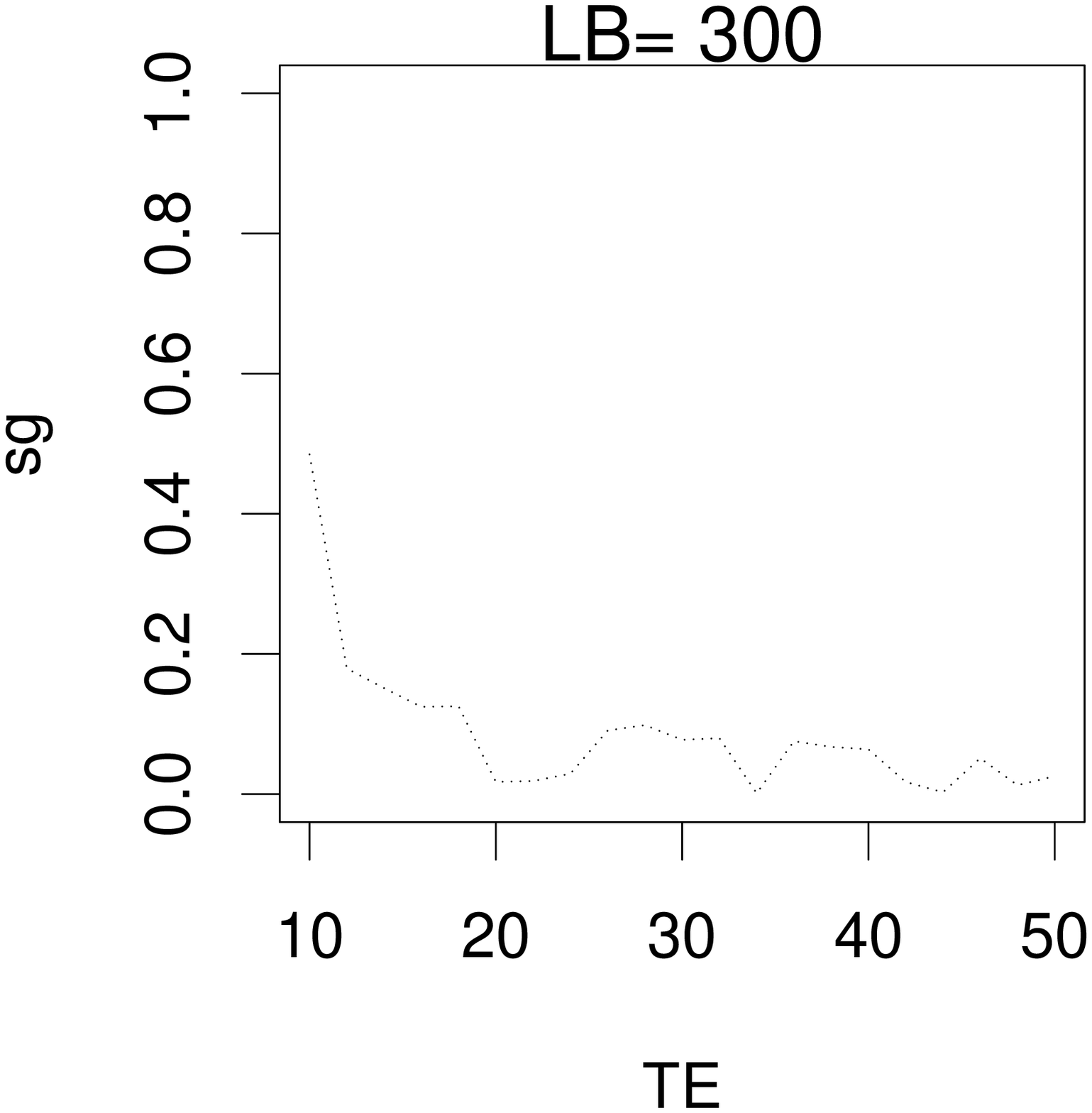}
  \includegraphics[scale=0.27]{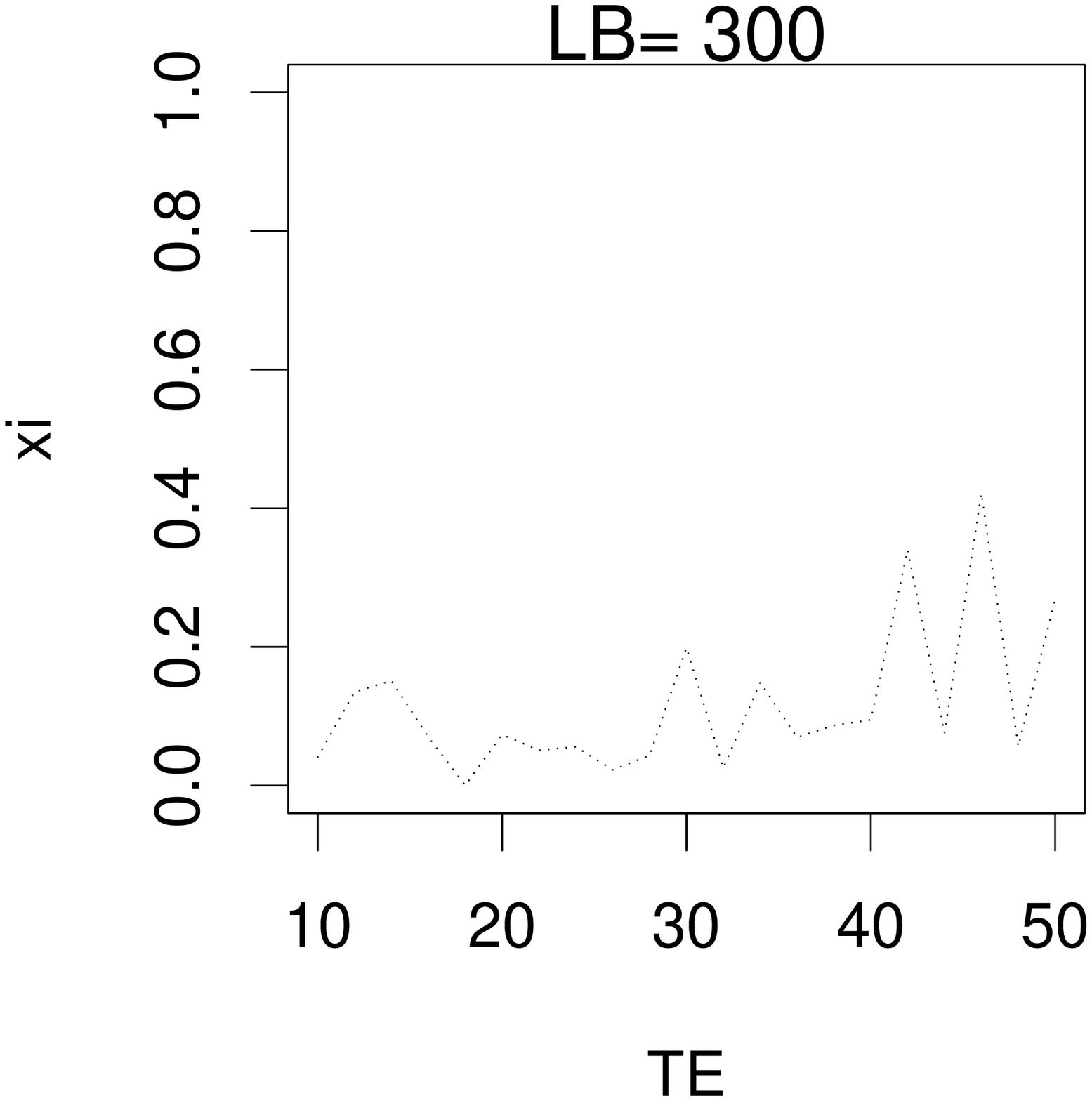}\\[5pt]
  \includegraphics[scale=0.27]{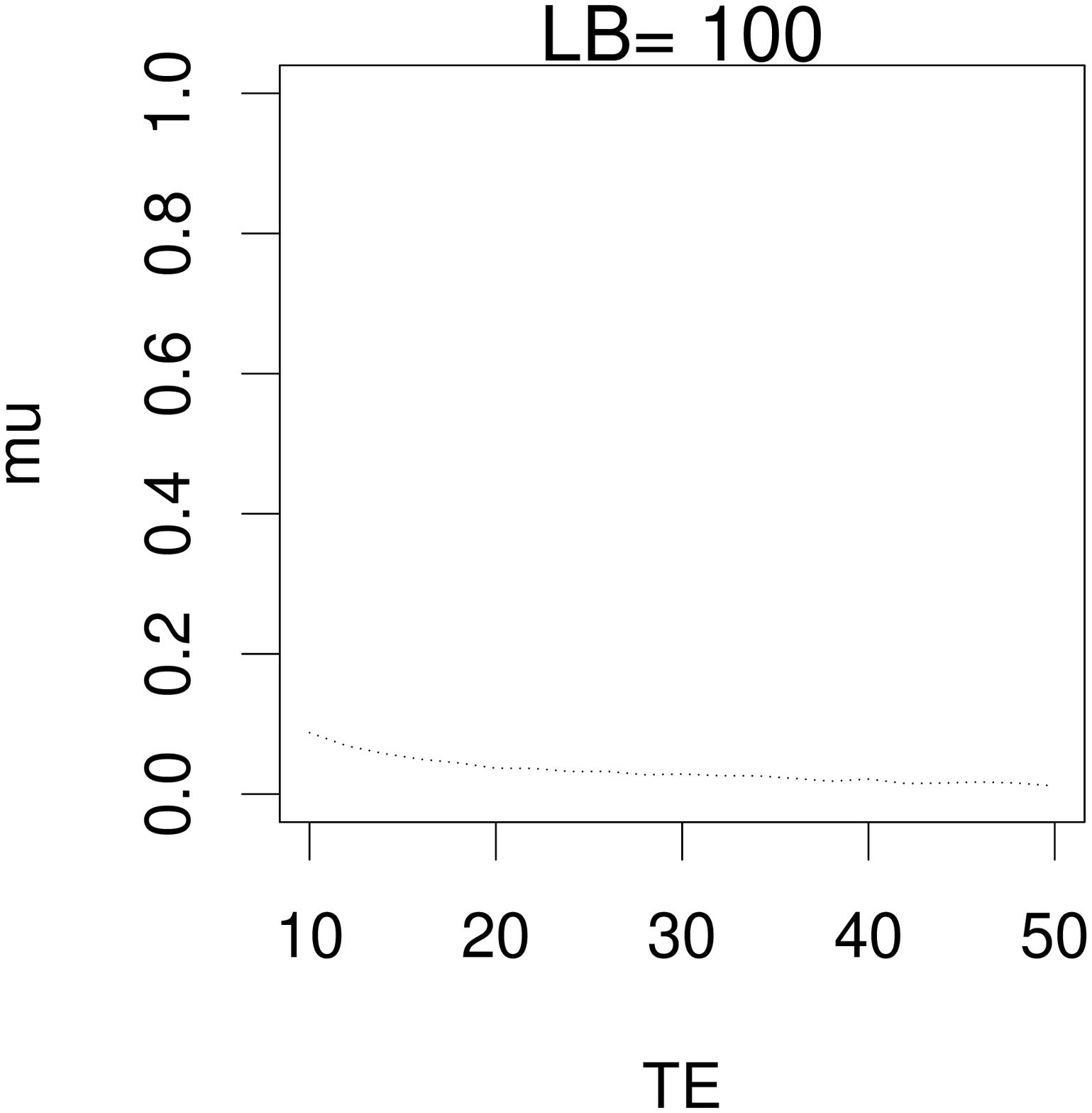}
  \includegraphics[scale=0.27]{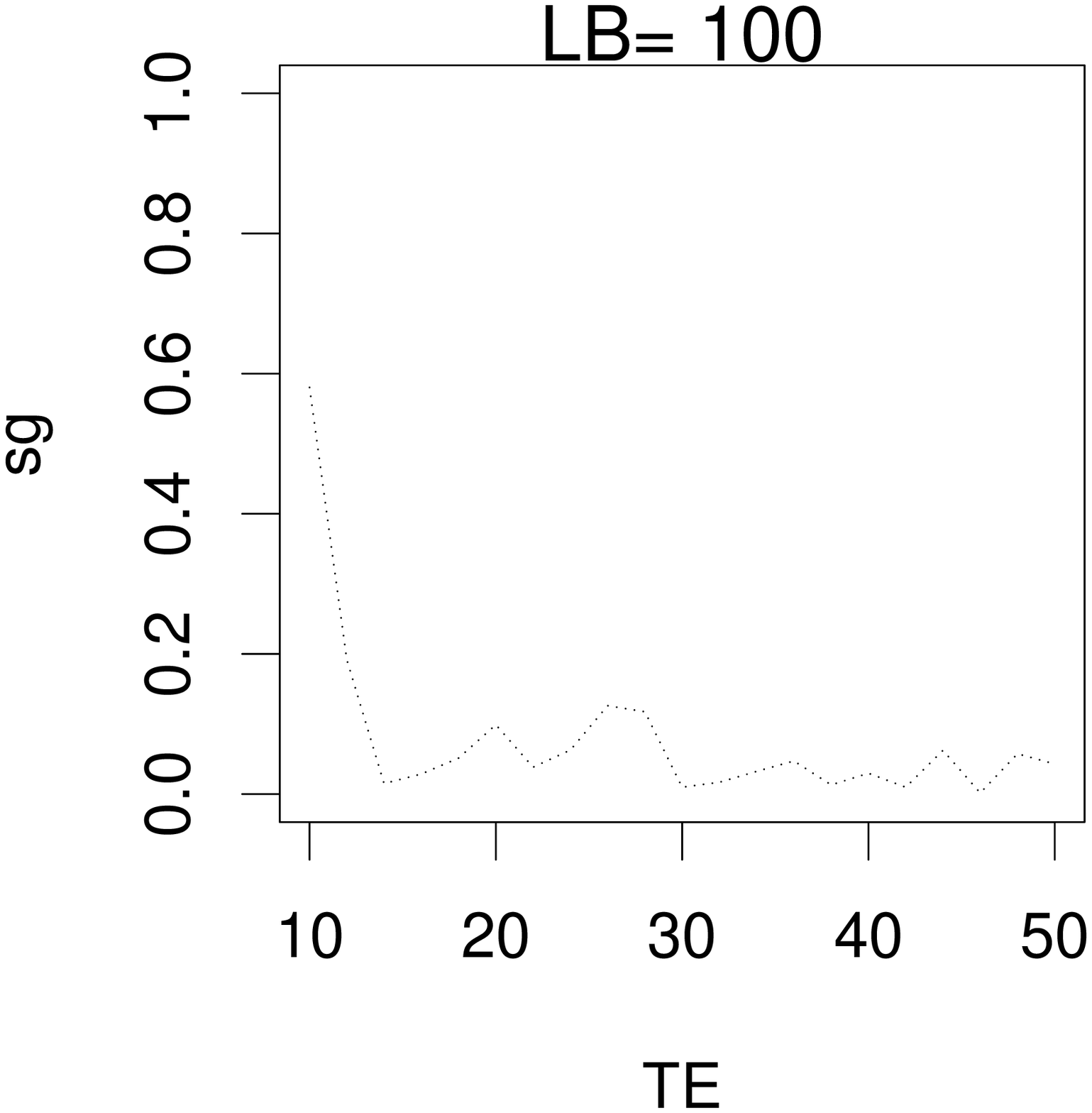}
  \includegraphics[scale=0.27]{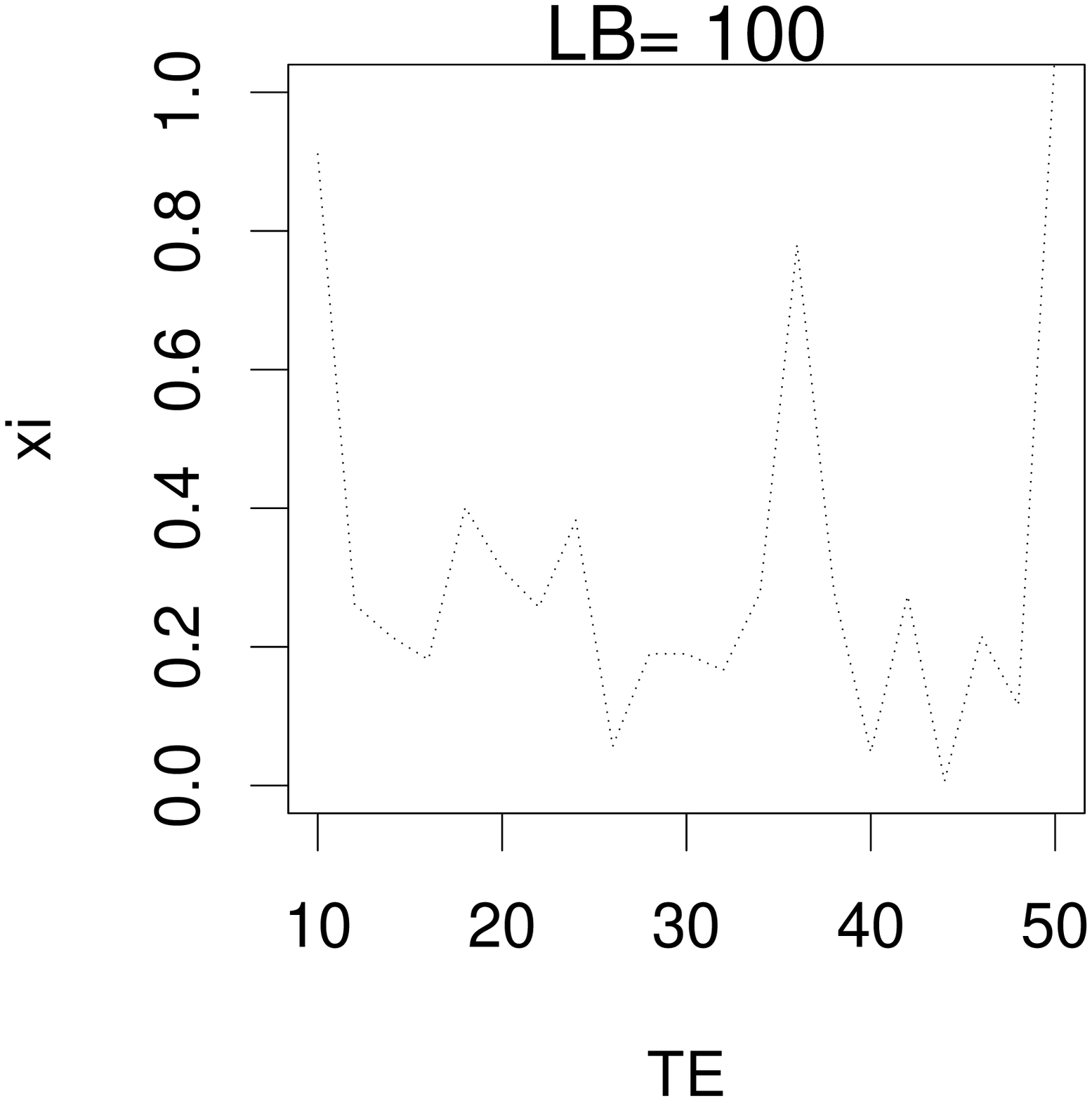}
  \caption{\baselineskip=24pt
    Same as \figref{GEV1000} for the estimates
    obtained with the stationary data in Part I, see text for details.  The
    time-dependent estimates of \figref{GEV1000} are plotted with dotted
    lines.  }
  \label{fig:cfrGEV}
\end{figure}

Denote by $G_{0,0}(\boldsymbol{w}^i)$ the time-independent GEV models
inferred in Part I from stationary sequences $\boldsymbol{w}^i$
of 1000 yearly maxima, computed with $T_E$ fixed at $T_E^i$.
Denote as $\tilde{\mu}^i$, $\tilde{\sigma}^i$, and $\tilde{\xi}^i$
the inferred values of the GEV parameters of $G_{0,0}(\boldsymbol{w}^i)$.
Since the graphs of the parameters $\tilde{\mu}^i$,
$\tilde{\sigma}^i$, and $\tilde{\xi}^i$ versus $T_E^i$ very closely
match those in \figref{GEV1000}, comparison with the stationary data
is presented under the form of relative differences (\figref{cfrGEV}).
To be precise, on the left column the absolute values of the ratios
\begin{equation}
  \label{eq:ratio}
  \frac{\mube^i(T_E^i)-\tilde{\mu}^i}{
      \mube^i(T_E^i)+\tilde{\mu}^i}
\end{equation}
are plotted against $T_E^i$ (similarly for the GEV parameters $\sigma$
and $\xi$). Remarkable agreement is obtained for the parameter $\mu$:
the relative differences less than $10\%$ and drop below $5\%$
for large $T_E$ and for all considered trend intensities.
Excellent agreement is also obtained for $\sigma$ (particularly for
large $T_E$) and for $\xi$ except for the fastest trend intensity
$\speedT$ $=$ $2/(100\, \textrm{years})$. In the latter case, indeed,
the sample uncertainty is as large as (or even larger than)
the estimates self.

We emphasize that inferring time-independent models $G_{0,0}(\boldsymbol{z}^i)$
from the non-stationary data $\boldsymbol{z}^i$
would induce very large errors (particularly in
the scale and shape parameters), compare \figref{diagnotrend}.
A much better (even surprising) agreement between the stationary and
non-stationary estimates is obtained with the procedure described
in the previous section:
first fitting the time-dependent model $G_{\pbe^i,\qbe^i}(\boldsymbol{z}^i)$
and then evaluating its parameters at the central point $T_E^i$.
There is agreement even in the estimates of the parameter $\xi$,
which is usually the most difficult one to infer.
Indeed, the inferred values are negative.
in the case of stationary time series, since the attractor is bounded
and since the energy observable $E(t)$ is a continuous function of the
phase space variables, the total energy is bounded on any orbit lying on
(or converging to) the attractor. Therefore, the total energy
extremes are \emph{necessarily} Weibull distributed
($\xi$ is negative) Part I.
Although this property is not bound to hold for non-stationary forcing,
it is still verified, see \tabref{best1000}.

\begin{figure}[p]
  \centering
  \psfrag{lTE}[]{$\log(T_E)$}
  \psfrag{lTEb}{$\log(T_E^b)$}
  \psfrag{lmu}[]{$\log(\mu)$}
  \includegraphics[width=0.32\textwidth]{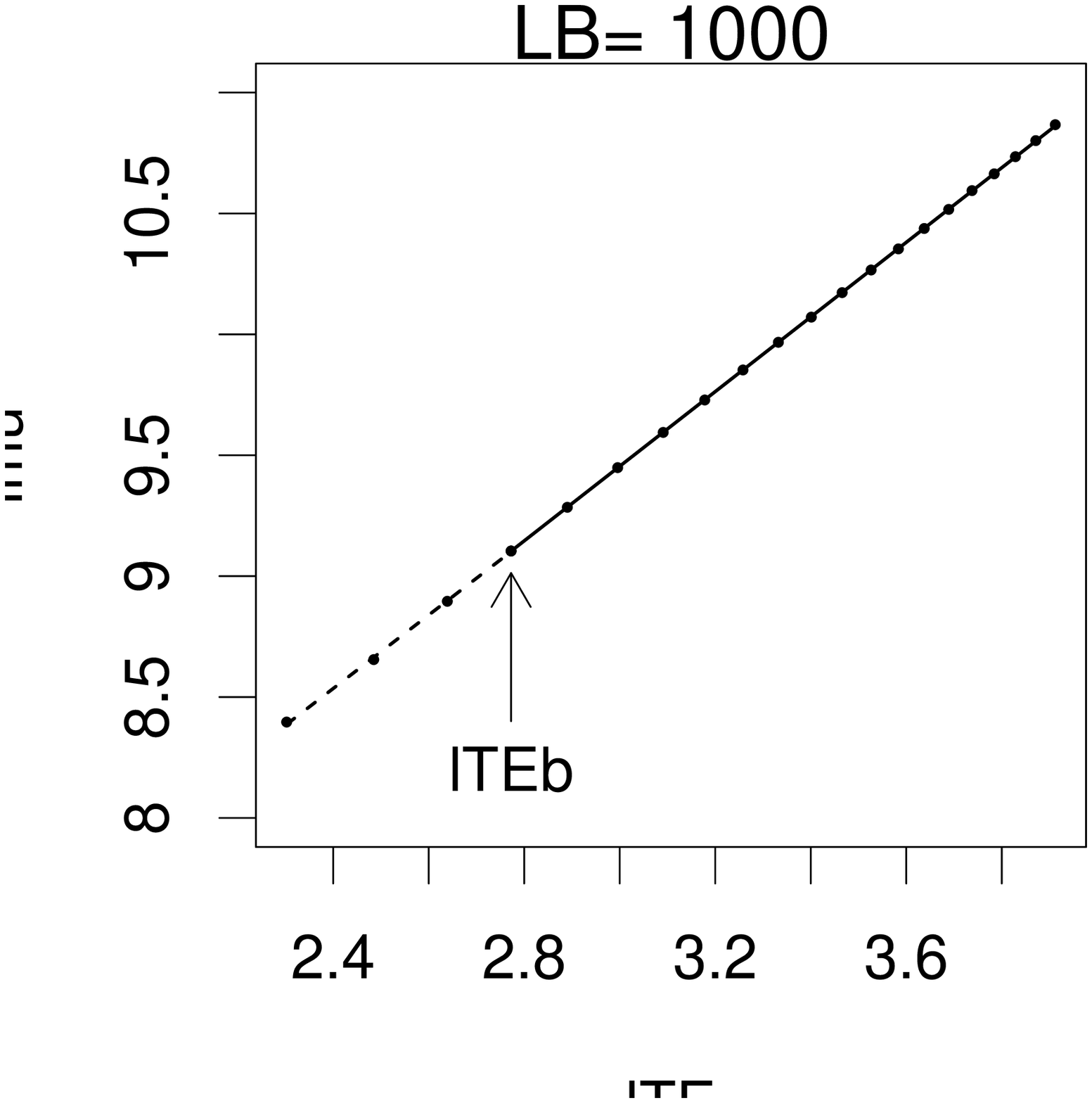}
  \psfrag{lmu}{}
  \includegraphics[width=0.32\textwidth]{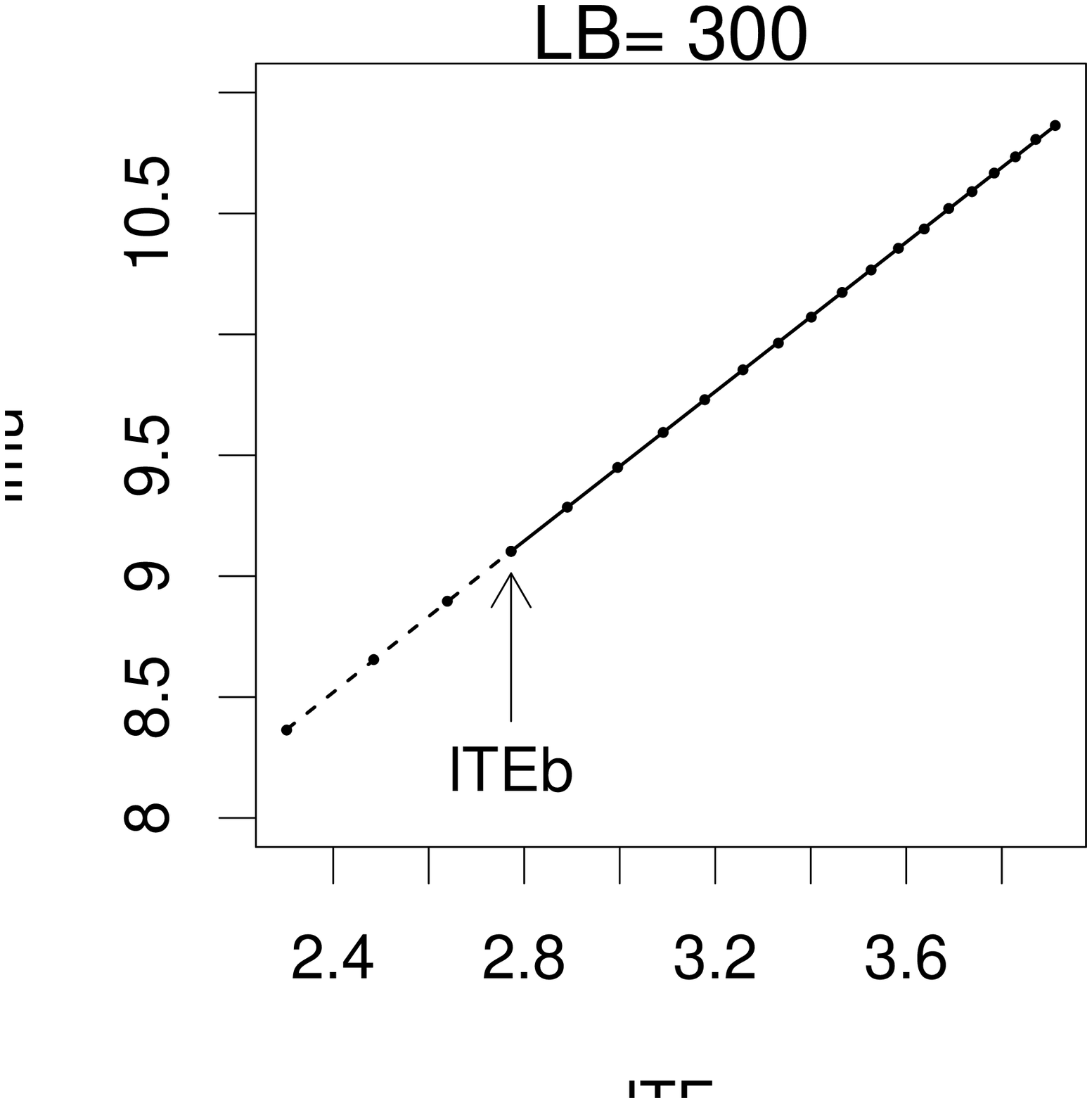}
  \includegraphics[width=0.32\textwidth]{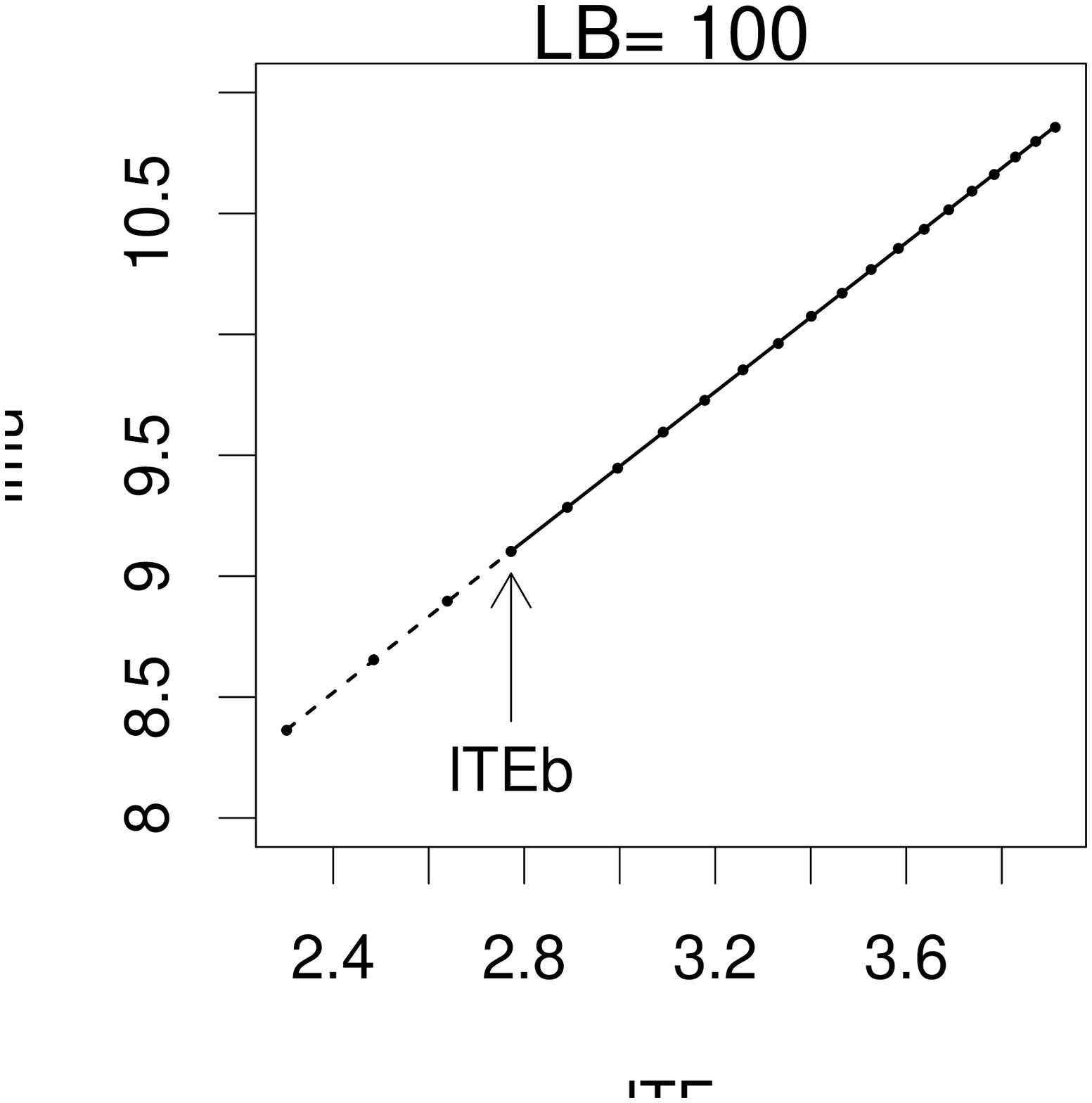}
  \caption{\baselineskip=24pt
    Power law fits of the inferred values of $\mube^i(T_E^i)$
    as a function of $T_E^i$ (see~\eqref{TEi}).
    From left to right: trend intensities of
    $2/(1000\, \textrm{years})$, $2/(300\, \textrm{years})$,
    and $2/(100\, \textrm{years})$ have been used.
    In each case, there are two intervals of $T_E$ characterized
    by different scaling law, separated by a point
    $T_E^b$, compare \tabref{powermuT}.
  }
  \label{fig:powermuT}
\end{figure}

\begin{figure}[p]
  \centering
  \psfrag{lTE}[]{$\log(T_E)$}
  \psfrag{lTEb}{$\log(T_E^b)$}
  \psfrag{lsg}[]{$\log(\sigma)$}
  \includegraphics[width=0.32\textwidth]{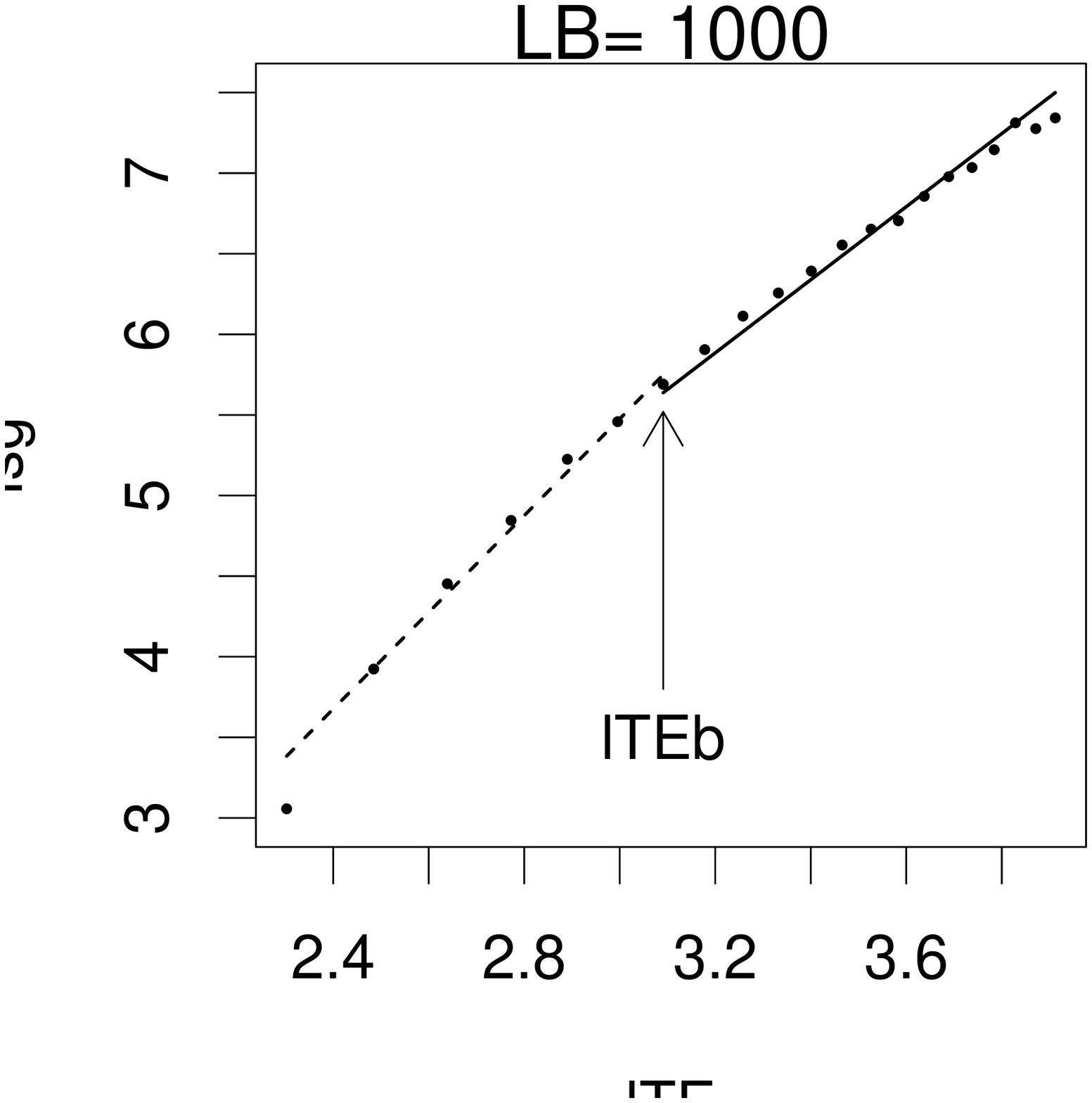}
  \psfrag{lsg}{}
  \includegraphics[width=0.32\textwidth]{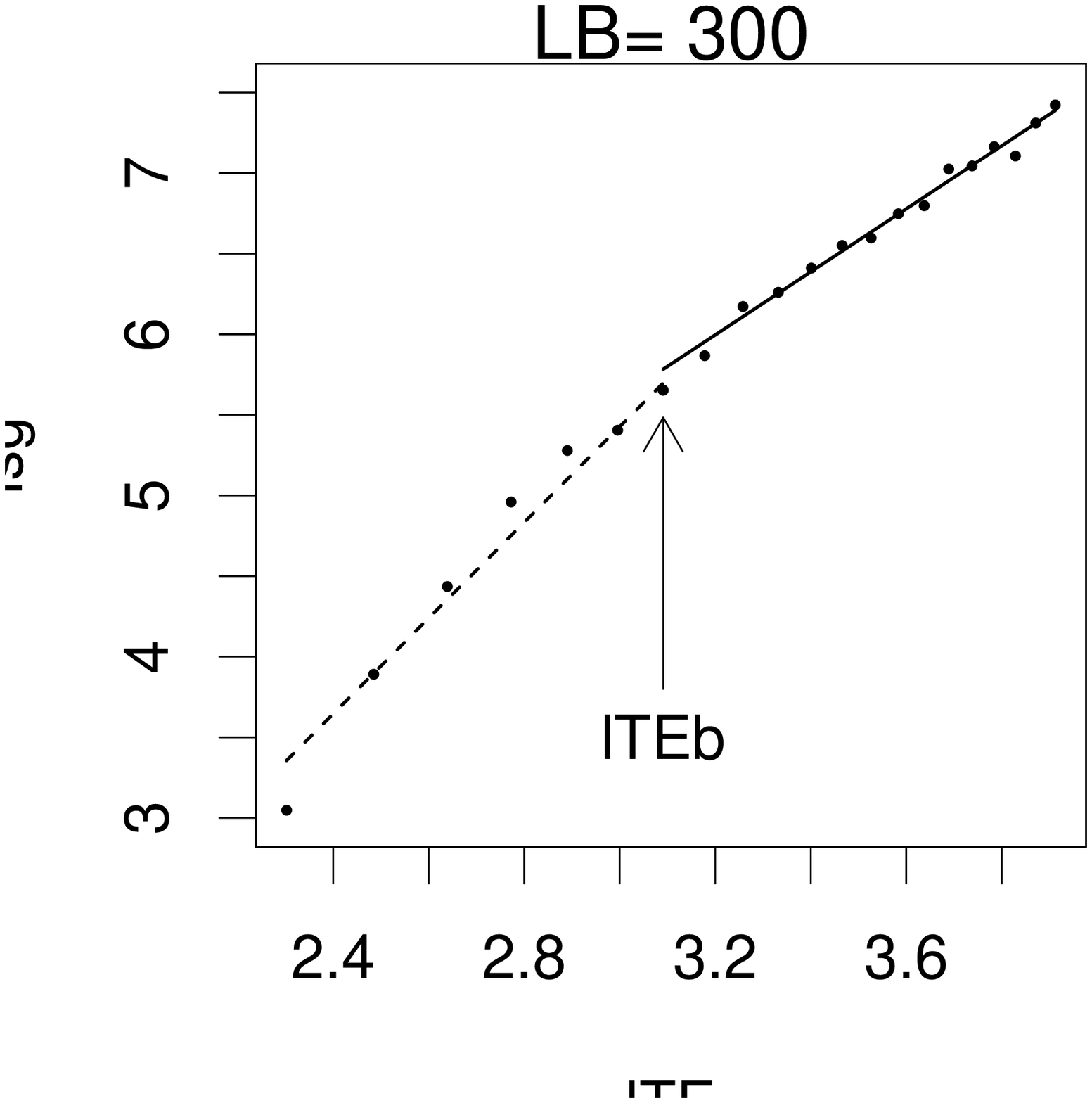}
  \includegraphics[width=0.32\textwidth]{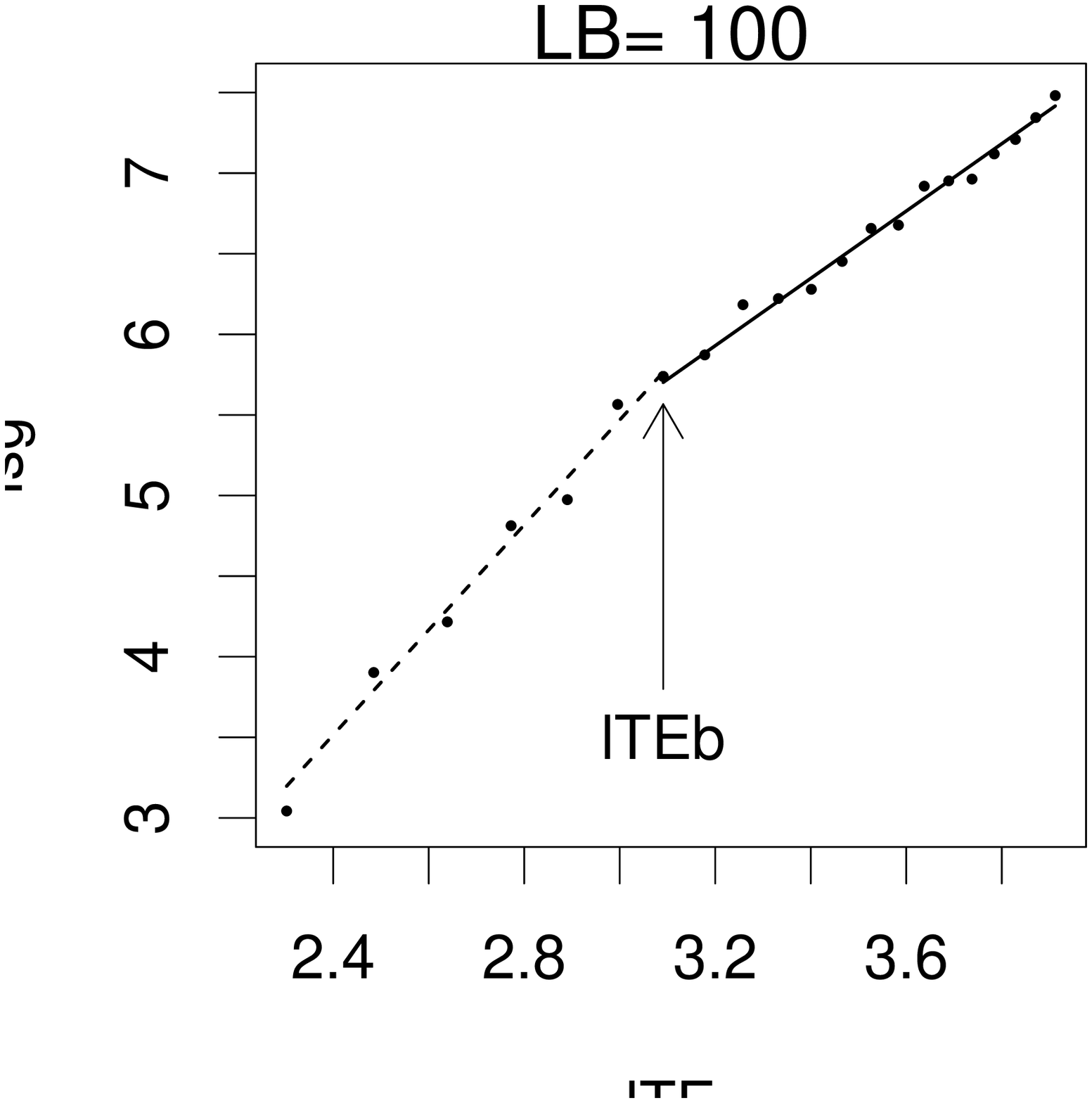}
  \caption{\baselineskip=24pt
    Same as \figref{powermuT} for the inferred values $\sibe^i(T_E^i)$.
  }
  \label{fig:powersigmaT}
\end{figure}

Two distinct power law regimes are identified for the GEV
parameters $(\mube^i,\sibe^i,\xibe^i)$ as functions of $T_E^i$,
having the form
\begin{equation}
  \label{eq:powermusigma}
  \mube^i(T_E^i) =\alpha_\mu  (T_E^i)^{\gamma_\mu}    \quad\text{and}\quad
  \sibe^i(T_E^i) =\alpha_\sigma (T_E^i)^{\gamma_\sigma},
\end{equation}
ee \figref{powermuT} and \figref{powersigmaT}.
The values of the fitted exponents $\gamma_\mu$ and $\gamma_\sigma$ in each
scaling regime are reported in \tabref{powermuT} and \tabref{powersigmaT},
respectively.
A similar power law dependence of the GEV parameters on $T_E$
was already observed in Part I for the stationary data
sets $\boldsymbol{w}^i$: indeed, the exponents obtained there
are very similar to those in \tabref{powermuT} and \tabref{powersigmaT},
particularly for large $T_E$.
The lack of a power-law scaling regime for the parameter $\sigma$
for small $T_E$ explains both the more pronounced differences between
the stationary and non-stationary estimates (\figref{cfrGEV})
and the necessity of including a quadratic term in $\mu$
and/or a linear term for $\sigma$ in the statistical model
to get acceptable inferences.
This highlights the strongly nonlinear behaviour of the
baroclinic model, whose response to changes of $T_E$
has different features depending on the considered range of variation.

Two factors explain the qualitative analogies and the quantitative agreements
between the time-dependent models discussed here and the stationary results of
Part I.  First of all, the trend intensity imposed on $T_E$ in all cases is
sufficiently slow with respect to the time of relaxation of the baroclinic
model to the statistics of extreme values of the total energy. For clarity, we
emphasize that the latter time scale is that used in \secref{GEVproblems} to
define the ``adiabatic'' hypothesis: it is the time necessary to obtain a good
sampling of the SRB measure on the attractor, provided that one may consider
the system as ``frozen'' (with constant $T_E$) for sufficiently long time
spans. We do not know whether this time scale bears any physical relation with
other time scales, such as those of baroclinic instability or low-frequency
variability (both have been described in~\cite{SM88} for the present model).
The second factor is that the system's statistical behaviour responds rather
smoothly to the imposed time-dependent variation of the parameter $T_E$. This
smooth dependence on $T_E$ of the statistical properties of the baroclinic
model was analyzed in detail in~\citet{LSV06,LSV06a} by considering not only
global physical quantities such as total energy and average wind profiles, but
also finer dynamical indicators, such as the Lyapunov exponents and dimension.
Both properties of smoothness and ``adiabaticity'' are of crucial importance
in order to justify the usage of non-stationary GEV models that are (locally)
smooth functions of time, such as our polinomial family $G_{p,q}$.

\section{Summary and Conclusions}
\label{sec:conclusions}

In this paper we have proposed a general, although not universal, framework
for the analysis of trends in extremes of climatic time series.  When all the
shortcomings which are present in datasets and observations have to be
considered, a rigorous definition of extremes and a neat, clean, and legible
approach to the evaluation of trends is necessary in order to get useful and
reliable information~\citep{ZHZK05}.  The time-dependent approach allows to
express the inferred GEV distributional parameters as functions of time.  As
expected, it is found that trend in the statistics of extreme values is
detectable in a reliable way, provided that the record of observations is
sufficiently long, depending on the time scale of the trend itself.  Trend
inference and assessment is much more problematic than in the statistically
stationary inference. First of all, one is faced with a serious conceptual
problem: there is no ``operational'' definition of probability, since, to say
it in loose words, the time series is not a sampling of a unique probability
distribution, as it is in the stationary case.  Even if one assumes that the
time series is a realization of a sequence of random variables (with different
distributions), the statistical properties of the sample are altered by any
operation such as resampling or taking shorter subsamples, which makes
sensitivity studies somewhat ambiguous. One must assume that the
distributions of the random variables vary slowly and smoothly with time,
so that the time series contains sufficient sampling information
on the ``local'' (in time) statistical behaviour.

In the present context, we have adopted GEV models whose parameters are
polynomial in time: the location parameter $\mu$ is at most quadratic with
respect to time and the scale parameter $\sigma$ is at most linear in time.
Since the relation between the macroscopic forcing $T_E$ and time is
invertible, the time dependence of the inferred GEV models can be expressed as
a relation between the GEV parameters and $T_E$, showing rather interesting
properties.  The location and scale parameters feature power-law dependence
with respect to $T_E$, while the shape parameter has in all cases a negative
value. As expected, both results are in agreement with what obtained in the
companion paper (Part I) for stationary data.  Since the parameter $T_E$
increases monotonically in the simulations with the baroclinic model, the
system certainly does not possess any invariant measure.  However, the results
suggest that, as $T_E$ increases, the system explores statistical states which
vary smoothly with $T_E$ and whose properties are locally quite similar to
those obtained in the stationary setting.  This is even captured for the
relatively fast trend intensity $\speedT=2/(100\, \textrm{years})$.  The
proposed explanation is that:
\begin{enumerate}
\item
  the system's statistical properties depend rather smoothly on $T_E$
  (also compare~\citep{LSV06,LSV06a});
\item
  the adopted time-scales of variation of $T_E$
  (\emph{i.e.}, the trend intensity $\Delta T_E$) are sufficiently slow
  compared to the relaxation time to the statistics of extreme values.
\end{enumerate}
The second condition, that was explained in more detail in
\secref{GEVproblems}, amounts to the heuristic statement that for sufficiently
short time spans the system's statistical properties can be considered
\emph{frozen} to those holding for a corresponding value of $T_E$. The
possibility of using GEV models that are locally smooth (polynomial functions
of time) depends essentially on these two conditions. For example, for a
system having several bifurcations as the control parameter is changed the
time-dependent GEV modeling would be much more complicated.  This problem is
currently under investigation.  However, even if the above two conditions do
hold, the inference of time-dependent GEV models is valid locally in time,
that is, if the sequences of maxima used for the inference span not too large
time periods.  For large time spans, indeed, the non-linear response of the
baroclinic model to variations in $T_E$ becomes dominant and polynomial GEV
models are no longer suitable. On the other hand, if the sequences of maxima
used for the inference are too short (depending on the trend intensity), wrong
trend estimates may be obtained.

We conclude by observing that the present and the companion paper (Part I) are
devoted not merely to the statistical inference of extremes and their trends
but also to explore the possibility of using extreme statistics in diagnosing
the dynamical state of a geophysical fluid.  Our analysis of the problem
reveals, in fact, that diagnostics which is based on ``universal'' (GEV
theorem), robust (smoothness properties), simple (power-law scaling),
controllable (low-dimensional parametric) statistical models can help very
much in setting up well targeted models of the general circulation,
see~\citep{LSV06,LSV06a}.

There are several ways in which we plan to extend the present study.  First of
all, we have considered a rather global indicator, the total energy of the
system. Other choices might be to analyze the wave kinetic energy, the
available energy or also the maximum vorticity on the domain of the model,
which might behave differently as $T_E$ is changed.  Moreover, there are
delicate issues connected with reducing the scale from a global indicator to a
local one, such as the value of the wind on a grid point.  This brings into
play all complications due to the multifractality and the spatial dependence
of the process. A further development of the present work is the usage of
extreme statistics as a dynamical indicator, in the sense of \emph{process
  oriented metrics}~\citet{LCdARS06}. All these issues are currently under
investigation.

\section*{Acknowledgments}
The authors are indebted to Nazario Tartaglione for useful conversations.
This work has been supported by MIUR PRIN Grant
``Gli estremi meteo-climatici nell'area mediterranea:
propriet\`a statistiche e dinamiche'', Italy, 2003.

\newpage
\bibliographystyle{ametsoc}


\begin{table}[ht]
  \centering
  \begin{tabular}{cccc}
    $\speedT$&$2/1000$&$2/300$&$2/100$\\
    \hline
    $\lengthts$&21000&6300&2100\\
    $\lengthbl$&1000&300&100\\
    \hline
  \end{tabular}
  \caption{\baselineskip=24pt
    The length $L$ of each of the three time series and the length
    $\lengthbl$ of each of the 21 the data blocks $B_i$
    (both are expressed in years), as a function of
    the intensity $\speedT$ of the trend~\eqref{TEt}
    imposed on the parameter $T_E$ of the baroclinic model.
  }
  \label{tab:data}
\end{table}

\begin{table}
  \centering
  \begin{tabular}{c|cccccc}
    $i$&$\mube_0$&$\mube_1$&$\mube_2$&$\sibe_0$&$\sibe_1$&$\xibe$\\
    \hline
    1 & 3585.5 & 1.40 & 5.8e-04 & 21.3 & 0 & -0.19\\
    2 & 5018.9 & 1.39 & 9.0e-05 & 39.3 & 0.022 & -0.17\\
    3 & 6497.8 & 1.61 & 0 & 85.9 & 0 & -0.15\\
    4 & 8106.6 & 1.77 & 0 & 105.6 & 0.043 & -0.15\\
    5 & 9851.7 & 1.84 & 0 & 155.5 & 0.060 & -0.16\\
    6 & 11709.6 & 1.96 & 0 & 206.8 & 0.055 & -0.17\\
    7 & 13733.7 & 1.79 & 2.1e-04 & 290.2 & 0.011 & -0.16\\
    8 & 15718.9 & 2.14 & 0 & 367.4 & 0 & -0.16\\
    9 & 17935.6 & 2.15 & 0 & 451.3 & 0 & -0.18\\
    10 & 20141.2 & 2.35 & 0 & 521.6 & 0 & -0.17\\
    11 & 22527.4 & 2.25 & 0 & 597.4 & 0 & -0.16\\
    12 & 24934.5 & 2.48 & 0 & 702.2 & 0 & -0.18\\
    13 & 27452.0 & 2.58 & 0 & 774.7 & 0 & -0.14\\
    14 & 30067.3 & 2.59 & 0 & 816.2 & 0 & -0.12\\
    15 & 32696.9 & 2.85 & 0 & 951.0 & 0 & -0.14\\
    16 & 35348.0 & 3.16 & 0 & 1072.7 & 0 & -0.17\\
    17 & 38382.8 & 3.06 & 0 & 1136.1 & 0 & -0.14\\
    18 & 41348.7 & 2.85 & 0 & 1267.9 & 0 & -0.11\\
    19 & 44351.3 & 3.16 & 0 & 1498.3 & 0 & -0.18\\
    20 & 47502.7 & 3.15 & 0 & 1444.3 & 0 & -0.11\\
    21 & 51288.4 & 2.26 & 0 & 1545.9 & 0 & -0.11\\
  \end{tabular}
  \caption{\baselineskip=24pt
    Best estimate GEV fits $G_{\pbe^i,\qbe^i}(\boldsymbol{z}^i)$
    with parameter vector as in~\eqref{beta}
    for the non-stationary time series with trend intensity
    $\speedT=2/(1000\, \textrm{years})$, see text for details.
  }
  \label{tab:best1000}
\end{table}

\begin{table}
  \centering
  \begin{tabular}{c|cccccc}
    $i$&$\mube_0$&$\mube_1$&$\mube_2$&$\sibe_0$&$\sibe_1$&$\xibe$\\
    \hline
    1 & 3583.7 & 4.69 & 0 & 21.1 & 0 & -0.22\\
    2 & 4988.2 & 5.00 & 0 & 48.9 & 0 & -0.17\\
    3 & 6490.9 & 5.41 & 0 & 68.8 & 0.104 & -0.13\\
    4 & 8050.0 & 6.18 & 0 & 142.6 & 0 & -0.19\\
    5 & 9798.6 & 6.52 & 0 & 196.4 & 0 & -0.17\\
    6 & 11735.9 & 6.45 & 0 & 222.6 & 0 & -0.15\\
    7 & 13750.8 & 6.20 & 0 & 285.4 & 0 & -0.19\\
    8 & 15763.2 & 6.97 & 0 & 353.0 & 0 & -0.15\\
    9 & 17844.1 & 7.86 & 0 & 479.5 & 0 & -0.19\\
    10 & 20108.0 & 7.57 & 0 & 524.3 & 0 & -0.13\\
    11 & 22485.1 & 7.75 & 0 & 607.9 & 0 & -0.14\\
    12 & 24954.8 & 8.26 & 0 & 700.3 & 0 & -0.14\\
    13 & 27386.1 & 9.06 & 0 & 733.9 & 0 & -0.18\\
    14 & 29928.5 & 10.07 & 0 & 852.9 & 0 & -0.15\\
    15 & 32724.5 & 8.88 & 0 & 888.0 & 0.054 & -0.16\\
    16 & 35671.6 & 9.36 & 0 & 1125.4 & 0 & -0.11\\
    17 & 38284.6 & 9.72 & 0 & 1147.4 & 0 & -0.12\\
    18 & 41436.1 & 9.96 & 0 & 1294.0 & 0 & -0.18\\
    19 & 44387.2 & 10.05 & 0 & 1220.3 & 0 & -0.06\\
    20 & 47706.3 & 10.77 & 0 & 1496.0 & 0 & -0.13\\
    21 & 50890.5 & 9.09 & 0 & 1673.7 & 0 & -0.14\\
  \end{tabular}
  \caption{\baselineskip=24pt
    As in \tabref{best1000} for trend intensity
    $\speedT=2/(300\, \textrm{years})$.
  }
  \label{tab:best300}
\end{table}

\begin{table}
  \centering
  \begin{tabular}{c|cccccc}
    $i$&$\mube_0$&$\mube_1$&$\mube_2$&$\sibe_0$&$\sibe_1$&$\xibe$\\
    \hline
    1 & 3572.1 & 14.22 & 0 & 21.0 & 0 & -0.25\\
    2 & 4986.8 & 14.91 & 0 & 49.6 & 0 & -0.27\\
    3 & 6485.6 & 16.40 & 0 & 67.7 & 0 & -0.12\\
    4 & 8118.5 & 17.11 & 0 & 123.1 & 0 & -0.14\\
    5 & 9905.1 & 17.29 & 0 & 144.9 & 0 & -0.07\\
    6 & 11679.7 & 19.66 & 0 & 261.3 & 0 & -0.15\\
    7 & 13628.2 & 21.49 & 0 & 309.9 & 0 & -0.19\\
    8 & 15542.3 & 24.39 & 0 & 354.5 & 0 & -0.13\\
    9 & 17757.6 & 25.41 & 0 & 484.3 & 0 & -0.22\\
    10 & 20246.2 & 19.26 & 0 & 503.4 & 0 & -0.20\\
    11 & 22556.3 & 23.61 & 0 & 533.9 & 0 & -0.07\\
    12 & 24848.3 & 25.25 & 0 & 632.3 & 0 & -0.24\\
    13 & 27441.5 & 27.18 & 0 & 780.3 & 0 & -0.27\\
    14 & 29638.6 & 35.68 & 0 & 791.0 & 0 & -0.03\\
    15 & 32617.7 & 28.06 & 0 & 1010.6 & 0 & -0.12\\
    16 & 35813.0 & 21.47 & 0 & 1046.2 & 0 & -0.12\\
    17 & 38422.4 & 28.22 & 0 & 1055.2 & 0 & -0.15\\
    18 & 41119.7 & 31.07 & 0 & 1238.6 & 0 & -0.18\\
    19 & 44510.3 & 27.28 & 0 & 1351.2 & 0 & -0.14\\
    20 & 47640.6 & 25.50 & 0 & 1548.8 & 0 & -0.12\\
    21 & 50454.4 & 28.14 & 0 & 1769.8 & 0 & -0.16\\
  \end{tabular}
  \caption{\baselineskip=24pt
    As in \tabref{best1000} for trend intensity
    $\speedT=2/(100\, \textrm{years})$.
  }
  \label{tab:best100}
\end{table}

\begin{table}[!h]
  \centering
  \begin{tabular}{|c||c|c|c|}
    \hline
    $\lengthbl$& $\gamma_{\mu,1}$& $T_E^{b}$& $\gamma_{\mu,2}$\\
    \hline \hline
    1000 & $1.5200 \pm 0.0041$ & 16& $1.5434 \pm 0.0042$\\
    300  & $1.5706 \pm 0.0091$ & 16& $1.5452 \pm 0.0072$\\
    100  & $1.5733 \pm 0.0125$ & 16& $1.5419 \pm 0.0129$\\
    \hline
  \end{tabular}
  \caption{\baselineskip=24pt
    Power law fits of the inferred location parameter $\mube^i(T_E^i)$
    as a function of $T_E^i$ (see~\eqref{TEi})
    of the form $\mube^i(T_E^i) =\alpha_\mu  (T_E^i)^{\gamma_\mu}$.
    Two distinct scaling regimes (with distinct exponents $\gamma_{\mu,1}$
    and $\gamma_{\mu,2}$) are identified and the corresponding
    adjacent intervals in the $T_E$-axis are separated by $T_E^{b}$.
  }
  \label{tab:powermuT}
\end{table}

\begin{table}[!h]
  \centering
  \begin{tabular}{|c||c|c|c|}
    \hline
    $\lengthbl$& $\gamma_{\sigma,1}$& $T_E^{b}$& $\gamma_{\sigma,2}$\\
    \hline \hline
    1000 & $3.5212 \pm 0.1600$ & 18& $2.1000 \pm 0.0725$\\
    300  & $3.9180 \pm 0.3142$ & 16& $2.1055 \pm 0.0580$\\
    100  & $3.2351 \pm 0.2154$ & 22& $2.0891 \pm 0.1797$\\
    \hline
  \end{tabular}
  \caption{\baselineskip=24pt
    Same as \tabref{powermuT} for the inferred
    scale parameter $\sibe^i(T_E^i)$.
  }
  \label{tab:powersigmaT}
\end{table}

\end{document}